\documentclass[numberedappendix]{emulateapj}
\usepackage{amssymb}
\usepackage{amsmath}
\newcommand{\code}[1]{{\tt #1}}
\newcommand{\msol}{\,\textrm{M}_\sun}                
\newcommand{\Chandra}{\emph{Chandra}}
\newcommand{\HST}{\emph{HST}}
\newcommand{\SExtractor}{\code{SExtractor}}
\newcommand{\betatot}{\gamma_{\textrm{tot}}}
\newcommand{\MLV}{\Upsilon_{*\textrm{V}}}

\newcommand{\MLVSPS}{\Upsilon_{*\textrm{V}}^{\textrm{SPS}}}

\newcommand{\ML}{\Upsilon_*}
\newcommand{\LV}{L_{\textrm{V}}}
\newcommand{\PaperII}{Paper~II}

\shorttitle{The Density Profiles of Galaxy Clusters. I. The Total Density Over Three Decades in Radius}
\shortauthors{Newman et al.}
\submitted{}
\begin{document}

\title{The Density Profiles of Massive, Relaxed Galaxy Clusters.\\
I. The Total Density Over Three Decades in Radius}
\author{Andrew B. Newman\altaffilmark{1},  Tommaso Treu\altaffilmark{2},
Richard S. Ellis\altaffilmark{1}, David J. Sand\altaffilmark{2,3},
Carlo Nipoti\altaffilmark{4}, Johan Richard\altaffilmark{5},
and Eric Jullo\altaffilmark{6}}

\altaffiltext{1}{Cahill Center for Astronomy and Astrophysics, California
Institute of Technology, MS 249-17, Pasadena, CA 91125, USA; anewman@astro.caltech.edu}
\altaffiltext{2}{Department of Physics, University of California, Santa Barbara, CA 93106, USA}
\altaffiltext{3}{Las Cumbres Observatory Global Telescope Network, Santa Barbara, CA 93117, USA}
\altaffiltext{4}{Astronomy Department, University of Bologna, via Ranzani 1, I-40127 Bologna, Italy}
\altaffiltext{5}{CRAL, Observatorie de Lyon, Universit\'{e} Lyon 1, 9 Avenue Ch.~Andr\'{e}, F-69561 Saint Genis Laval Cedex, France}
\altaffiltext{6}{Laboratoire d'Astrophysique de Marseille, Universit\'{e} d'Aix-Marseille and CNRS, UMR7326, 38 rue F.~Joliot-Curie, F-13388 Marseille Cedex 13, France}

\begin{abstract}
Clusters of galaxies are excellent locations to probe the distribution of baryons and dark matter (DM) over a wide range of scales. We study a sample of seven massive ($M_{200} = 0.4 - 2 \times 10^{15} \msol$), relaxed galaxy clusters with centrally-located brightest cluster galaxies (BCGs) at $z = 0.2 - 0.3$. Using the observational tools of strong and weak gravitational lensing, combined with resolved stellar kinematics within the BCG, we measure the total radial density profile, comprising both dark and baryonic matter, over scales of $\simeq 3 - 3000$~kpc. We present Keck spectroscopy yielding seven new spectroscopic redshifts of multiply imaged sources and extended stellar velocity dispersion profiles of the BCGs. Lensing-derived mass profiles typically agree with independent X-ray estimates within $\simeq 15\%$, suggesting that departures from hydrostatic equilibrium are small and that the clusters in our sample (except A383) are not strongly elongated or compressed along the line of sight. The inner logarithmic slope $\betatot$ of the total density profile measured over $r/r_{200} = 0.003-0.03$, where $\rho_{\textrm{tot}} \propto r^{-\betatot}$, is found to be nearly universal, with a mean $\langle\betatot\rangle = 1.16 \pm 0.05~{\rm (random)}~{}^{+0.05}_{-0.07}$~(systematic) and an intrinsic scatter $\sigma_{\gamma} < 0.13$ (68\% confidence). This is further supported by the very homogeneous shape of the observed velocity dispersion profiles, which are mutually consistent after a simple scaling. Remarkably, this slope agrees closely with high-resolution numerical simulations that contain only DM, despite the significant contribution of stellar mass on the scales we probe. The Navarro--Frenk--White profile characteristic of collisionless cold DM is a better description of the \emph{total} mass density at radii $\gtrsim 5-10$~kpc than that of DM alone. Hydrodynamical simulations that include baryons, cooling, and feedback currently provide a poorer match. We discuss the significance of our findings for understanding the physical processes governing the assembly of BCGs and cluster cores, particularly the influence of baryons on the inner DM halo.
\end{abstract} 

\section{Introduction}
\label{sec:intro}

In a cold dark matter (CDM) universe, dark matter (DM) halos are expected to be nearly self-similar, and their detailed structure can be followed in large numerical simulations based only on gravity \citep[e.g.,][]{Dubinski91,NFW96,Moore98,Ghigna00,Springel05,Diemand05,Navarro10,Gao12}. A key result of cold, collisionless gravitational collapse is the formation of a central density cusp with a characteristic profile $\rho_{\textrm{DM}} \propto r^{-1}$. At large radii the density falls as $\rho_{\textrm{DM}} \propto r^{-3}$. These slopes are characteristic of the Navarro--Frenk--White (NFW) profile, which provides a reasonable description of results from $N$-body simulations. With improved resolution, recent simulations have elucidated deviations from this simple functional form \citep[e.g.,][]{Merritt06,Navarro10,Gao12}, showing that halo profiles are not strictly self-similar and that the density slope likely becomes slightly shallower at very small radii.

Real halos also contain baryons that may significantly modify the structure of the DM. Cooling allows baryons to condense toward the center, which makes the DM more concentrated \citep{Blumenthal86,Gnedin04,Sellwood05,Gustafsson06,Pedrosa09,Abadi10,SommerLarsen10}. Additional baryonic effects have been proposed to reduce the central concentration, even producing DM cores. These include heating of the central cusp via dynamical friction with infalling satellites \citep[e.g.,][]{ElZant01,ElZant04,Nipoti04,RomanoDiaz08,Jardel09,Johansson09,DelPopolo12}, feedback from supernovae in low-mass galaxies \citep{Governato10,Pontzen12,Brooks12}, and active galactic nucleus (AGN) feedback in clusters of galaxies \citep{Peirani08,Martizzi12}. Much effort have been devoted to understanding the net result of these competing effects on halos using comprehensive hydrodynamical simulations over a range of mass scales \citep[e.g.,][]{Duffy10,Gnedin11}, but due to the difficulty of realistically treating all the relevant physics, predictions for halos with baryons remain unclear.

Understanding the relative distribution of dark and baryonic matter is important for several reasons. If CDM is an accurate description, then the structure of real halos can inform us about the assembly of galaxies, groups, and clusters through the imprint of baryons on their halos. For instance, dark and baryonic density profiles can inform us about the relative importance of dissipational and dissipationless assembly processes \citep[e.g.,][]{Lackner10}. The observation from that massive ellipticals have nearly isothermal total mass profiles within their effective radii -- with very little scatter -- is a strong constraint on their formation and evolution \citep{Koopmans09}. On the other hand, the structure of halos may constrain DM particle scenarios in which the inner halo is distinct from CDM \citep[e.g.,][]{Spergel00,Dave01,KuziodeNaray10,Maccio12}, if the baryonic effects can be better understood. Central densities are also relevant for indirect DM searches, since the rate of gamma ray production from annihilation scales as $\rho_{\textrm{DM}}^2$.

Determining precise and robust mass profiles is challenging, particularly if the goal is to separate the dark and baryonic components. Low surface brightness and dwarf spheroidal galaxies are often considered ideal targets for DM studies, since the mass fraction of baryons is minimal, and observations indicate that many of these galaxies have a DM core rather than the expected cusp \citep[e.g.,][]{Simon05,deBlok08,Wolf12}. Due to their shallow potential wells, however, these are fragile systems and may be disrupted by supernovae (see references above).

Galaxy clusters are also promising systems for detailed study of of mass distributions. 
Owing to the wide range of observational tools that can be brought to bear, the mass in individual clusters can be measured in detail over a very wide range of scales. Clusters are DM-dominated outside of the very central regions and are the only systems that can be individually mapped to their virial radius, using weak gravitational lensing. In selected clusters, strong lensing provides exquisite mass measurements that are independent of the dynamical state. X-ray emission from the hot intracluster medium (ICM) can also be used to derive mass profiles under the assumption of hydrostatic equilibrium. Each of these tools is valid over a specific radial interval. Weak lensing cannot reach within $\sim100$~kpc. The strong lensing zone is usually confined to roughly $30-150$~kpc (partly due to the difficulty in locating central images superposed on cluster galaxies).  X-ray emission is difficult to interpret within $\simeq 50$~kpc due to gas cooling and substructure, while temperature measurements become prohibitive at $\gtrsim 700$~kpc. Therefore, combining several mass probes is necessary to derive comprehensive constraints.

X-ray and lensing studies have shown that NFW profiles can generally provide adequate descriptions of cluster halos at radii $r \gtrsim 50$~kpc \citep[e.g.,][]{Kneib03,Broadhurst05,Mandelbaum06,Schmidt07,Okabe10,Umetsu11,Coe12,Morandi12}. Several studies have questioned whether the relationship between halo mass and concentration, derived based on NFW models, follows that in simulations. Many lensing clusters have surprisingly high concentrations \citep[e.g.,][]{Kneib03,Broadhurst08,Zitrin11}. Interpreting this requires careful study of possible measurement biases or selection effects \citep{Hennawi07,Meneghetti10b}. Measuring the \emph{shape} of the radial density profile to test whether the NFW form (or the result of numerical simulations generally) is valid over the full range of scales -- for any mass and concentration -- is more challenging, but possibly more profound. The tools mentioned so far cannot test for deviations from an NFW profile in the inner halo with much statistical power, even when multiple clusters are stacked \citep[e.g.,][]{Schmidt07,Umetsu11}, except possibly in rare lensing configurations \citep[e.g.,][]{Limousin07}. More constraints on smaller scales are necessary to provide a lever arm long enough to measure the inner density slope and probe the innermost decade in radius now resolved in the best simulations. Here the stellar mass in the brightest cluster galaxy (BCG) is significant, and there has often been confusion in the literature about whether the \emph{total} density or only that of the \emph{dark matter} is being reported and compared to simulations.

In relaxed clusters hosting a BCG that is closely aligned with the center of the halo, the kinematics of the stars trace the total gravitational potential \citep{MiraldaEscude95,Natarajan96}. Spectroscopy using $8-10$~m telescopes can reach from the stellar-dominated regime to the regime where DM is dynamically significant, even at the cosmological distances of lensing clusters ($z \gtrsim 0.2$). \citet{Sand02,Sand04} showed that by combining strong lensing with stellar kinematics, the contribution of the stellar mass can be constrained, which allows the DM halo to be isolated and its inner slope measured. Particularly strong results were obtained in clusters presenting radial arcs. In five of the six clusters they studied, the inner logarithmic density slope $\beta = - d \log \rho_{\textrm{DM}} / d \log r$ was found to be $\beta < 1$, shallower than a standard NFW profile. \citet{Sand08} improved the analysis in two clusters by moving beyond axisymmetric lens models, which had been suggested as a source of systematic bias \citep{Meneghetti07}, and found similar results on the DM slope. \citet[][N09]{N09} additionally incorporated weak lensing constraints in A611, providing the first cluster mass profile over three decades in radius. \citet[][N11]{N11} presented very extended stellar kinematics in A383 and additionally used X-ray observations to assess the non-spherical geometry of the cluster along the line of sight (l.o.s.). In both A611 and A383 we confirmed a shallow inner density slope $\beta < 1$ for the DM.

\begin{deluxetable*}{lccccccccc}[th!]
\tablewidth{\linewidth}
\tablecolumns{10}
\tablecaption{Cluster Sample and Alignment between BCG and Mass Centers}
\tablehead{\colhead{} & \colhead{} & \colhead{} & \colhead{} & \colhead{} & \multicolumn{3}{c}{BCG Offset (kpc) from} & \colhead{} & \colhead{} \\
\colhead{} & \colhead{} & \colhead{} & \colhead{BCG Peculiar} & \colhead{Source of} & \colhead{X-Ray} & \multicolumn{2}{c}{Lensing Center} & \colhead{Cool} & \colhead{$L_{\textrm{X}}$} \\
\colhead{Name} & \colhead{$\langle z \rangle$} & \colhead{$N_{\textrm{gal}}$} & \colhead{Velocity (km~s${}^{-1}$)} & \colhead{Galaxy Redshifts} & \colhead{Centroid} & \colhead{$\Delta x$} & \colhead{$\Delta y$} & \colhead{Core?} & \colhead{$(10^{37}$ W)}}
\startdata
MS2137.3-2353 &   0.314 & \ldots   & \ldots             & \ldots                & $4^\dagger$ & $1.2 \pm 0.8$ & $0.1 \pm 0.6$ & Yes${}^a$ & 11.10 \\
A963          &   0.206 & \ldots   & \ldots             & \ldots                & 6   & \ldots & \ldots & No${}^b$  & 5.03 \\
A383          &   0.190 & $26$     & $-261 \pm 187$     & This work             & 2   & $-2.7 \pm 0.6$ & $2.9 \pm 1.1$ & Yes${}^b$ & 4.12  \\
A611          &   0.288 & $236$    & $-67 \pm 68$       & This work             & 1   & $-1.3 \pm 0.9$ & $4.2 \pm 0.8$ & No${}^b$  & 5.33  \\
A2537         &   0.294 & $273$    & $-325 \pm 311$     & \citet{Braglia09}     & $13^\dagger$ & $-0.4 \pm 1.2$ & $5.2 \pm 1.5$ & No${}^c$ & 9.37 \\
              &         &            &                  & and this work         &      &  &  &  & \\
A2667         &   0.233 & $22$     & $438 \pm 730$      & \citet{Covone06}      & 3   & $-6.5 \pm 3.6$ & $4.1 \pm 2.9$ & Yes${}^b$ & 11.97 \\
A2390         &   0.229 & $52$     & $270 \pm 218$      & \citet{Yee96}         & 2   & $4.9 \pm 6.7$ & $-0.2 \pm 3.5$ & Yes${}^b$ & 14.81
\enddata
\tablecomments{Redshifts $\langle z \rangle$ are the biweight mean of $N_{\textrm{gal}}$ cluster galaxies identified with an iterative $2.5\sigma$ clip applied. The uncertainty is provided on the peculiar velocity of the BCG [$v_{\textrm{BCG}} = c(z_{\textrm{clus}}-z_{\textrm{BCG}})/(1+z_{\textrm{clus}})$]. For A2537 the redshift and $v_{\textrm{BCG}}$ are given relative to the main peak (Figure \ref{fig:velocities}). Where no redshift survey is available, the redshift of the BCG is given instead. Offsets between the BCG and the X-ray centroid measured in the central 1~arcmin are from \citet{Sanderson09a} and \citet{Richard10}, except those marked ${}^{\dagger}$ which are original to this work. Offsets between the BCG and lensing center are discussed in Section~\ref{sec:alignment}; $\Delta x > 0$ and $\Delta y > 0$ denote offsets west and north of the BCG. $L_{\textrm{X}}$ is the X-ray luminosity in the $0.1-2.4$~keV band within $R_{500}$ from \citet{Piffaretti11}. Sources of cool core classification: ${}^a$ \citet{Donnarumma09}, ${}^b$ \citet{Richard10}, ${}^c$ \citet{Rossetti11}. \label{tab:sample}}
\end{deluxetable*}

In this paper we present strong lensing, weak lensing, and resolved stellar kinematic observations for a sample of seven massive, relaxed galaxy clusters. The clusters span the redshift range $z = 0.2 - 0.3$ and have virial masses $M_{200} = 0.4 - 2 \times 10^{15} \msol$. Taken together, these data span scales of $\simeq 3 - 3000$~kpc, which is well matched to the dynamic range achieved in modern $N$-body simulations of clusters. We use these data to constrain the \emph{total density distribution} over three decades in radius, providing a benchmark for high-resolution simulations.
We focus on the shape of total density profile in this paper and show that it is in surprising agreement with numerical simulations that contain only DM. In \PaperII~of the series, we consider the DM and stellar mass profiles separately. 

The plan of the paper follows. In Section~2 we introduce the cluster sample and describe its characteristics. In Section 3, technical aspects of the weak lensing analysis, based primarily on Subaru~imaging, are presented along with shear profiles and two-dimensional (2D) mass maps. Section 4 describes our strong lensing interpretations, including seven new spectroscopic redshifts of multiply imaged sources. In Section 5 we present \emph{Hubble Space Telescope} (\HST) surface photometry of the BCGs and stellar population synthesis (SPS) models. Spectroscopy of the BCGs and the derived kinematic measures are discussed in Section 6. In Section 7, we present the mathematical framework used to derive our mass profiles. Section 8 compares our lensing-derived mass profiles to independent X-ray measures, in order to assess the possible influence of projection effects on our results. Finally, in Section 9 we present the total mass profiles derived for the full sample, focusing particularly on the total inner slope, and in Section 10 we discuss our results in the context of recent simulations. Section 11 summarizes our findings. Readers interested only in the results and not the technical aspects may wish to begin in Section 8.

Throughout we adopt a $\Lambda$CDM cosmology with $\Omega_m = 0.3$, $\Omega_{\Lambda} = 0.7$, and $H_0 = 70$~km~s${}^{-1}$~Mpc${}^{-1}$. At $z = 0.25$, $1'' = 3.91$~kpc. Magnitudes are reported in the AB system.

\section{The Cluster Sample}
\label{sec:thesample}

\begin{figure*}
\centering
\includegraphics[width=0.45\linewidth]{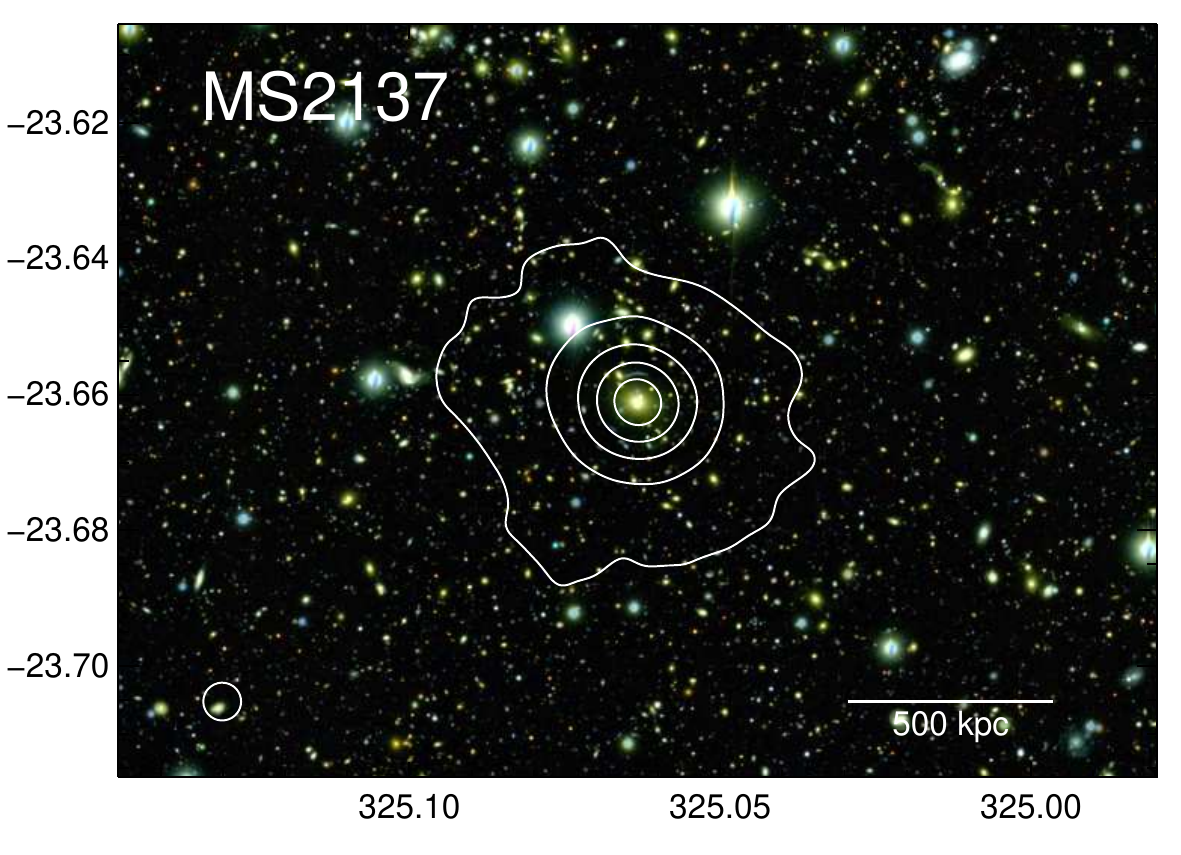}
\includegraphics[width=0.45\linewidth]{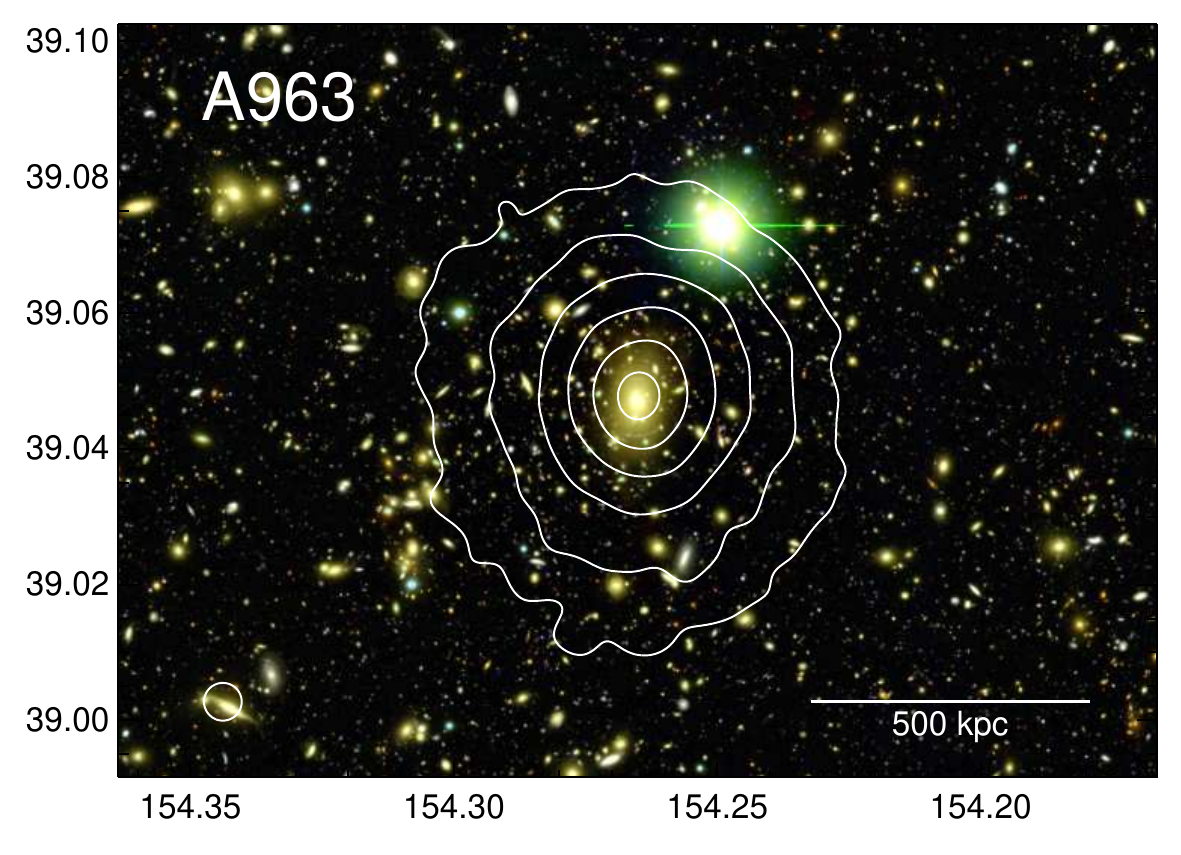} \\
\includegraphics[width=0.45\linewidth]{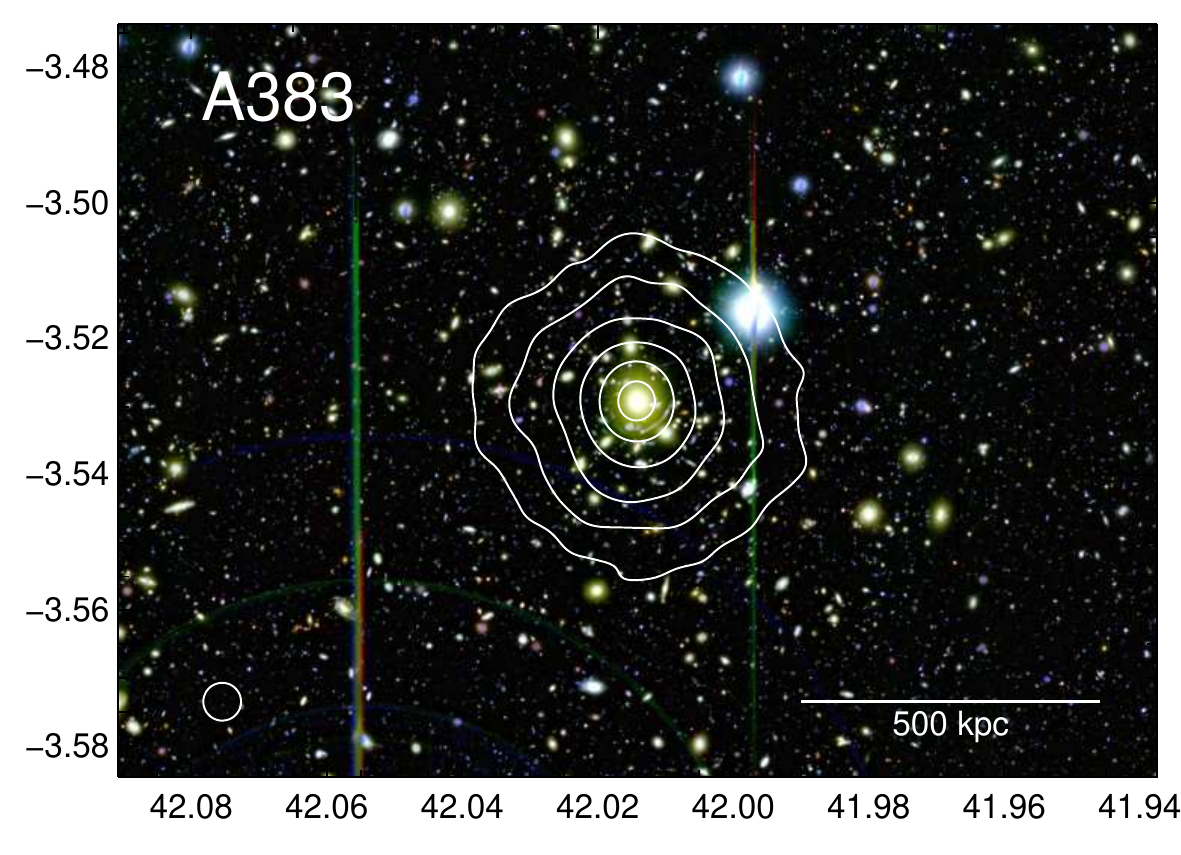} 
\includegraphics[width=0.45\linewidth]{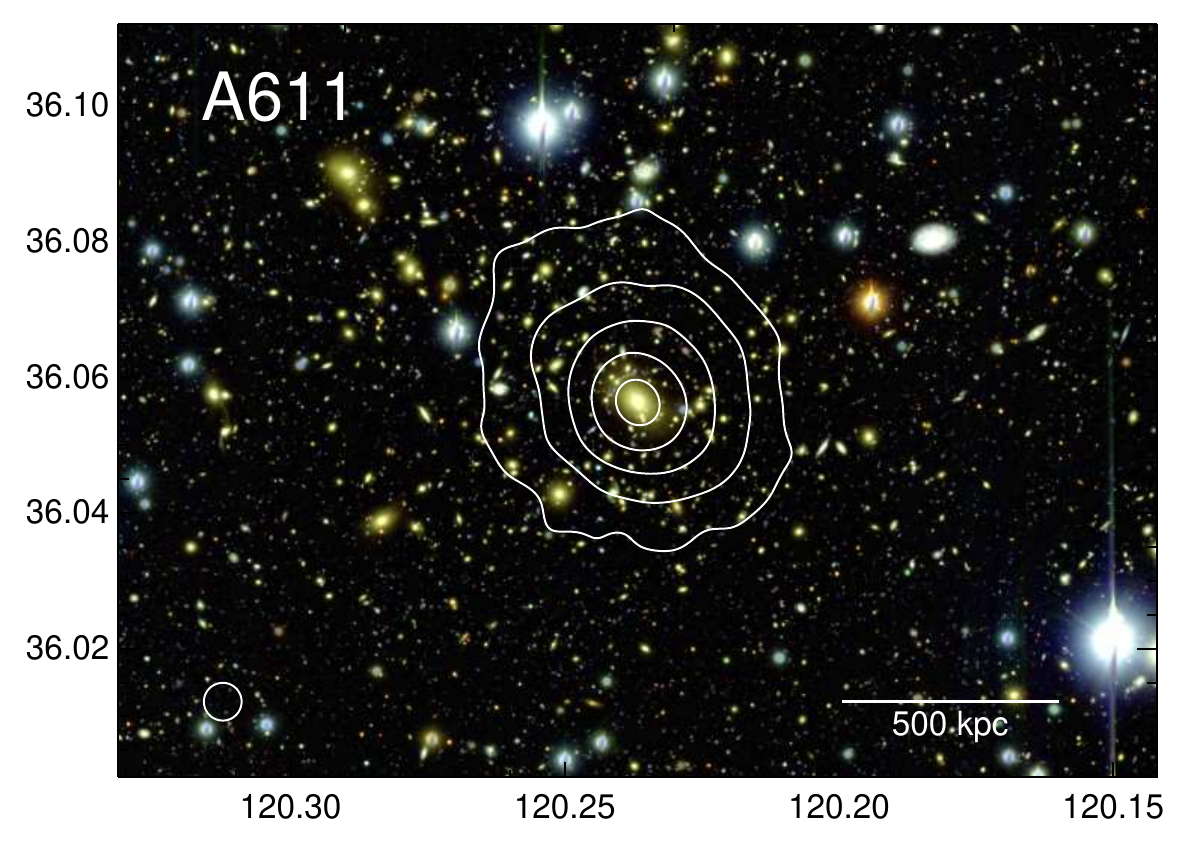} \\
\includegraphics[width=0.45\linewidth]{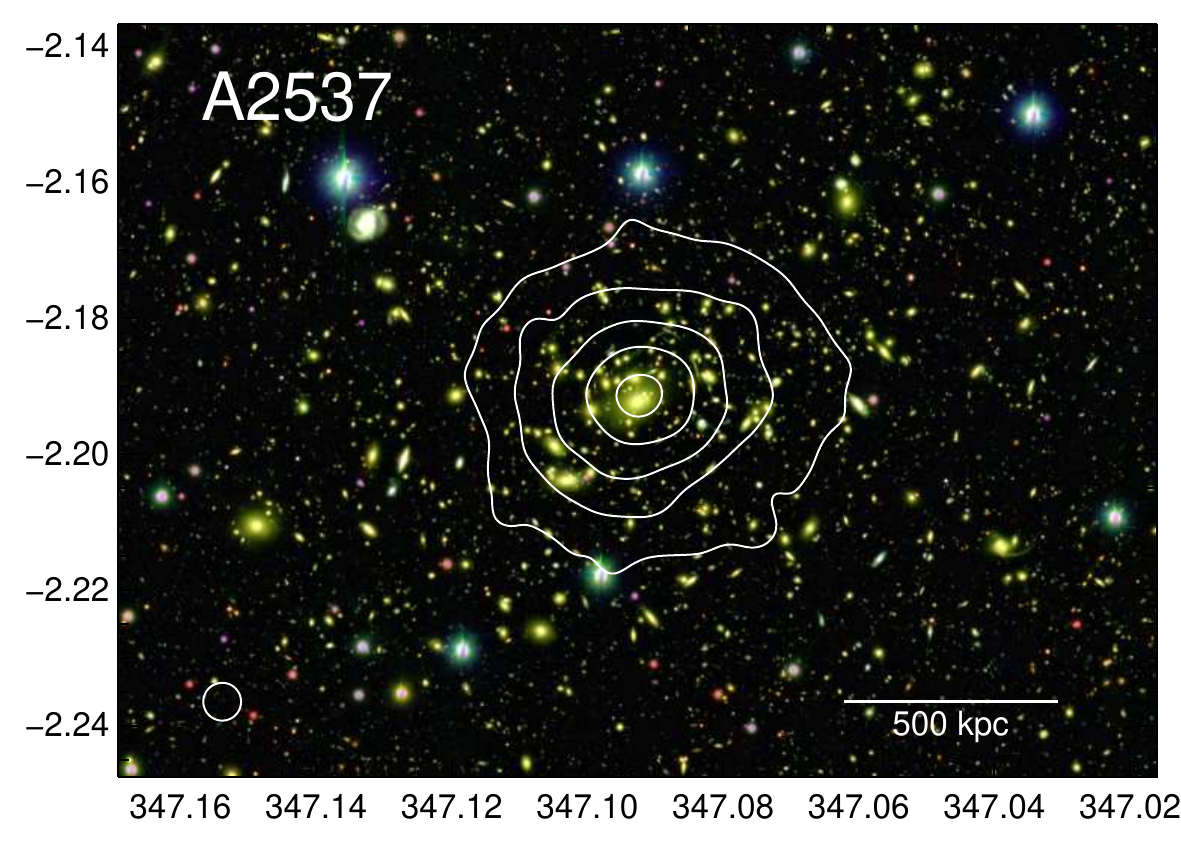}
\includegraphics[width=0.45\linewidth]{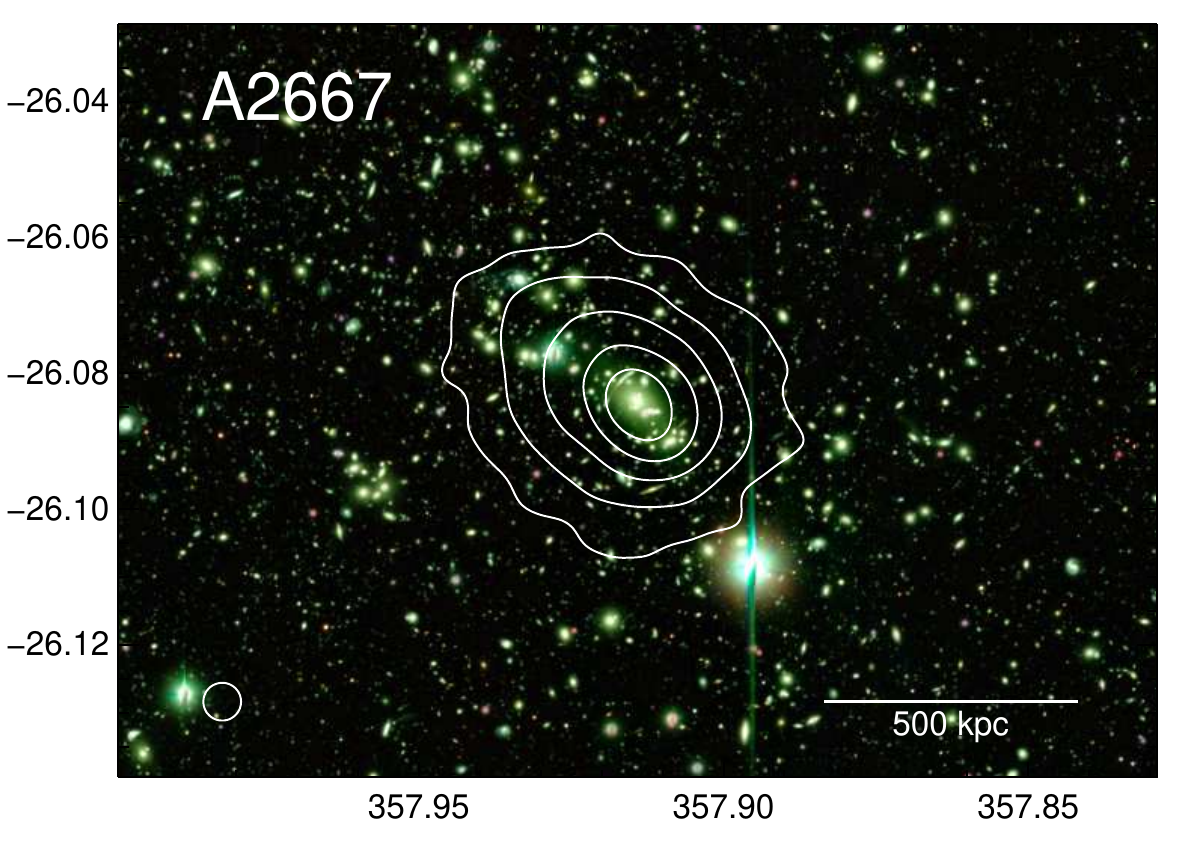} \\
\includegraphics[width=0.45\linewidth]{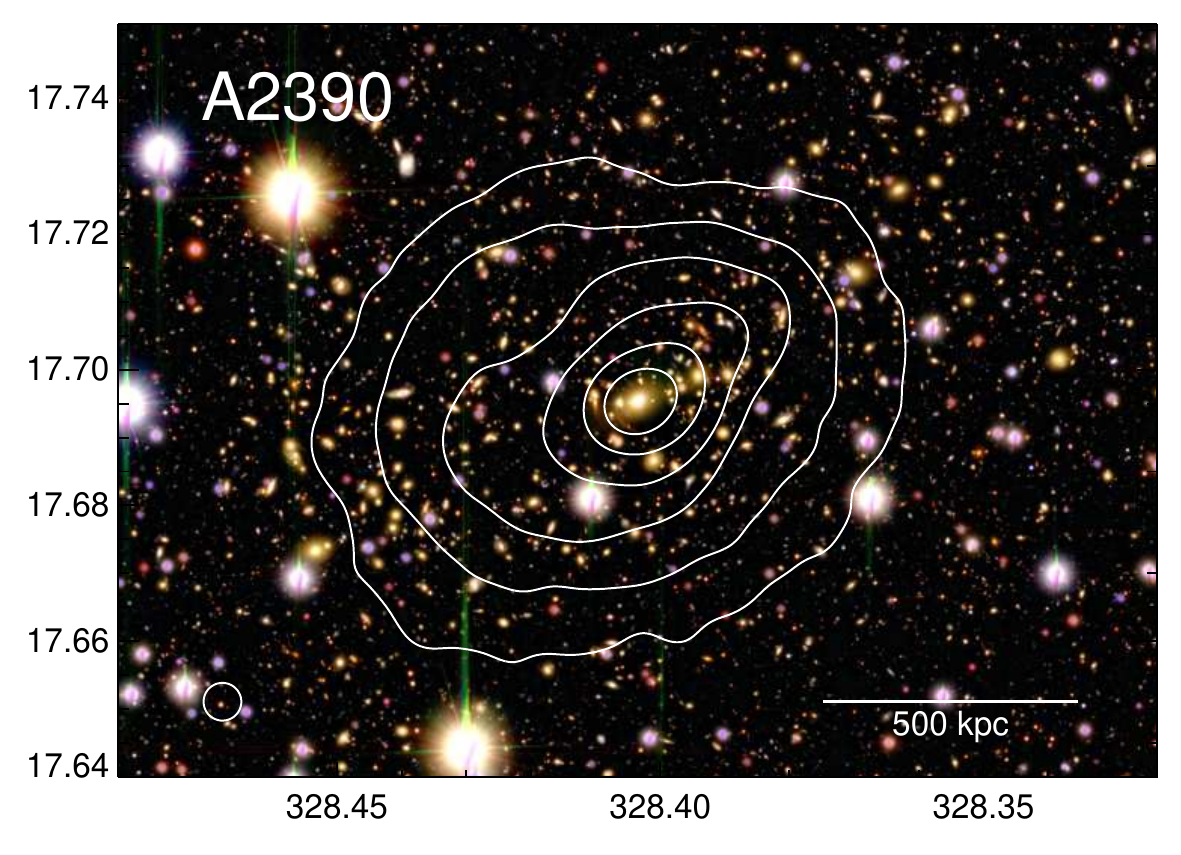}
\caption{Color composites of the central regions of each cluster based on the imaging data introduced in Section~\ref{sec:wlredux} are displayed with an arcsinh stretch \citep{Lupton04}. Only a small portion of the total field of view is shown. The \Chandra~X-ray emission in the $0.8-7$~keV band is overlaid, smoothed with a Gaussian kernel whose size (FWHM of $20\arcsec$) is indicated in the lower left of each panel. Contour levels are equally spaced logarithmically but are otherwise arbitrary. Axes show the R.A.~and declination.\label{fig:opticalXray}}
\end{figure*}

\begin{figure}
\centering
\includegraphics[width=0.9\linewidth]{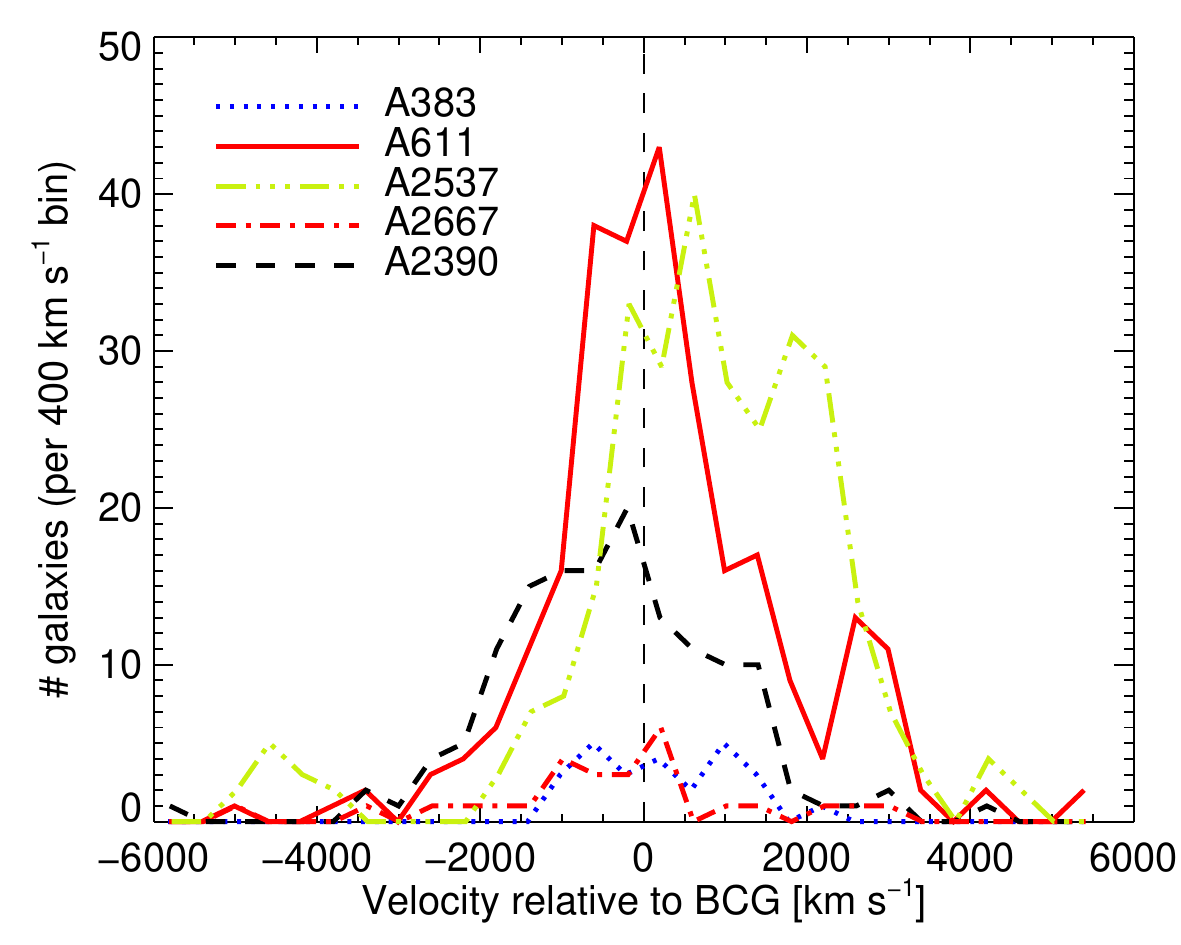}
\caption{Histogram of velocities of cluster galaxies relative to the BCG, $\Delta v =c (z-z_{\textrm{BCG}})/(1+z_{\textrm{BCG}})$, based on the sources listed in Table~\ref{tab:sample}. The available data are consistent with the BCGs being at rest in the cluster potentials. A2537 has a bimodal velocity structure: the BCG coincides with the primary peak, but there is a second peak at $\Delta v \simeq 2000$~km~s${}^{-1}$ as discussed in the text.
\label{fig:velocities}}
\end{figure}

Our goal is to fit simple parametric models to lensing and kinematic data on scales ranging from $\simeq 3$ to $3000$~kpc and to compare our results to simulations. This requires selecting a sample of clusters that are reasonably relaxed and symmetric, both to ensure that our models are adequate and to make clean comparisons with theory. Furthermore, our use of stellar kinematics to trace the mass distribution on small scales requires that the centers of the BCG and DM halo are well aligned. Table~\ref{tab:sample} introduces the sample of seven massive clusters, which range in redshift from $z=0.19$ to 0.31. As we describe below, A611, A383, MS2137, A963, and A2667 are well relaxed clusters, A2390 is likely only slightly perturbed, and A2537 shows signs of a more complex mass distribution.

Optical images of the central $\simeq 1$~Mpc of each cluster are shown in Figure~\ref{fig:opticalXray} with X-ray contours overlaid. The X-ray data were obtained from the \Chandra~archive,\footnote{Observation IDs 3194, 2320, 4974, 903, 2214, 4962, 9372, and 4193.} and point sources were removed using the \code{CIAO} tools. We first discuss A611, A383, MS2137, A963, and A2667, which are prototypically relaxed clusters, and reserve A2390 and A2537 for individual comments below. The X-ray emission in these five clusters is regular, symmetric, and well aligned with the BCG, and is extended along the same directions as both the BCG and the surface density in our lens models. The alignment is quantified in Table~\ref{tab:sample}, which shows that the X-ray centroid is typically within a few kpc of the BCG, comparable to the measurement uncertainty (A.~Sanderson, private communication, 2012). Similar small offsets between the BCG and center of mass are derived from lens models, which we discuss further in Section~\ref{sec:alignment}.

It is unlikely that we have simply selected clusters in which the BCG is offset primarily along the l.o.s., given that these clusters exhibit many characteristics that are known to be correlated with a relaxed state and a centrally-located BCG: a large luminosity gap between the BCG and the second rank galaxy, a low substructure fraction, and the presence of a cooling core \citep{Sanderson09a,Smith10,Richard10}. 
Furthermore, the available redshifts in the fields of A383, A611, and A2667 (see sources Table~\ref{tab:sample}) are consistent with a unimodal velocity distribution in which the BCG is at rest in the cluster potential, as shown in Figure~\ref{fig:velocities}.

A2390 shows slightly more complicated X-ray emission that is characterized primarily by a low-level extension to the northwest on $\sim200$~kpc scales, in the same location as an enhancement of cluster galaxies. The extension has long been noted \citep{Kassiola92,Pierre96,Frye98}. As we discuss in Section~\ref{sec:additionalcomponents}, our strong lensing model does not demand a major additional mass concentration in this region, provided an elliptical halo is used. Further, the X-ray and galaxy distributions are regular on larger scales, the BCG is well aligned with the X-ray and lensing centers (Table~\ref{tab:sample}), the velocity distribution of cluster galaxies is unimodal and centered on the BCG (Figure~\ref{fig:velocities}), and there is a strong cooling core \citep{Richard10}. From this we infer that A2390 is likely to be only mildly unrelaxed.

Finally, we consider A2537, which is the most likely disturbed cluster in our sample. The X-ray emission is regular and symmetric, but centered slightly north of the BCG (13 kpc). There is no cool core \citep{Rossetti11}. The curvature of the arcs suggests that a second mass concentration may be present to north (Section~\ref{sec:additionalcomponents}). Crucially, the distribution of cluster galaxy velocities appears bimodal (Figure~\ref{fig:velocities}), with the main peak centered on the BCG and a second peak at $\Delta v \simeq 2000$~km~s${}^{-1}$. Galaxies in the high-velocity tail do not appear spatially distinct from the remainder. It is possible A2537 has not fully relaxed from a merger near the l.o.s.~(perhaps similarly to Cl0024+1654; \citealt{Czoske02}). Throughout, we bear in mind the uncertain dynamical state of this cluster when interpreting our results.

\section{Weak lensing\label{sec:wl}}

We begin our discussion of the data forming the basis of our analysis on the largest scales. These are probed by weak gravitational shear, the systematic distortion in the shapes of background sources by the cluster. Weak-lensing analyses present a number of technical challenges. Proper handling of the point-spread function (PSF) of the instrument used for the observations is essential, since it induces spurious shear of comparable magnitude to the real signal and varies across the focal plane. Additionally, galaxies located behind the cluster must be isolated in order to avoid dilution of the shear signal by unlensed cluster galaxies and those in the foreground: this requires multi-color photometry. In Section~\ref{sec:wldata} we introduce the imaging data, primarily from the Subaru telescope, and its reduction. In Section~\ref{sec:shear} we briefly describe our technique for extracting the shear signal, which was discussed more extensively in N09, and verify our method using simulated data. Section~\ref{sec:photz} describes the photometric redshift measurements used to select background sources and tests of their validity. Finally, in Section~\ref{sec:wlresults} we present 2D mass maps and tangential shear profiles.

\subsection{Data reduction and catalog construction\label{sec:wlredux}}
\label{sec:wldata}

The imaging data used in our weak-lensing analysis are listed in Table~\ref{tab:wlobs}. Most observations were conducted with SuprimeCam \citep{Miyazaki02} at the Subaru telescope, either by the authors or using archival data. Its $30'$ field of view is well matched to our sample. In a few cases, additional color information is provided from our own observations at the Magellan Observatory or via archival data from the Canada--France--Hawaii telescope (CFHT).

The data were reduced following the procedures described in N09 that use the \code{IMCAT}\footnote{\url{http://www.ifa.hawaii.edu/$\sim$kaiser/imcat/}}-based pipeline developed by \citet{Donovan} and \citet{Capak07}. In particular, we note that the sky subtraction scheme described in these works is effective at removing small-scale structure from scattered light. Halos around bright stars were carefully masked. All filters observed for a given cluster were reduced onto a stereographic projection with a common tangent point and pixel scale of $0\farcs2$. Absolute astrometry was tied to the USNO-B \citep{USNOB} or Sloan Digital Sky Survey (SDSS) DR7 or DR8 \citep{SDSS7} catalogs. The frame-to-frame scatter in the final positions of bright stars was typically $3-5$~mas per coordinate. Object detection and shape measurements were conducted in the $R$-band image ($I$ band in A963) in the native seeing.
We used \SExtractor~\citep{SExtractor} for detection, adopting a low threshold ({\tt DETECT\_THRESH} $ = 0.75$, {\tt DETECT\_MINAREA} $ = 9$); further selection criteria are described in Section \ref{sec:shear}. Colors were measured in $2\arcsec$ apertures by running \SExtractor~in dual-image mode on PSF-matched mosaics.

For all clusters except A2667 and MS2137, photometric zeropoints were determined through comparison with stellar photometry in the SDSS. This has the merit of uniform and accurate calibration when including archival data for which conventional standard star images may not be available and observing conditions are uncertain. Galactic extinction was then corrected using the \citet{Schlegel98} dust maps. Transformations of stellar colors from the SDSS to the SuprimeCam filter system were taken from \citet{Capak07}, \citet{Yagi10}, and \citet{Shim06} where possible. For the remaining filters, 
transformations were derived from fits to synthetic photometry of stars in the \citet{Pickles85} spectrophotometric library, based on filter and instrument response curves provided by the observatories. We verified that this method yields transformation equations consistent with the empirical equations referenced above, and also with zeropoints derived from a Landolt standard field (N09) within a few percent. $BVRIz$ photometry in MS2137, which is outside the SDSS DR8 footprint, was calibrated through alignment with the stellar locus in A611 and A2390, taking advantage of the feature in the $VRI$ color-color diagram \citep[e.g.,][]{High09}. Zeropoints for A2667 were taken from observations of other clusters on the same night, with small shifts applied based on the stellar locus. Below we evaluate the accuracy of this calibration based on the derived photometric redshifts.

\begin{deluxetable*}{llllccccc}
\tablecolumns{9}
\tablewidth{\linewidth}
\tablecaption{Imaging Observations for Weak-lensing Analysis\label{tab:wlobs}}
\tablehead{\colhead{Cluster} &
\colhead{Instrument} &
\colhead{Filter} &
\colhead{Dates of Observation} &
\colhead{Exposure} &
\colhead{Seeing} &
\colhead{Depth} & 
\colhead{$\langle D_{ls}/D_s \rangle$} &
\colhead{$n_{\textrm{bkg}}$} \\
\colhead{} &
\colhead{} &
\colhead{} & 
\colhead{} & 
\colhead{Time (ks)} &
\colhead{(FWHM, $\arcsec$)} &
\colhead{(mag, $5\sigma$)} &
\colhead{} &
\colhead{(arcmin${}^{-2}$)}}
\startdata
MS2137 & SC & $B$                    & 2007-08-13 & 1.4 & 0.75 & 26.5 & &\\
       & SC & $V^{\dagger}$           & 2007-11-14 & 1.2 & 1.01 & 25.8 & &\\
       & SC & $\mathbf R^{\dagger}$   & 2007-11-13 & 2.4 & 0.68 & 26.2 & 0.563 & 17.8\\
       & SC & $I^{\dagger}$           & 2007-11-14 & 1.9 & 0.95 & 25.4 & &\\
       & SC & $z+$                   & 2007-07-18 & 1.6 & 0.81 & 24.8 & & \\ \hline

A963 & CFHT 12K   & $B$            & 1999-11-15, 17 & 7.2 & 0.90 & 26.3 & &\\
     & SC & $V$            & 2000-11-28     & 1.8 & 0.69 & 25.8 & &\\
     & SC & $R$            & 2000-11-24, 25 & 3.4 & 0.69 & 26.1 & &\\
     & SC & $\mathbf I$    & 2003-04-08     & 3.0 & 0.64 & 25.8 & 0.693 & 22.8 \\ \hline

A383  & MegaPrime & $u^*$       & 2003-12-20, 23 and 2004-01-21 & 9.2 & 1.21 & 26.1 & &\\
      & SC     & $B$         & 2002-09-10 and 2008-01-10    & 7.5 & 0.85 & 27.0 & &\\
      & SC     & $V$         & 2008-01-09     & 2.4 & 0.64 & 26.5 & &\\
      & SC     & $\mathbf R^{\dagger}$ & 2007-11-13     & 2.4 & 0.56 & 26.3 & 0.731 & 22.7 \\
      & SC     & $i+$        & 2005-10-02     & 2.4 & 0.62 & 25.8 & &\\
      & SC     & $z+$        & 2002-09-11     & 1.5 & 0.59 & 24.9 & &\\ \hline

A611 & SC & $B^{\dagger}$         & 2007-11-13 & 2.4 & 0.62 & 26.7 & &\\
     & SC & $V^{\dagger}$         & 2007-11-14 & 1.2 & 0.56 & 26.2 & &\\
     & SC & $\mathbf R^{\dagger}$ & 2007-11-13 & 2.4 & 0.68 & 26.2 & 0.600 & 18.3 \\
     & SC & $I^{\dagger}$         & 2007-11-14 & 2.4 & 0.66 & 25.8 & &\\ \hline
     
A2537 & IMACS & $B^{\dagger}$          & 2009-08-23 & 5.0 & 0.99 & 26.0 &  &\\
      & SC    & $\mathbf R^{\dagger}$  & 2007-11-13 & 2.4 & 0.56 & 26.4 & 0.600 & 9.9\\
      & IMACS & $I^{\dagger}$          & 2009-08-23 & 4.1 & 0.88 & 25.5 &  & \\ \hline

A2667 & SC     & $V^{\dagger}$          & 2007-11-14 & 1.2 & 0.88 & 26.0 &  &\\
      & SC     & $\mathbf R^{\dagger}$  & 2007-11-13 & 2.4 & 0.66 & 26.4 & 0.686 & 14.6 \\
      & IMACS & $I^{\dagger}$           & 2009-08-23 & 1.7 & 0.91 & 25.2 & &\\ \hline

A2390 & SC & $B$               & 2004-09-15 and 2005-11-30 & 2.2 & 0.80 & 26.7 & &    \\
      & SC & $V$               & 2004-07-18  & 0.8 & 0.63 & 26.1 &  & \\
      & SC & $\mathbf R$       & 2004-09-15  & 2.3 & 0.64 & 26.1 & 0.686 & 15.1 \\
      & SC & $I$               & 2004-09-18  & 6.3 & 0.74 & 26.1 & &  \\
      & SC & $z+$              & 2004-09-16  & 1.7 & 0.97 & 24.9 & & 
\enddata
\tablecomments{SC denotes SuprimeCam. ${}^{\dagger}$ indicates observations conducted by the authors.
The remainder were obtained from the Subaru and CFHT archives. The filter used for detection
and shear measurement is in bold. Depth is measured by the median magnitude of all $5\sigma$ detections
within a $2\arcsec$ diameter aperture. The surface density of sources selected for the shear analysis $n_{\textrm{bkg}}$, along
with their mean lensing distance ratio $\langle D_{ls} / D_s \rangle$, are listed.}
\end{deluxetable*}

\begin{figure}
\centering
\includegraphics[width=\linewidth]{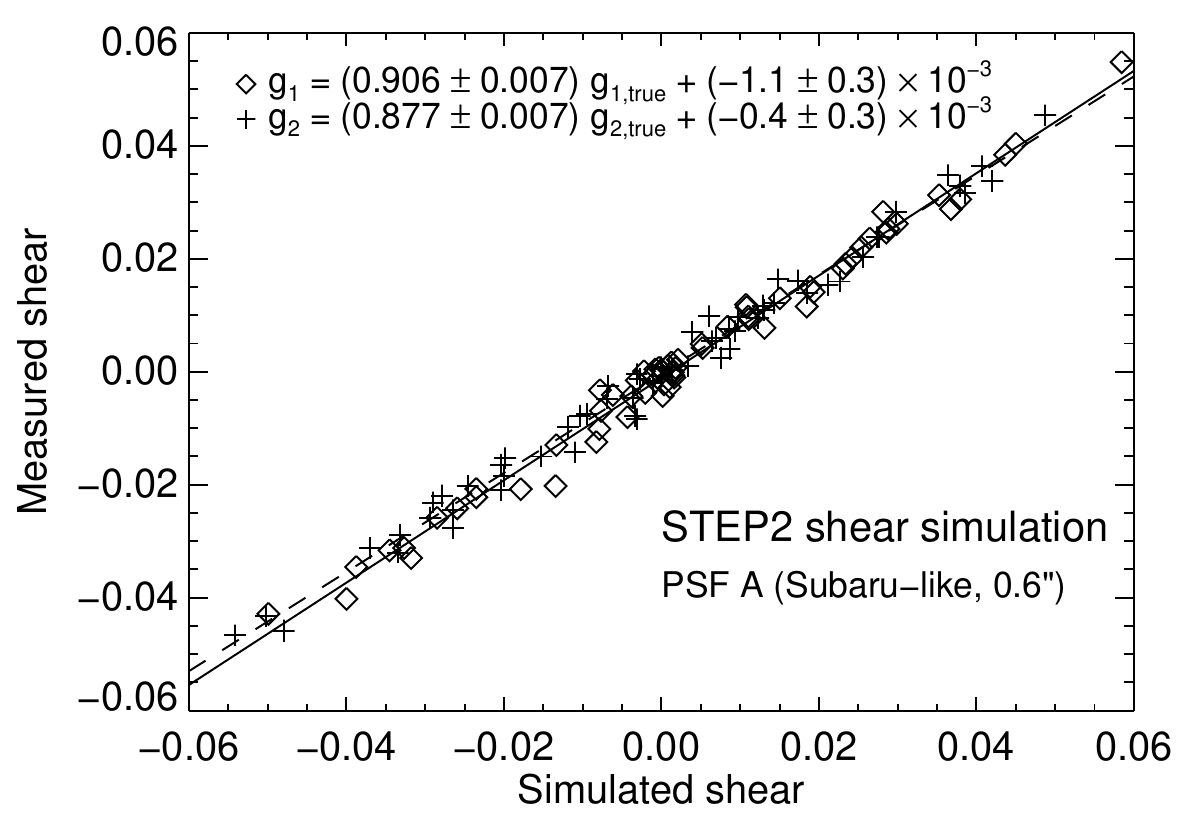}
\caption{Validation of our shear measurement method using images from the STEP2 simulation, designed to mimic our typical SuprimeCam imaging. As described in \citet{STEP2}, shot noise is reduced using rotated image pairs. We determine a calibration factor $g_\textrm{meas} = 0.89 g_\textrm{true}$ comparable to other techniques, with a negligible additive bias. Similar results hold for other PSFs. \label{fig:step}}
\end{figure}

\subsection{Shear measurement and source selection\label{sec:shear}}

Galaxy shape measurements were performed based on the \citet[][KSB95]{KSB95} method as implemented in the \code{IMCAT} software package. The details of this procedure, including modeling and correction for the PSF, were described in the Appendix of N09. We have implemented a few minor changes to this procedure. First, the stellar anisotropy kernel $q_{\alpha}^*$ has been computed for a grid of Gaussian window functions of varying widths, rather than a single size. The smooth variation of $q_{\alpha}^*$ across the detector was well fit by an fifth-degree polynomial in the pixel coordinates $x$ and $y$. (We refer to N09 for a demonstration of the quality of the PSF correction, which is similar for other clusters.) When raw ellipticities are corrected for the PSF anisotropy, the fitted $q_{\alpha}^*$ are interpolated to match the window function width appropriate to each galaxy, which we take as its \code{SExtractor} \texttt{FLUX\_RADIUS}, or $r_h$. Second, rather than fitting the shear polarizability $P^{\gamma}$ as a function of galaxy properties, we use the individual measurements for each galaxy. Third, selected galaxies are equally weighted in our shear analysis. We found that these small modifications led to slightly better performance (a calibration factor closer to unity) when the shear pipeline is tested on simulated data, as described below.

From the \SExtractor~catalog described in Section~\ref{sec:wlredux}, we selected resolved, well detected galaxies for our shear analysis via the following criteria: (1) $\textrm{S/N} > 7$, where $\textrm{S/N}$ is the detection significance defined in \citet{Erben01} measured with a window function having $\sigma = r_h$; (2) $1.15 r_h* < r_h < 6$~pixels, where $r_h^*$ is the median stellar \texttt{FLUX\_RADIUS}, to avoid unresolved and very large galaxies; (3) $|e| < 1$, $|g| < 1.5$, ${\rm tr}~P_{sm} > 0$, and $0.15 < P^{\gamma} < 2$, to exclude sources with pathological moments; (4) \texttt{MAG\_AUTO} $ > 21$; (5) to eliminate blended and asymmetric galaxies, a distance of at least 6 pixels to the nearest object, a distance of at least $3(r_{h,1} + r_{h,2})$ to any other object $> 3$~mag brighter, and a shift of less than 1 pixel between centroids measured with and without the window function ($d$ in N09); and finally (6) a photometric redshift selection described below.

We verify and calibrate the shear pipeline using simulated images from the STEP2 project \citep{STEP2}, which were designed to mimic the depth, sampling, and PSF typical of SuprimeCam data (Figure~\ref{fig:step}). For PSF A (FWHM = $0\farcs6$), we find a linear relation between simulated and recovered shear with a slope of 0.89, averaged between shear components, and negligible additive bias. Very similar results hold for PSF C ($0\farcs8$), leading to a mean calibration factor $m_{\textrm{WL}} = g_{\textrm{meas}}/g_{\textrm{true}} = 0.89 \pm 0.01$. This is typical of other authors and methods. Although STEP2 does not extend to the shears $g = 0.2-0.3$ that we measure near cluster centers, the tight linearity in Figure~\ref{fig:step} gives us confidence that the shear pipeline is working well and that an extrapolation of the calibration factor to higher shear is reasonable.

\begin{figure*}
\centering
\includegraphics[width=0.33\linewidth]{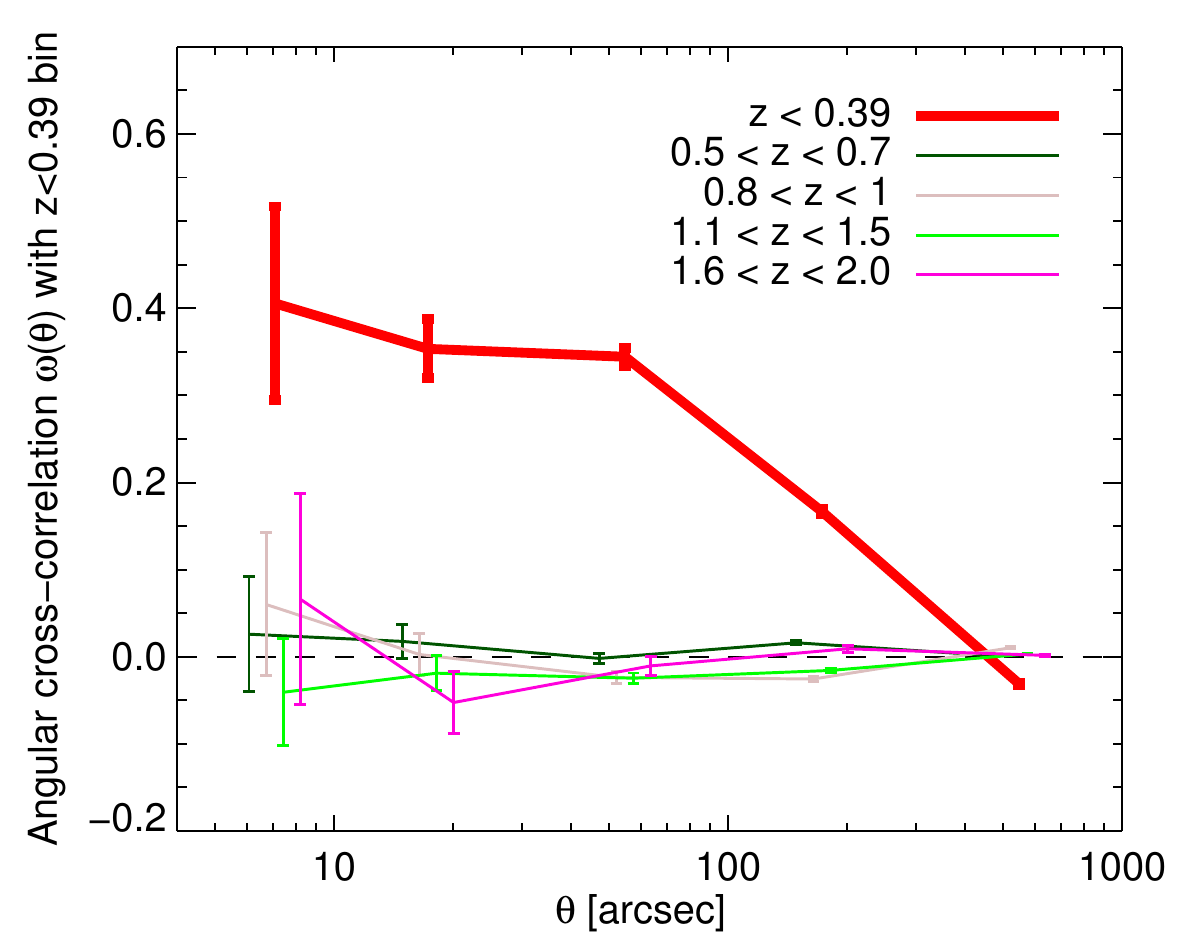} \hfill
\includegraphics[width=0.33\linewidth]{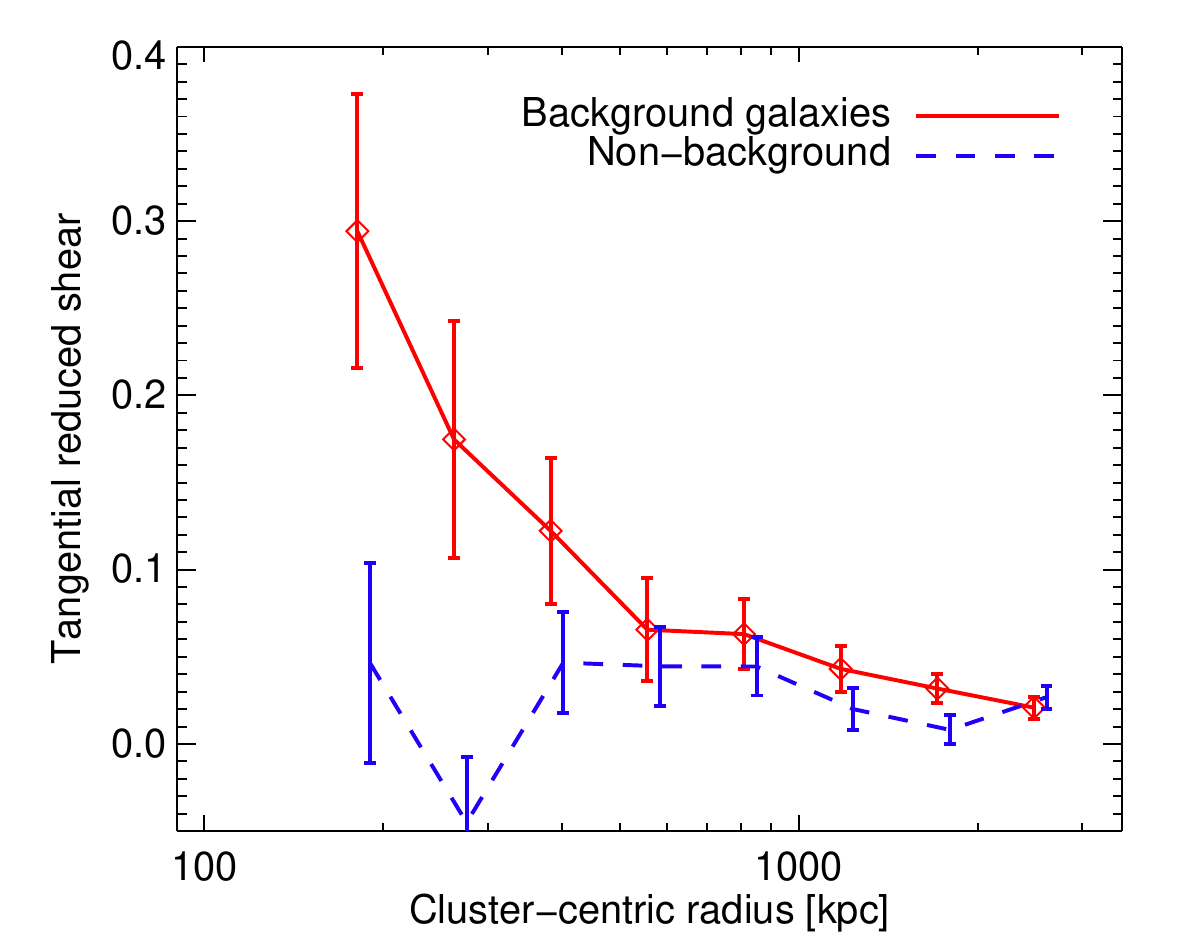} \hfill
\includegraphics[width=0.33\linewidth]{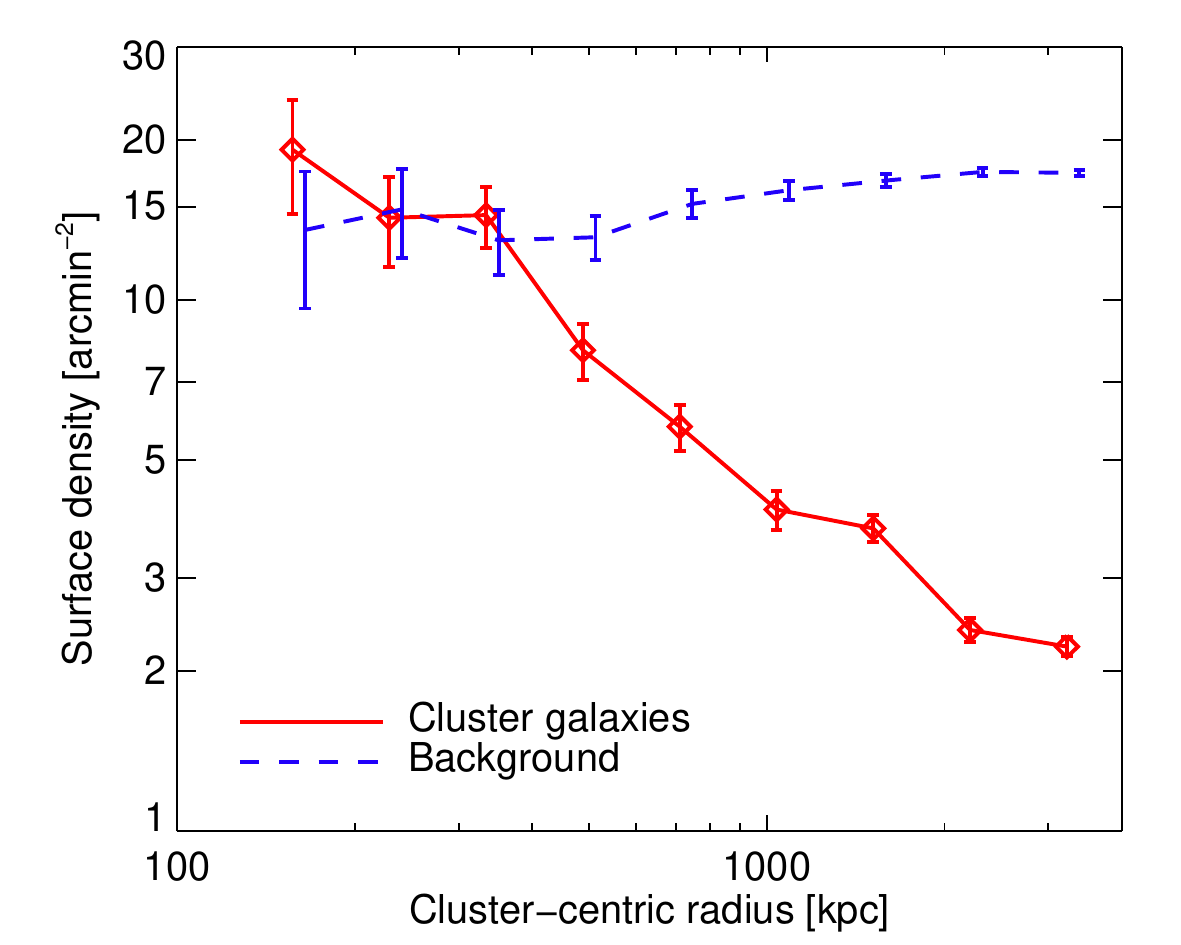}
\caption{Tests of our background galaxy selection for the weak lensing analysis, demonstrated in A611. \textbf{Left:} angular correlation between galaxies with $z_{\textrm{phot}}$ securely below the selection threshold $z_{\textrm{min}} = 0.39$, and those in other bins of $z_{\textrm{phot}}$. The auto-correlation at $z < z_{\textrm{min}}$ and the lack of any cross-correlation with other $z_{\textrm{min}} < z_{\textrm{phot}} < 2$ galaxies indicate low contamination. \textbf{Middle:} tangential shear profile measured using selected background sources (red), which shows the expected rise toward the center, and using excluded foreground and cluster galaxies (blue), which shows a flat, low signal. \textbf{Right:} the surface density profile of galaxies with secure $z_{\textrm{phot}}$ near the cluster redshift (red) shows the expected rise, while the density of selected background sources (blue) is flat or declining toward the center. \label{fig:photztest}}
\end{figure*}

\subsection{Photometric redshifts\label{sec:photz}}

We estimate photometric redshifts of all sources in order to select those located behind the clusters. This technique makes use of all the information available in the multi-color photometry. We use the \code{BPZ} \citep{BPZ} software (version 1.99.3) with its CWWSB4 set of eight templates and the default prior. \code{BPZ} provides both a marginalized redshift probability density $P(z)$ as well as a point estimator $z_b$. We use both and define $z_{\textrm{phot}} = z_b$ below. For five of the seven clusters in our sample, the spectroscopic redshift and the peak $z_{\textrm{phot}}$, as measured from bright galaxies in the cluster core, agreed with a scatter of $\sigma_{\Delta z/(1+z)} = 0.02$. This supports the quality of the photometric calibration described in Section~\ref{sec:wlredux}. In A2537 and A2667, the peak $z_{\textrm{phot}}$ is too high by $\simeq0.1$. This is not surprising, since these clusters are the only two observed through only three filters, and these do not closely bracket the 4000~\AA~break.

Two criteria were used to select background galaxies. Firstly, we required $z_{\textrm{min}} < z_{\textrm{phot}} < z_{\textrm{max}}$, where we define $z_{\textrm{min}} = z_{\textrm{clus}} + 0.1$ and $z_{\textrm{max}} = 2$ by default. (For the special cases of A2537 and A2667 discussed above, we conservatively take $z_{\textrm{min}} = 0.55$ and $z_{\textrm{min}} = 0.50$, respectively.) Second, we eliminated sources with a significant low-redshift solution by requiring that the probability that $z > z_{\textrm{clus}} + 0.1$, determined by integrating $P(z)$, is $> 90\%$. Adopting a higher threshold generally had little effect on the resulting shear profiles, but reduced the surface density of selected sources. A2667 showed the greatest possibility of residual dilution, consistent with the more limited photometry described above, but we show in Section~\ref{sec:losellip} that the shear profile is consistent with the strong-lensing and X-ray mass measurements where they overlap.

Dilution of the shear signal from cluster or foreground sources is probably the main systematic error in cluster weak-lensing analyses. Therefore, we conducted several astrophysical tests to assess the reliability of our background galaxy identification. These are illustrated in Figure~\ref{fig:photztest} for the case of A611. Firstly, we looked for an angular clustering signal between galaxies identified as in the cluster or the foreground, and those in several bins of higher redshift. The cross-correlation signal (left panel) is low or absent at $z < 2$, while the auto-correlation in the foreground bin is prominent. If we admit sources with $z_{\textrm{phot}} \gtrsim 2$, a significant clustering with low-redshift sources arises from confusion between the photometrically inferred Balmer and Lyman breaks; this motivates our choice of $z_{\textrm{max}} = 2$. Second, we examined the radial shear profile, which shows a well-defined rise when using our selected background sample and a flat, low signal when using the remainder of sources (middle panel), as expected if they are mostly unlensed. Finally, we investigated the surface density of sources as a function of cluster-centric radius (right panel). The density of cluster galaxies rises rapidly towards the center, while that of background sources is flat or declines. These tests give us confidence that the photometric redshifts are effective at isolating lensed sources.

In our shear analysis, we incorporate the individual $z_{\textrm{phot}}$ measurements of the background sources. However, as a check of our $z_{\textrm{phot}}$ distribution, we computed the mean distance ratio $\langle D_{ls}/D_s \rangle$ that determines the lensing efficiency (Table \ref{tab:wlobs}). We then selected galaxies from the COSMOS survey with a matching magnitude distribution in the detection band and with similar $z_{\textrm{min}}$ and $z_{\textrm{max}}$ cuts.\footnote{In detail, we increased $z_{\textrm{min}}$ by 0.1 to account for the effects of our $P(z)$ cut that could not be directly mimicked in COSMOS. The COSMOS broadband photometry was linearly interpolated to the central wavelength of our detection band when necessary.}
The $\langle D_{ls} / D_s \rangle$ determined from the 30-band $z_{\textrm{phot}}$ in COSMOS \citep{Ilbert09} agreed with our determinations with a scatter of only 3\%, suggesting that errors in the mean distance to the background sources have a minimal effect on our analysis.

\subsection{Results}
\label{sec:wlresults}

\begin{figure*}[t]
\centering
\includegraphics[width=\linewidth]{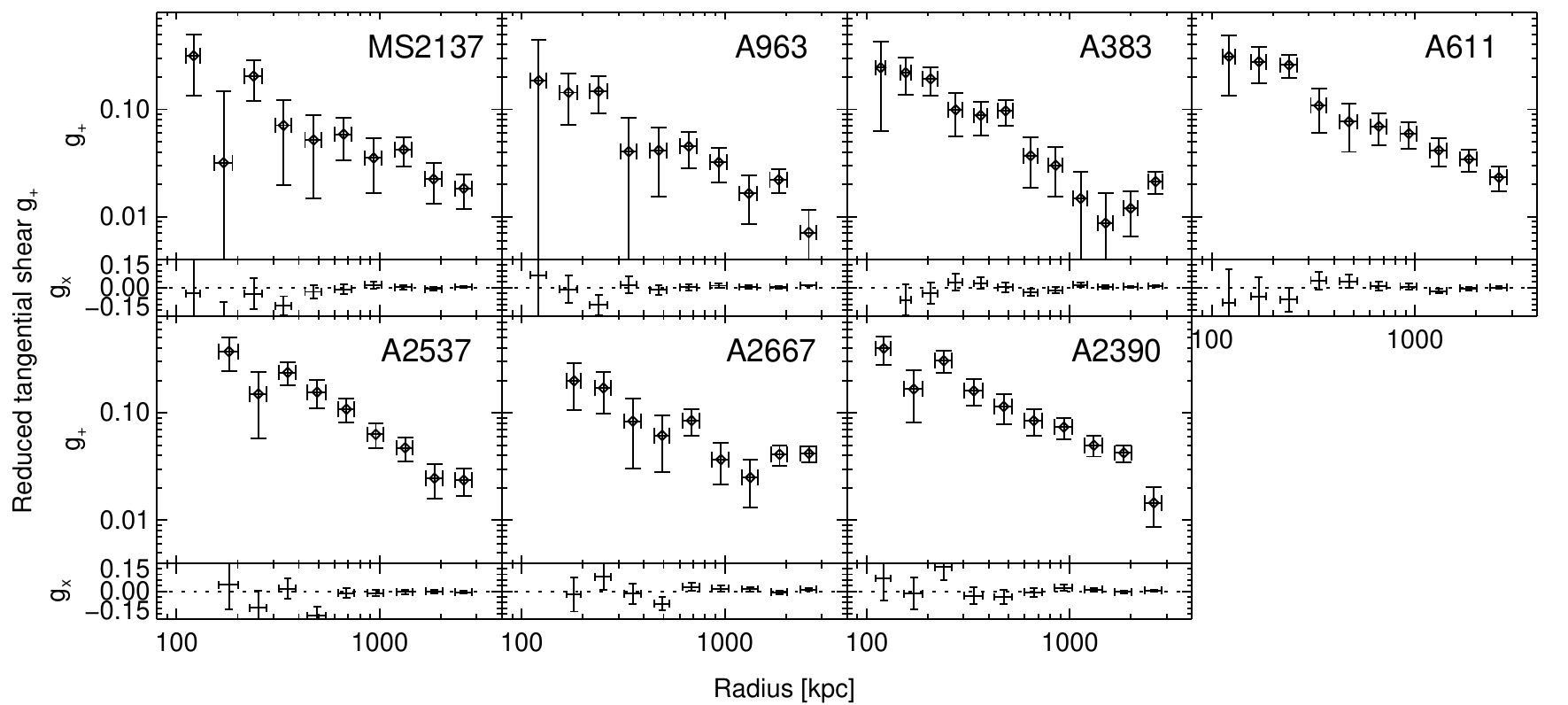}
\caption{Tangential reduced shear $g_+$ measured in annular bins. The $B$-mode component $g_{\times}$, which should vanish in the absence of systematic errors, is also plotted and is consistent with zero. Measurements have been calibrated on simulated STEP2 data as described in Section \ref{sec:shear}. Horizontal error bars indicate the dispersion in radius within each bin. The rising signal seen at large radii in A2667 and A383 is ascribed to secondary mass peaks as discussed in the text.\label{fig:tanshearprof}}
\end{figure*}

The mean distortion of background galaxies is a measure of the reduced shear $g = \gamma / (1 - \kappa)$, where $\gamma$ and $\kappa$ are the shear and convergence \citep[e.g.,][]{Schneider06}. Figure~\ref{fig:tanshearprof} displays the azimuthally-averaged tangential reduced shear for all seven clusters. In general we select galaxies with $100~\textrm{kpc} < R < 3~\textrm{Mpc}$ for the shear analysis. At smaller radii there are few sources and contamination from cluster galaxies is most severe, while the outer limit corresponds roughly to the SuprimeCam field of view. In A2667 and A2537, where our photometry is less extensive, we require $R > 150$~kpc to account for the greater possibility of dilution at small radii.  In all clusters, a smoothly rising tangential shear profile is observed, with no clear evidence for dilution from contaminating foreground sources. A significant $B$-mode signal, which should not arise physically and is thus often used a diagnostic of systematic errors, is not detected.

\begin{figure*}
\centering
\includegraphics[width=0.4\linewidth]{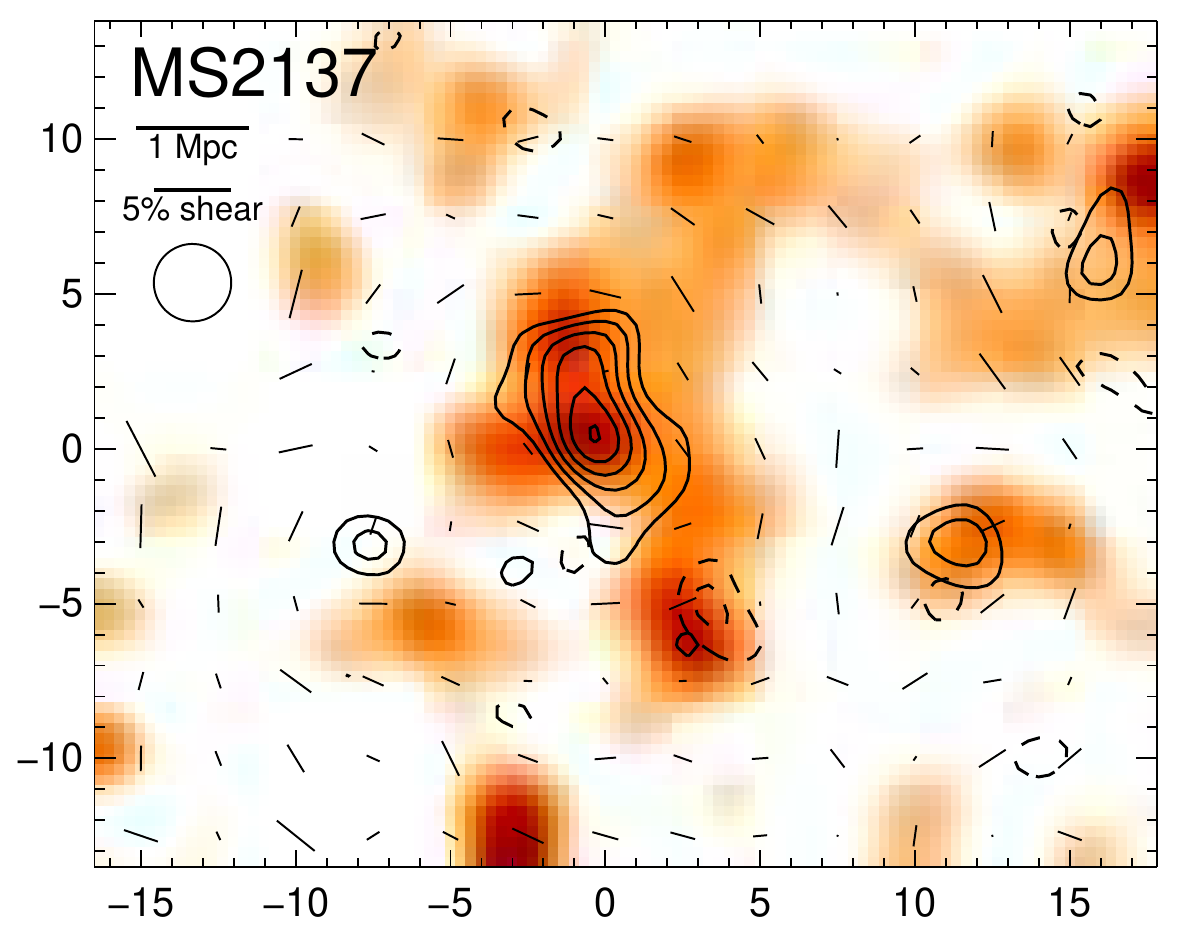}
\includegraphics[width=0.4\linewidth]{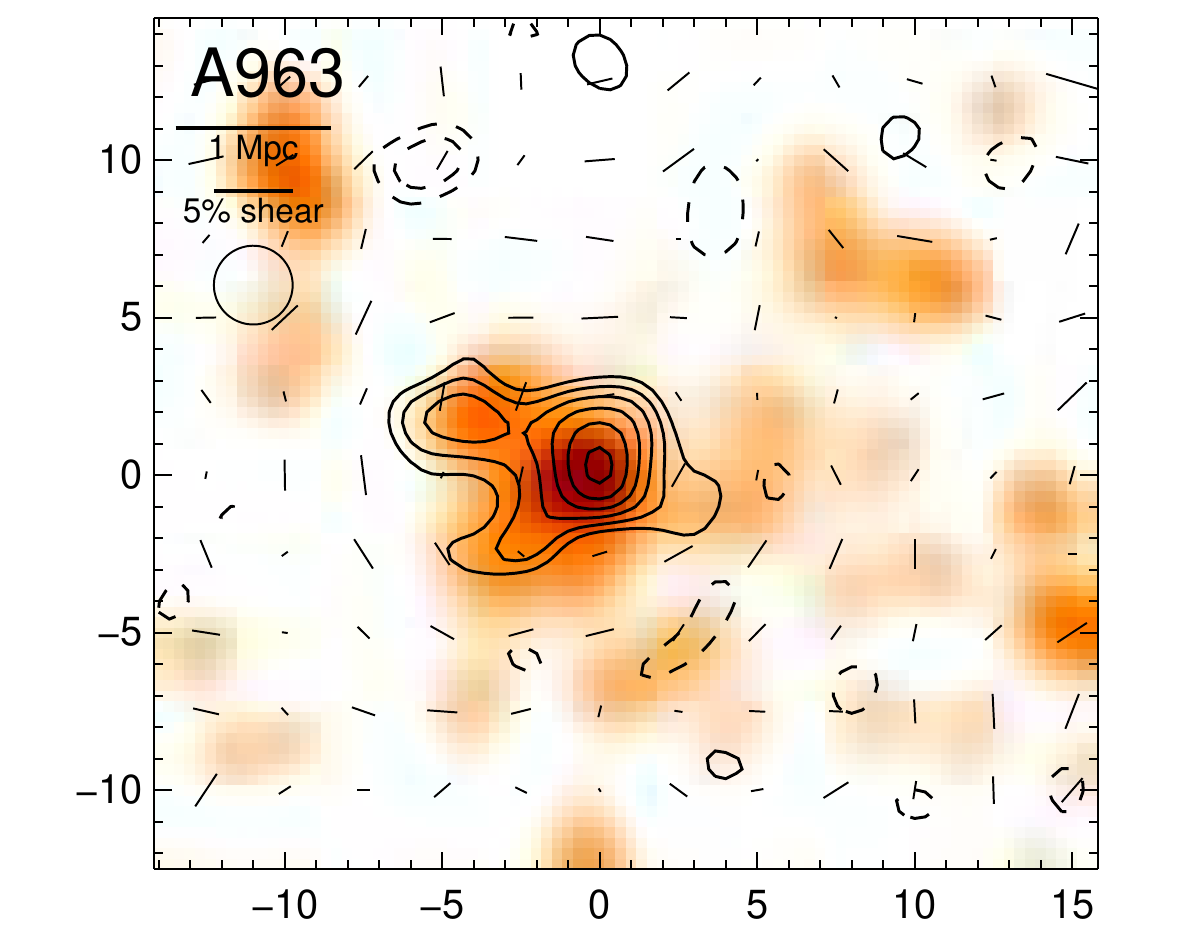} \\
\includegraphics[width=0.4\linewidth]{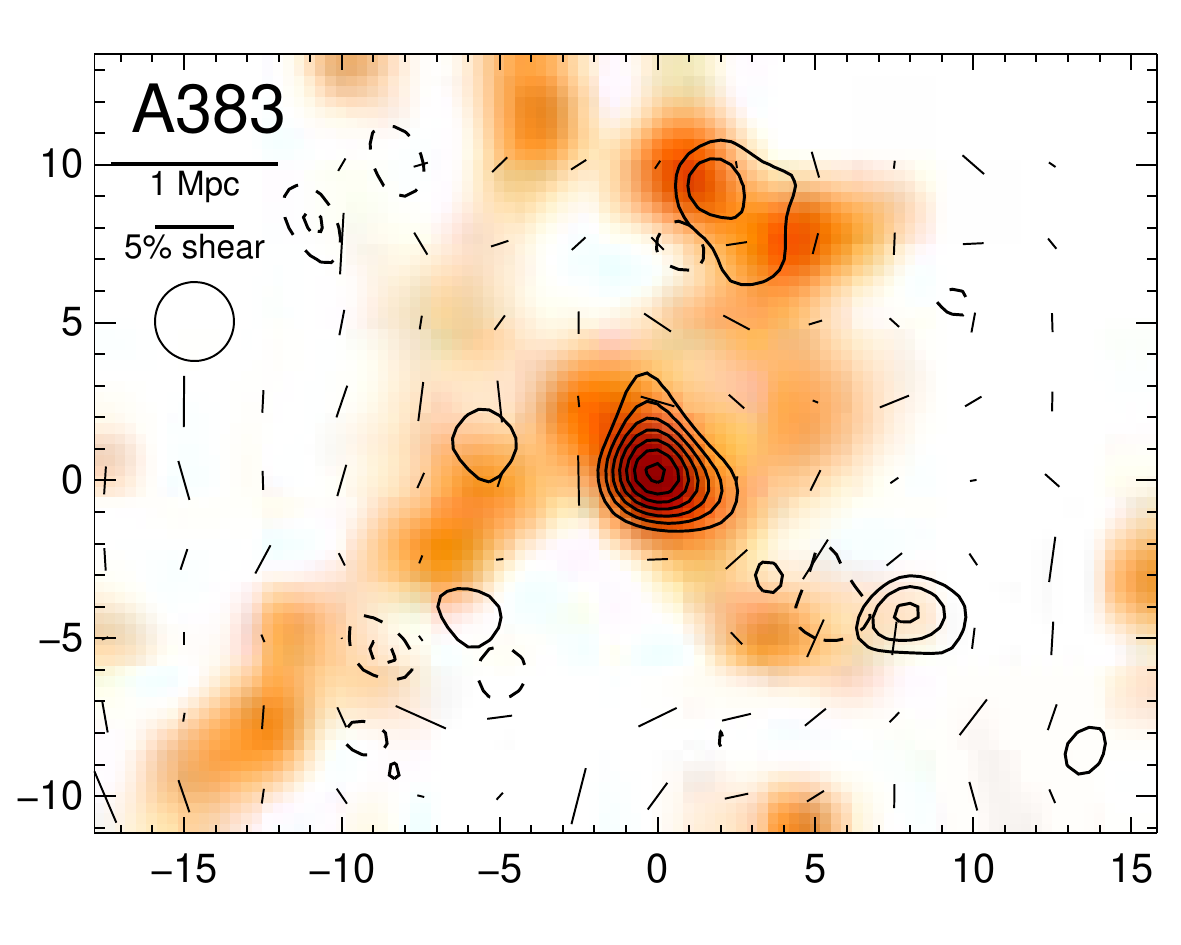}
\includegraphics[width=0.4\linewidth]{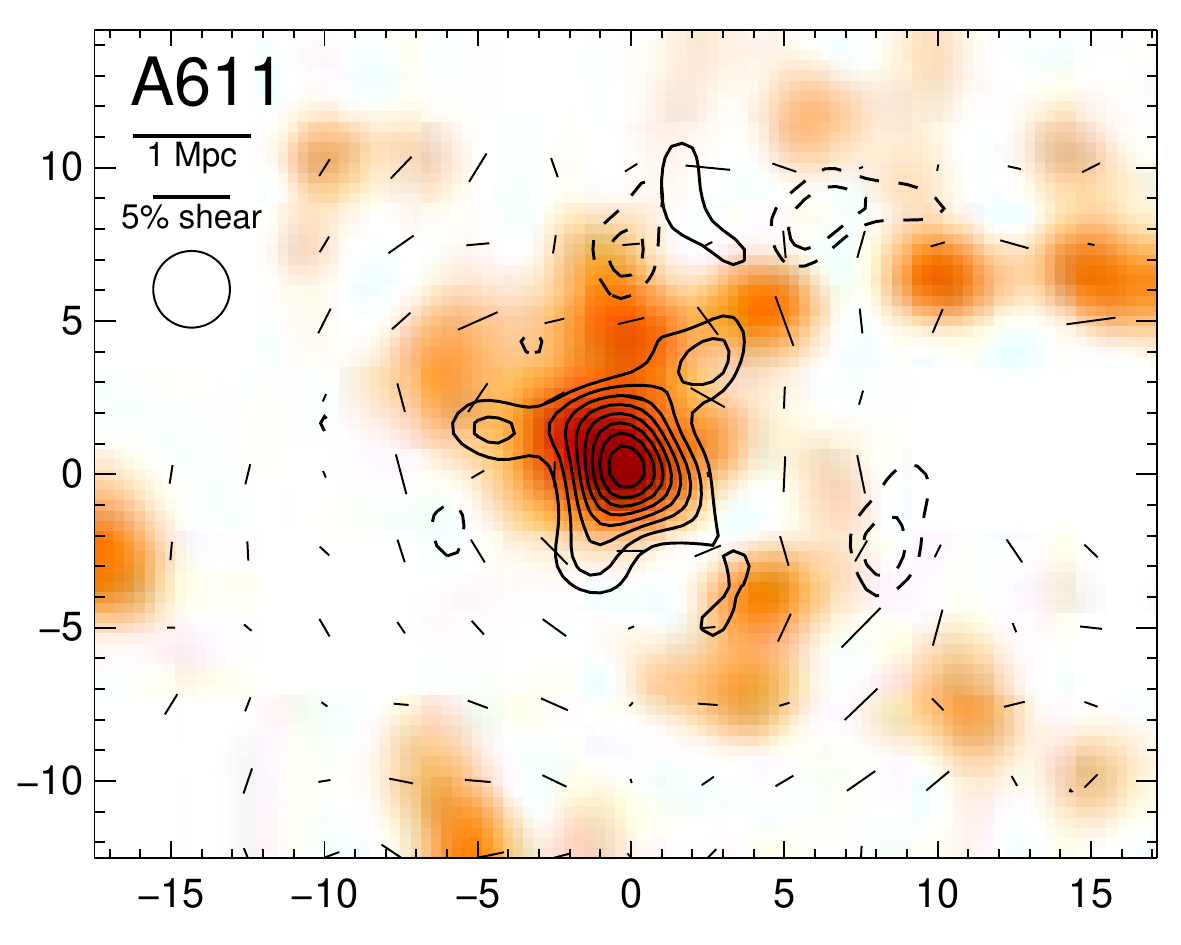} \\
\includegraphics[width=0.4\linewidth]{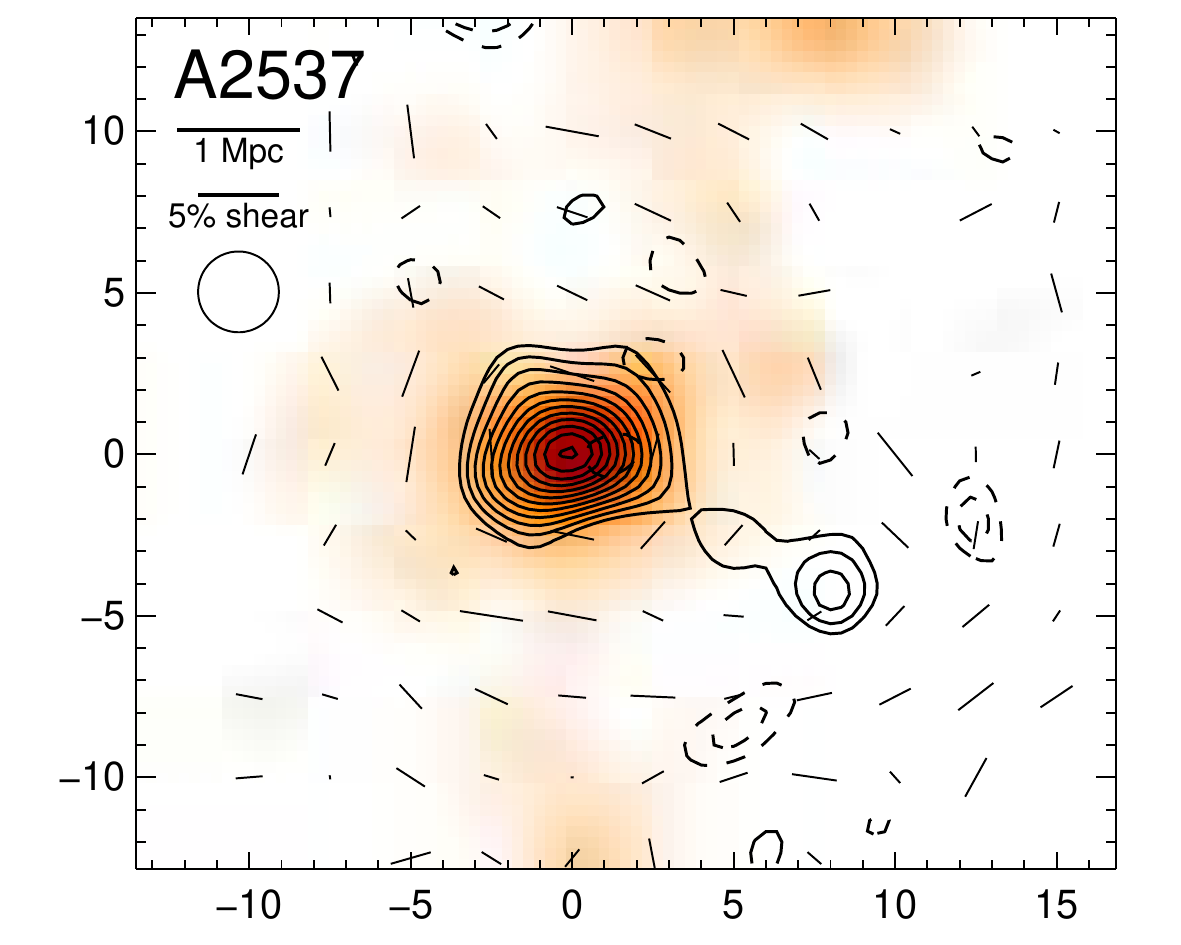}
\includegraphics[width=0.4\linewidth]{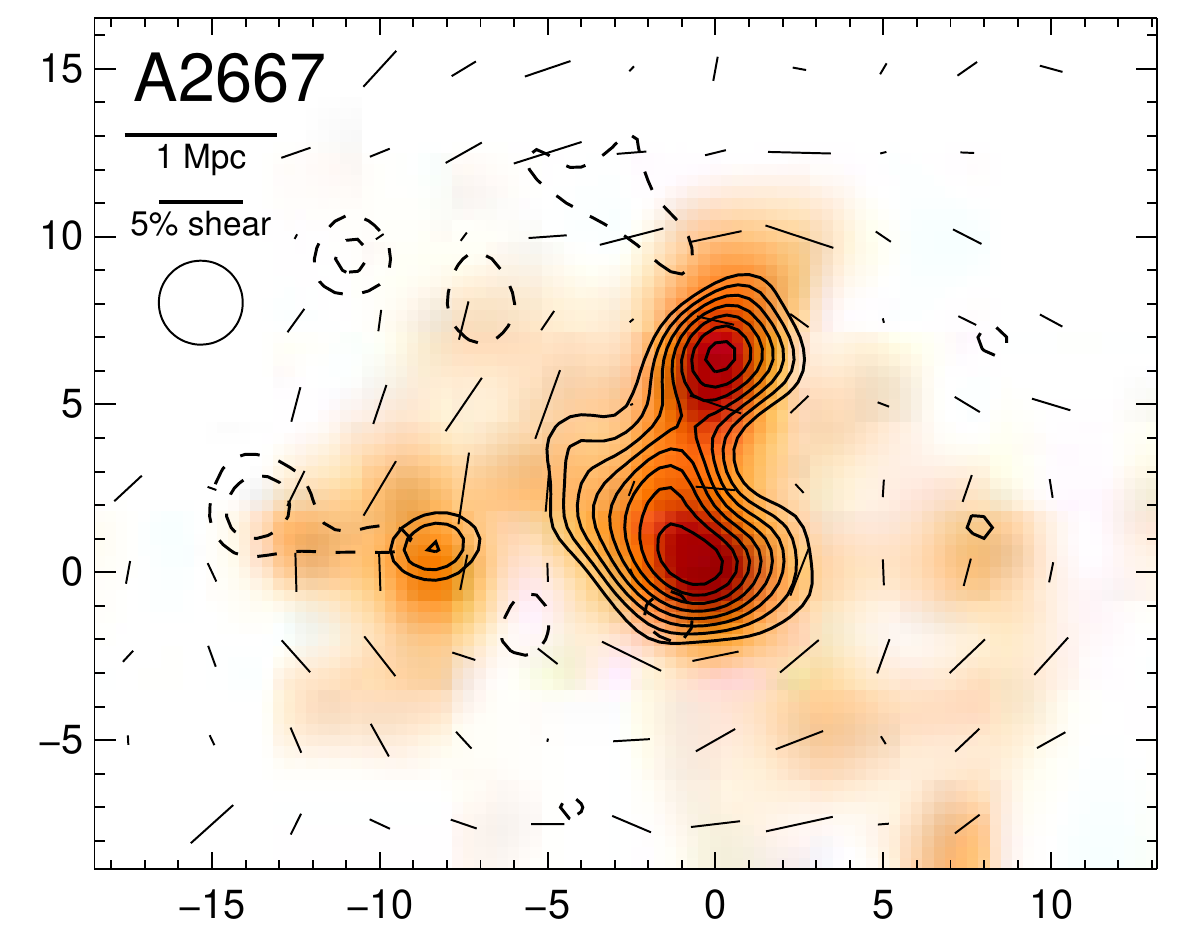} \\
\includegraphics[width=0.4\linewidth]{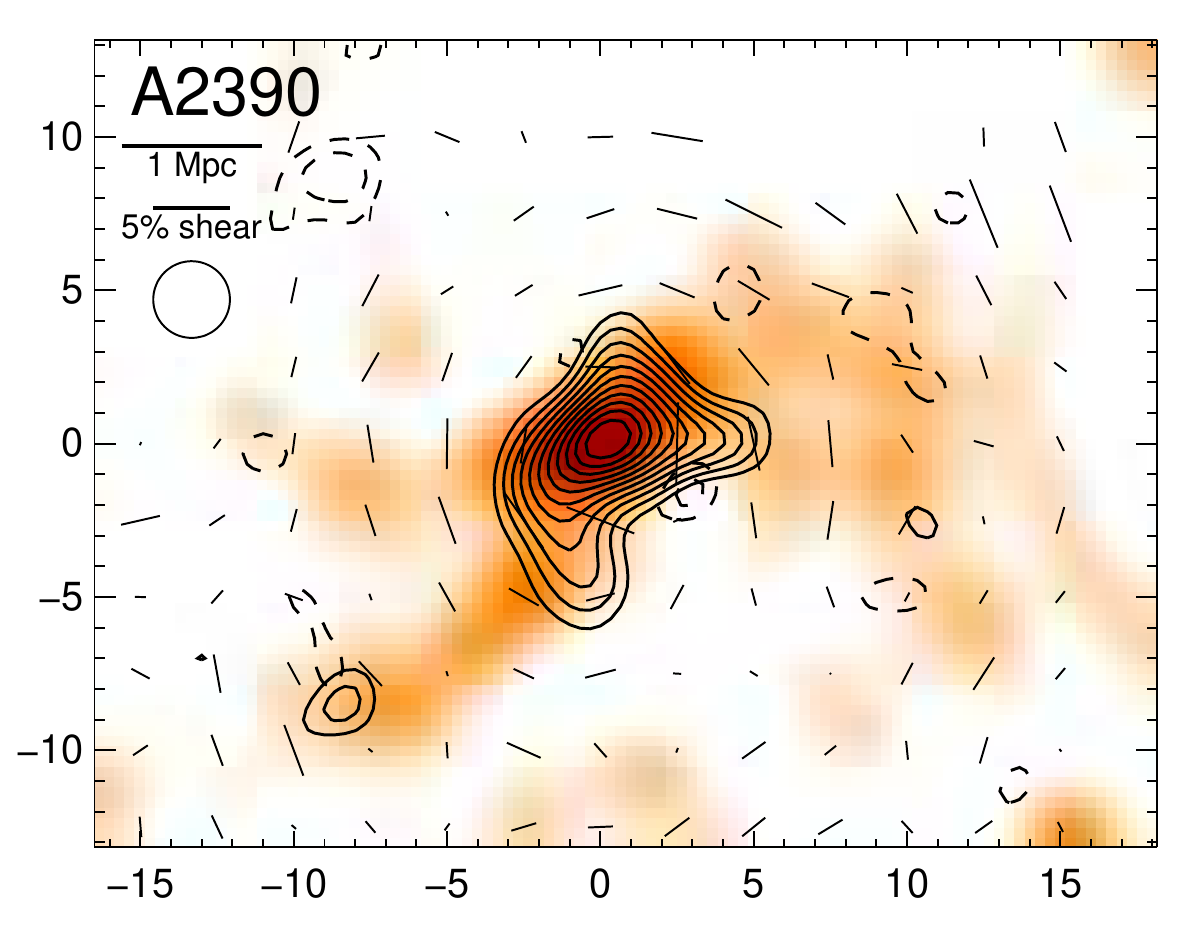}
\caption{Surface density $\kappa$ contours, derived from our weak-lensing analysis, are overlaid on the smoothed $I$-band light from galaxies near the cluster redshift, identified as described in Section~\ref{sec:galcat}. Contours begin at $3\sigma$ and increase by $1\sigma$. Dashed contours show the absolute value of the $\kappa$ map derived instead from the $B$-mode signal, which should be null and is used to assess the noise and systematics. The smoothed shear field is also overlaid. All quantities are smoothed by a Gaussian with FWHM = $150\arcsec$ as indicated by the circle in the upper left.\label{ref:kappa} Axes indicate the distance from the BCG in arcmin, with north up and east left.\label{fig:kappa}}
\end{figure*}

For each cluster we also produced 2D surface density maps following \citet{Kaiser93}, which are shown in Figure~\ref{fig:kappa}. To increase the surface density of sources, we loosened the $P(z)$ selection criterion described in Section~\ref{sec:shear}; this has no effect on our quantitative results, which do not rely on the 2D maps. In general mass and light are well aligned, and any other structures in the fields are detected at marginal significance. (This can be seen by noting that the dashed contours show the mass reconstructed using the $B$-mode signal: all such peaks are spurious and give an indication of the number of noise peaks of a given significance expected in this field of view.) These mass maps are useful for investigating the upturn or plateau in the radially-averaged shear signal seen at large radii in A383 and A2667. The upturn in A383 is likely related to substructures near the virial radius, and following N10, we therefore restrict to $R < 1.5$~Mpc in this cluster.

In A2667, the radial shear profile shows a high plateau to  $R > 3$~Mpc, which is explained in the mass map by a second large mass concentration clearly detected $6\farcm2 = 1.4$~Mpc north of the main, strong-lensing cluster. The secondary clump detected in the lensing map is exactly aligned with an excess of bright red galaxies near the cluster redshift (Figure~\ref{fig:kappa}). The brightest of these galaxies has a redshift $z = 0.2042$ from the 2dF survey \citep{Colless01}, corresponding to a comoving distance of 100 Mpc along the l.o.s. This suggests the second clump is slightly in the foreground of A2667. In our weak-lensing study, we model both mass concentrations simultaneously, and results for the main cluster are independent of the redshift of the second peak.

\section{Strong lensing}
\label{sec:SL}

We now turn to the identification of sources multiply imaged by the clusters. Every cluster in our sample has been imaged by \HST, and every one except A2537 has been the subject of an earlier lensing study, as described below. We refer to and build upon these models.  In Sections $4.1-4.7$ we consider each cluster individually, and in Section 4.8 we describe the construction of catalogs of cluster galaxies relevant as perturbers in our strong lens models.

The positions of the multiple images are illustrated in Figure~\ref{fig:slimages} and tabulated in the Appendix. We have retained the nomenclature of various authors; however, in all cases the final number or letter distinguishes multiple images of the same source. In several cases, we have added new spectroscopic redshifts based on the observations detailed in Section~\ref{sec:vdobs}. These spectra are shown in Figure~\ref{fig:arcredshifts}.

\begin{figure*}
\centering
\includegraphics[width=0.495\linewidth]{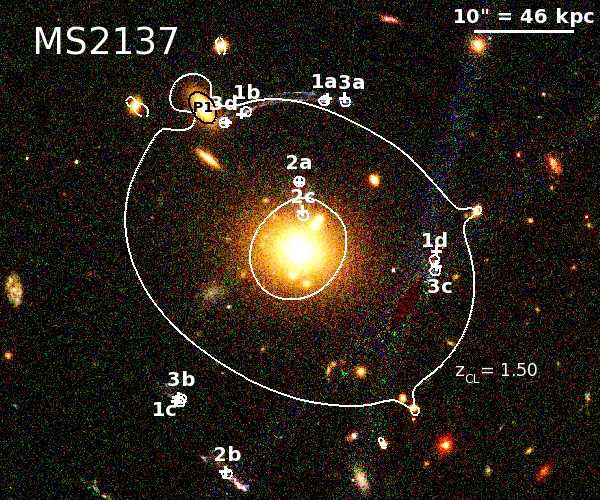} \hfill
\includegraphics[width=0.495\linewidth]{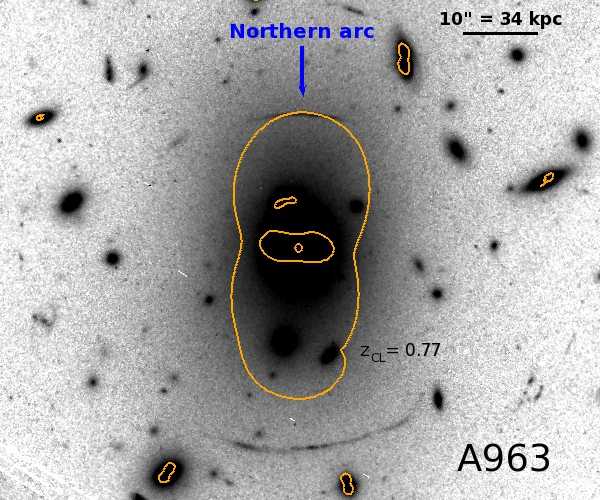} \\[0.1cm]
\includegraphics[width=0.495\linewidth]{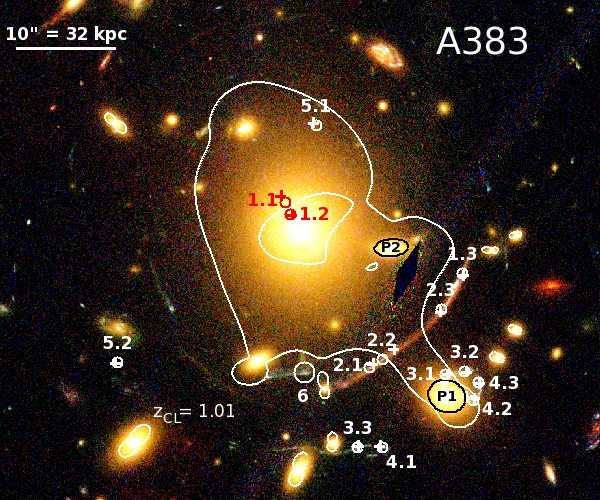} \hfill
\includegraphics[width=0.495\linewidth]{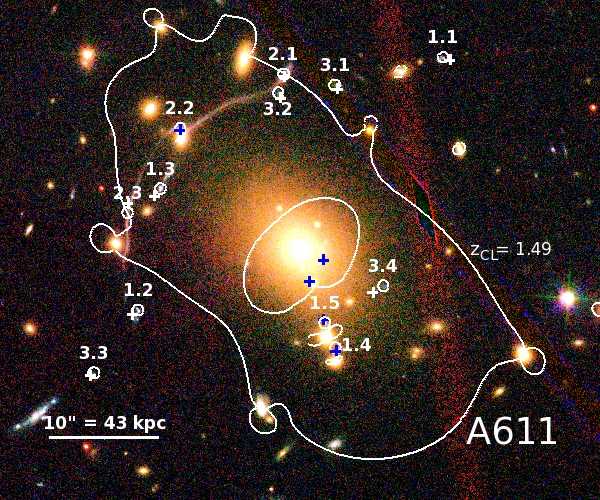} \\
\includegraphics[width=0.495\linewidth]{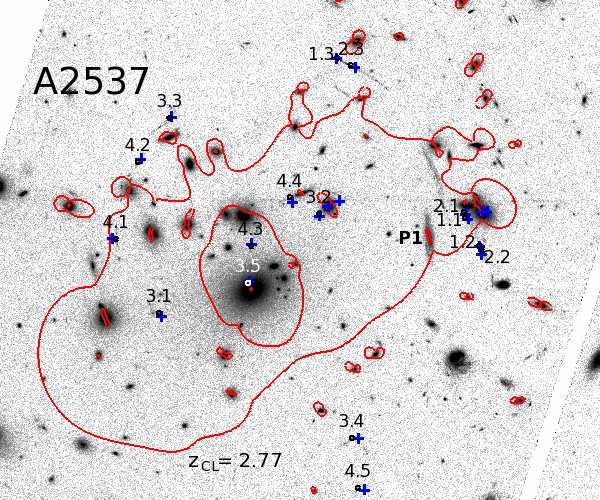} \hfill
\includegraphics[width=0.495\linewidth]{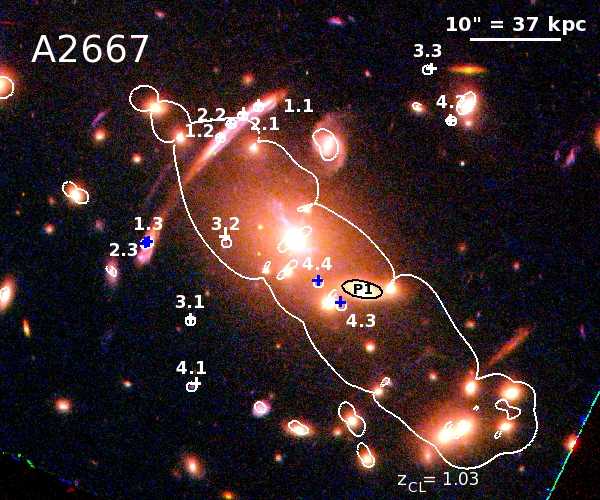}
\caption{\emph{HST} images of the central cluster cores, with multiply imaged sources identified (circles). Where possible we show color composite images, using data from the sources in Table~\ref{tab:bcgphot} or from the CLASH survey (A611, MS2137, A383). Reconstructed image positions based on the models described in Section~\ref{sec:totdens} are indicated by crosses (colors vary for clarity); critical lines are also overlaid at the redshifts $z_{\textrm{CL}}$ indicated in each panel. Individually optimized perturbing galaxies are denoted P1, P2, etc. \label{fig:slimages}}
\end{figure*}

\begin{figure}
\centering
\includegraphics[width=0.99\linewidth]{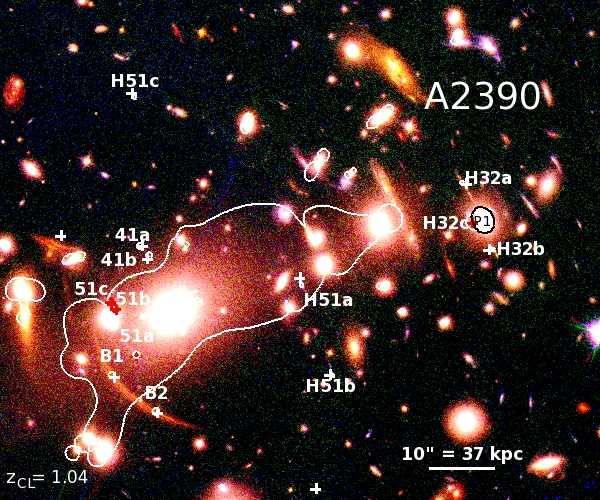}
\addtocounter{figure}{-1}
\caption{Continued}
\end{figure}

\begin{figure*}
\centering
\includegraphics[width=0.33\linewidth]{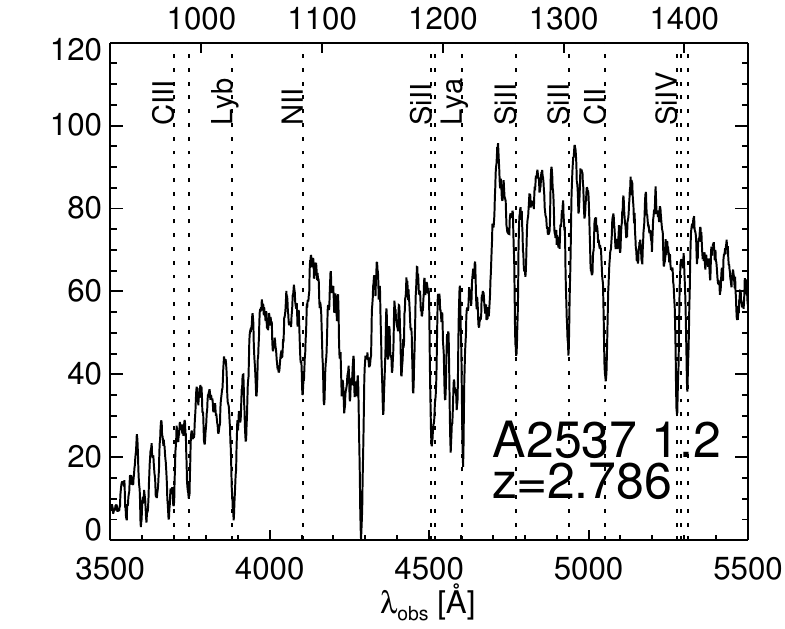}
\includegraphics[width=0.33\linewidth]{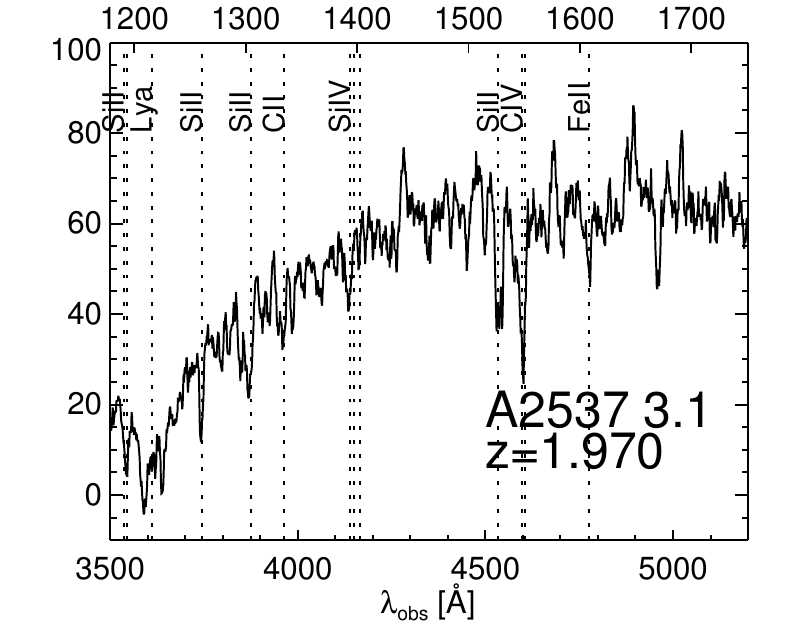}
\includegraphics[width=0.33\linewidth]{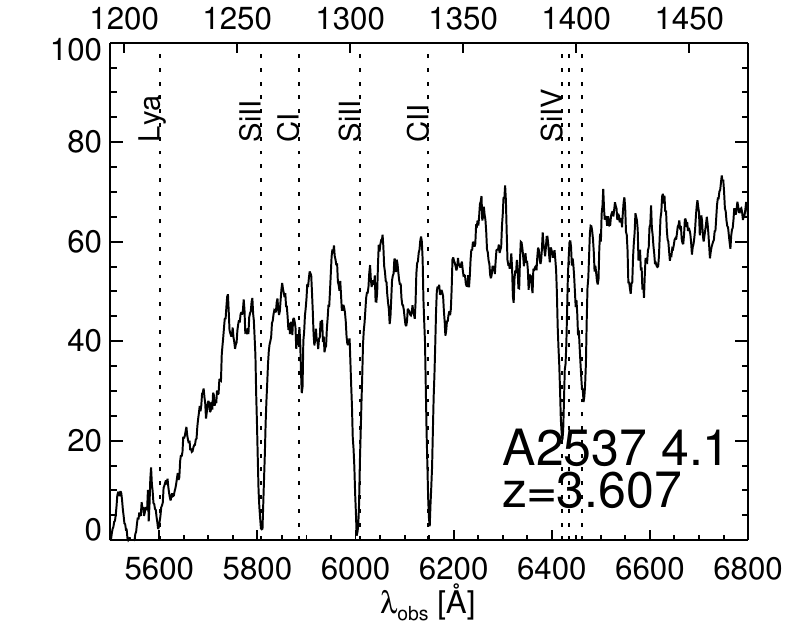} \\
\includegraphics[width=0.33\linewidth]{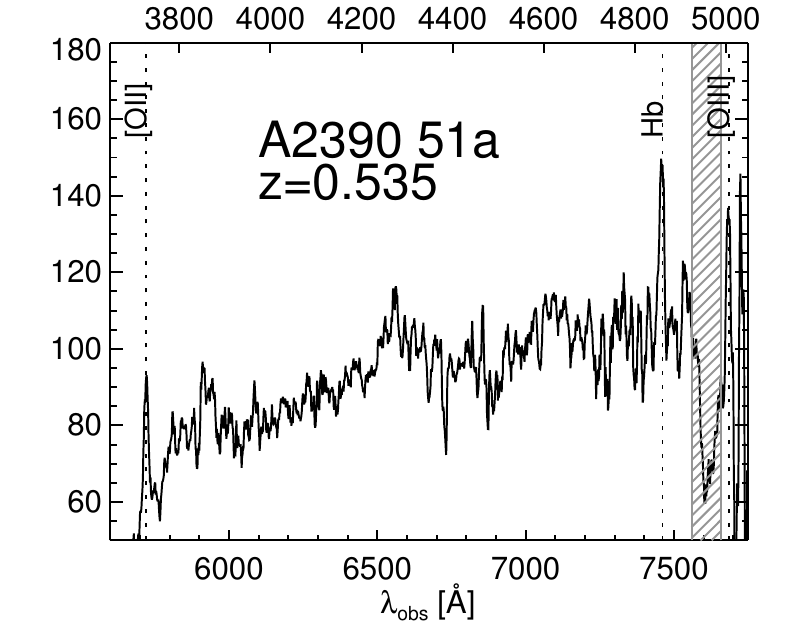}
\includegraphics[width=0.33\linewidth]{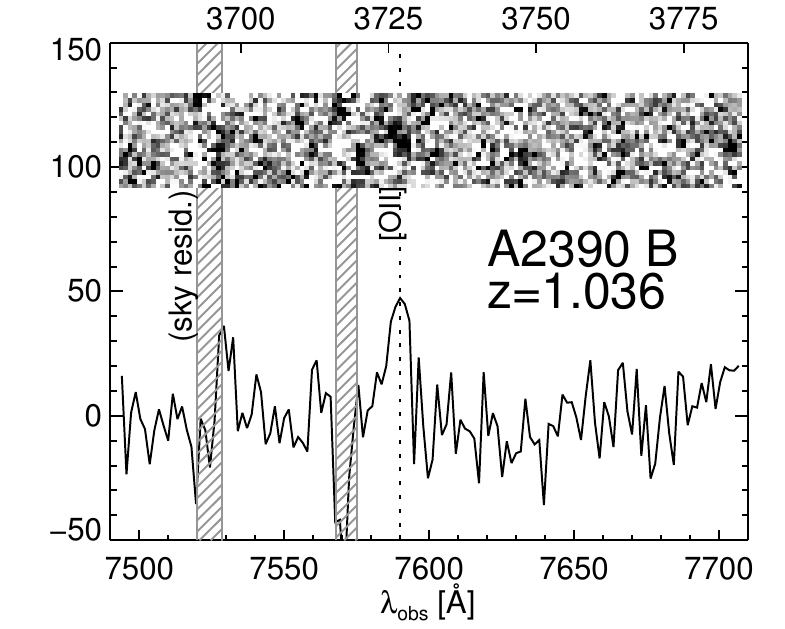}
\includegraphics[width=0.33\linewidth]{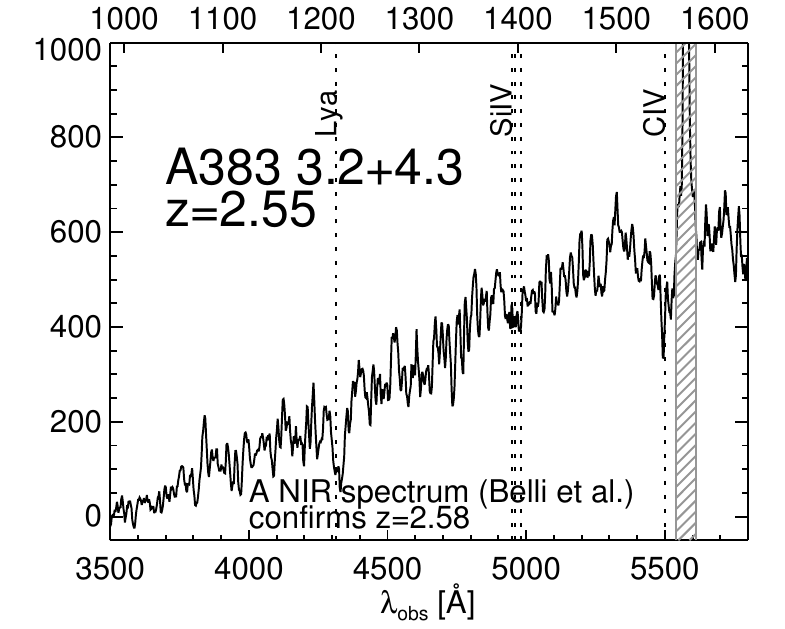} \\
\includegraphics[width=0.33\linewidth]{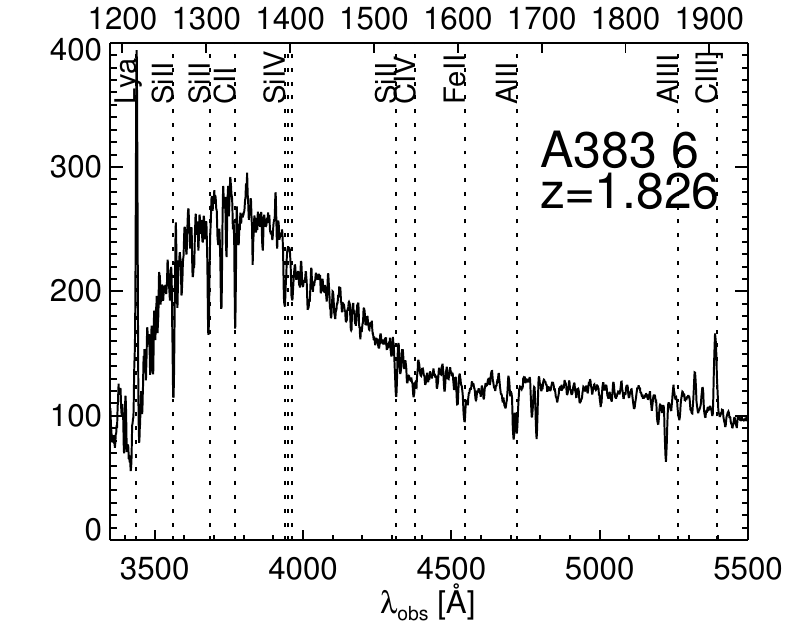}
\caption{Spectra of multiply imaged sources obtained in new observations described in Section~\ref{sec:vdobs}.
The axis at the top of each plot indicates the rest-frame wavelength. Selected lines are identified, and areas of residual sky emission or absorption are hatched.
The spectra are not flux calibrated, and the flux units are arbitrary. 
Multiple features are identified in each spectrum, resulting in a unique redshift determination with
the exception of A2390 B. The identification of the single weak emission line in the latter case as [\ion{O}{2}] is supported
by photometric redshift estimates of this red arc.
\label{fig:arcredshifts}}
\end{figure*}

\subsection{MS2137}
This famous cluster presents tangential and radial arcs at $z = 1.501$ and 1.502, respectively \citep{Sand02}. We incorporate two additional images to the model of \citet{Sand08}: a fourth counterimage 3d to system 3, and the mirror image (2c) of the radial arc. The latter was not included in our previous analyses due to the difficulty of securing a clear identification in the light from the BCG, but the counterimage is clear in recent, deeper imaging from the CLASH survey \citep{CLASH}.

\subsection{A963}
\label{sec:A963SL}
A set of merging images forms the ``northern arc'' at $z = 0.771$ \citep{Ellis91}. Since conjugate points could not be clearly identified, we incorporate this arc as constraint on the position of the critical line, following \citet[][R10]{Richard10}, which is assumed to pass through the arc.

\subsection{A383}
\label{sec:A383lensing}
The model follows N11, which built upon \citet{Sand04,Sand08} and \citet{Smith05}. We add the pair of $z=6.027$ images (system 5) later identified spectroscopically by \citet{Richard11}, along with minor shifts to other image positions made based on deeper imaging from CLASH. The radial and tangential arc system at $z = 1.01$ \citep[systems 1 and 2,][]{Smith01,Sand04} and a complex system with a redshift $z = 2.55$ (system 3, N11) strongly constrain the mass model. Subsequent near-infrared observations confirmed the latter redshift via H$\alpha$ and [\ion{O}{3}] emission lines and provided a more precise value $z=2.58$ (S.~Belli et al., in preparation). We have not included system 6 as a constraint due to its peculiar and unexpected symmetry \citep[see discussion by][]{Morandi12}, but do report a spectroscopic redshift $z = 1.826$.

\subsection{A611}
\label{sec:A611multiples}
We adopt the model of N09, comprising a five-image system with an originally-reported redshift of $z = 2.06$ (system 1), a giant tangential arc at $z=0.908$ (system 2), and a four-image system with no spectroscopic redshift (system 3). These redshifts were published in R10. A subsequent near-infrared spectrum of system 1 revealed an unambiguous redshift of $z = 1.49$ via H$\alpha$, H$\beta$, and [\ion{O}{3}] emission (S.~Belli et al., in preparation). This shows that the redshift $z=2.06$ in R10 resulted from a misidentification of the single rest-UV emission line \ion{C}{3}] $\lambda 1909$ as \ion{C}{4} $\lambda 1549$. We return to the impact of this on mass models in Section~\ref{sec:previouslensing}. Additionally, the counterimage of the faint Ly$\alpha$ emitter identified in R10, whose position was suggested by the original lens model, is a less likely identification in models based on the new redshift.
  Thus, we do not include this system as a constraint. We located probable central counterimages of systems 1 and 3 well within the BCG light (see N09, Figure~6) based on predictions of the lens models. Although we have conservatively not imposed their positions as constraints, we verified that including the central image of system 1 (the more reliable identification) would not significantly influence our results.

\subsection{A2537}
\label{sec:A2537_SL}
This cluster displays many spectacular arcs that have so far not been modeled in the literature. We identify four systems with new spectroscopic redshifts of $z = 1.970, 2.786,$ and 3.607 (Figures~\ref{fig:slimages} and \ref{fig:arcredshifts}). Several conjugate images were initially identified on the basis of similar morphology to construct a preliminary lens model, which was iteratively refined to locate the positions of the other images. Image systems 1 and 2 are located within a three-fold ``naked cusp'' arc at $z = 2.786$. Systems 3 and 4 form five-fold images at $z = 1.970$ and $z=3.607$, respectively, both containing central images within the radial critical line. We discuss the inclusion of galaxy P1 as a perturber in our lens model in Section~\ref{sec:modeling}. 

\subsection{A2667}
Our model is based on that of \citet[][C06]{Covone06}. It consists of an extremely bright giant tangential arc at $z = 1.034$ \citep{Sand05} and two systems with no spectroscopic redshifts named B and D in C06 (3 and 4, respectively, in our nomenclature). Based on interim lens modeling, we identified two additional counterimages 4.3 and 4.4 shown in Figure~\ref{fig:slimages}. The giant arc is incorporated via two features (systems 1 and 2) located as flux maxima and minima.

\subsection{A2390}
The lens model is based on those presented in \citet{Jullothesis} and R10. It contains two arcs at $z = 4.05$, the H3 and H5 systems of \citet{Pello99}. (For reasons discussed in Section~\ref{sec:inferringmodels}, we do not include all the detectable conjugate points within these arcs as constraints.) The 41a/b system was previously identified on the basis of clear mirror symmetry but has no spectroscopic redshift. We secured a new spectroscopic redshift $z=0.535$ for the 51a/b system near the cluster center, as well as a redshift $z=1.036$ for the giant red arc (system B) to the southeast of the BCG based on very weak [\ion{O}{2}] emission (Figure~\ref{fig:arcredshifts}). Two conjugate points in the red arc were identified as flux minima in an \emph{HST}/WFC3-IR F125W image (proposal ID 11678). The lens model predicts a counter-image to the northeast of the BCG, which we locate but do not include as a constraint due to uncertainty in its precise position (it appears to be superposed on a singly imaged portion of the galaxy).

\subsection{Cluster galaxy identification}
\label{sec:galcat}

Strong lens models must account for mass in cluster galaxies, which perturb the positions of critical lines locally. We initially identified likely cluster galaxies as those with photometric redshifts near that of the cluster ($|\Delta z| < 0.15$). In A2537 and A2667, for which only two colors are available, we instead identified the locus of the cluster in the color-color plane. Absolute magnitudes in the $r$ band were estimated and compared to $M_{r,*} = -21.38$ \citep{Rudnick09}, appropriate to cluster galaxies at the redshifts of our sample. Only galaxies brighter than $0.1 L_{r,*}$ were considered, unless they fell close to a multiple image. Early-type galaxies with $L \simeq 0.1L_*$ have $\sigma \approx 90$~km~s${}^{-1}$ using the scaling relations we introduce in Section~\ref{sec:modeling}, which corresponds to deflection angle of $\simeq 0\farcs15$ in the singular isothermal sphere approximation, well within the uncertainty of $\sigma_{\textrm{pos}} = 0\farcs5$ that we assign to the image positions.
The radial extent of the sample was limited to extend safely beyond the strong-lensing zone.
This catalog was manually refined in some cases.
Although initially based on our multi-color ground-based catalogs, the parameters of the galaxies (center, ellipticity, P.A., flux) were refined using the \HST~imaging. The final catalogs contain $\simeq 10-60$ galaxies, varying with the richness of the cluster and the extent of the strong lensing zone.

\section{BCG photometry}

In order to model the distribution of stellar mass in the BCG and to interpret our kinematic observations, the luminosity profile of the galaxy must be known. Furthermore, we wish to relate the stellar mass-to-light ratios derived in our models to estimates from SPS, particularly in \PaperII. In this section, we present fits to the surface brightness profiles and broadband colors of the BCGs.

\subsection{Surface brightness profiles}
\label{sec:sbprofiles}

\begin{figure}
\centering
\includegraphics[width=\linewidth]{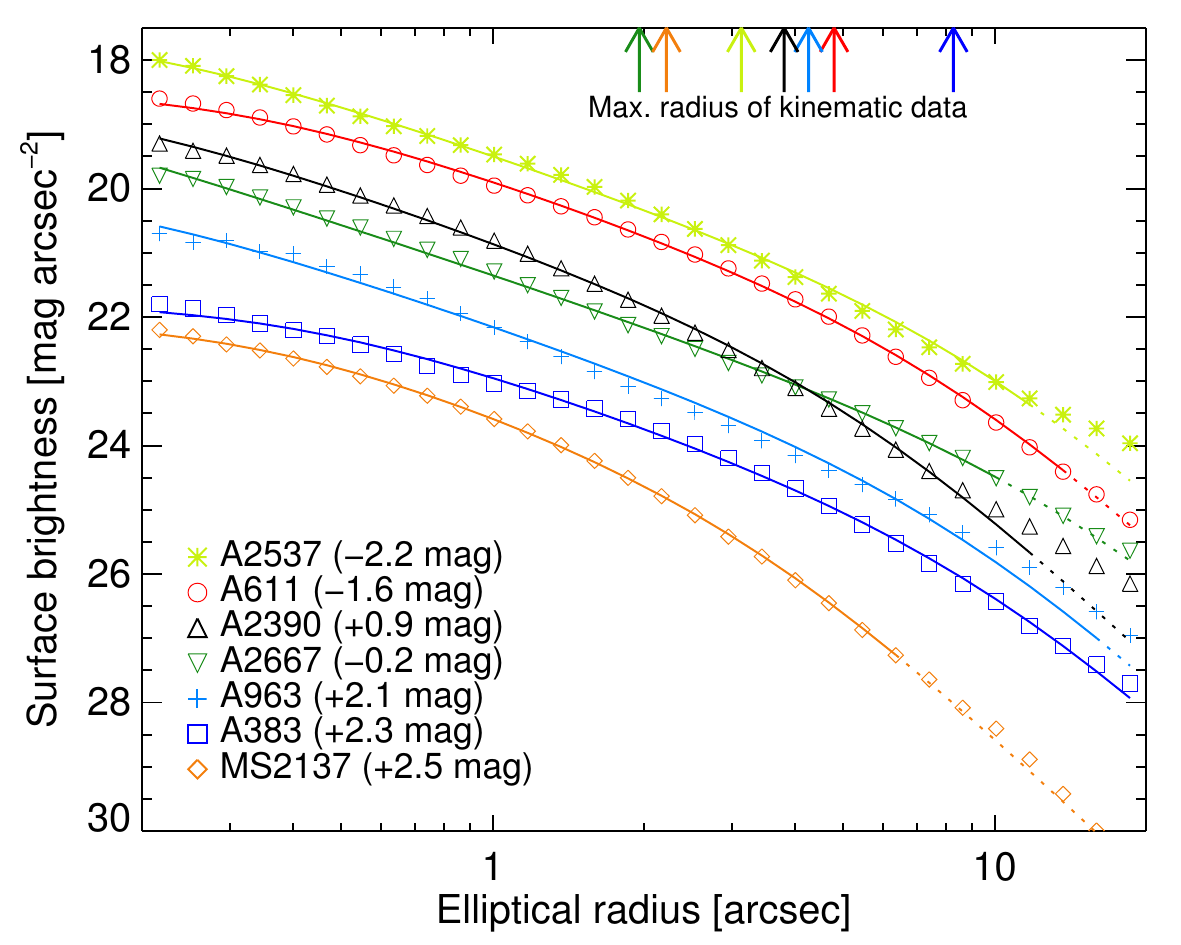}
\caption{Surface brightness profiles of BCGs, measured in \emph{HST} imaging through the filters indicated
in Table~\ref{tab:bcgphot}. Data are shown as diamonds, with formal errors usually smaller than the symbol size.
These are vertically offset as shown in the caption for clarity.
dPIE fits are drawn as solid lines throughout the radial interval most relevant for dynamical modeling and dotted outside. The critical interval is estimated approximately as where the surface brightness exceeds 10\% of that at the outer limit of the kinematic data (indicated by top arrows). \label{fig:bcg_sb}}
\end{figure}

\begin{deluxetable*}{lcccccccc}
\tablecolumns{9}
\tablewidth{\linewidth}
\tablecaption{\emph{HST} surface photometry of BCGs\label{tab:bcgphot}}
\tablehead{\colhead{Cluster} &
\colhead{Instrument/Filter} &
\multicolumn{5}{c}{dPIE Fit Parameters} & 
\colhead{$\LV$} & 
\colhead{Proposal}
\\
\multicolumn{2}{l}{} & 
\colhead{$r_{\textrm{cut}}$ (kpc)} &
\colhead{$r_{\textrm{core}}$ (kpc)} &
\colhead{$b/a$} &
\colhead{P.A.} &
\colhead{Mag.} & 
\colhead{($10^{11} L_{\odot}$)} & 
\colhead{ID}}
\startdata
MS2137 & ACS/F625W   & $18.7\pm2.6$ & 1.4  & 0.89 & 75    & 17.31 & 3.20 & 12102 \\
A963   & WFPC2/F702W & $35.6\pm4.6$ & 0.47  & 0.81 & 6.4  & 15.41 & 4.61 & 8249 \\
A383   & ACS/F606W   & $38.2\pm3.0$ & 1.2   & 0.89 & 8.7  & 15.81 & 4.06 & 12065 \\
A611   & ACS/F606W   & $46.2\pm3.4$ & 1.2  & 0.73 & 42.3  & 16.81 & 5.47  & 9270 \\
A2537  & ACS/F606W   & $52.7\pm6.5$ & 0.75  & 0.74 & $-58.5$ & 16.90 & 5.86 & 9270 \\
A2667  & WFPC2/F606W & $68.8\pm10.6$ & 0.26 & 0.69 & 40.4 & 16.33 & 3.89 & 8882 \\
A2390  & ACS/F850LP  & $24.4\pm2.9$ & 0.44 & 0.73 & $-50.6$ & 15.79 & 2.92 & 10504
\enddata
\tablecomments{Uncertainties in $r_{\textrm{cut}}$ include random and systematic errors assessed by varying
the background. Errors in $r_{\textrm{core}}$, $b/a$ and P.A.~(measured in degrees east of north)
are negligible for our analysis. Circularized radii are reported. The rest-frame $\LV$ is corrected for Galactic extinction; the observed magnitude is not. The uncertainty in the observed magnitude and in $\LV$ assuming a dPIE model is $\simeq 0.1$ mag.}
\end{deluxetable*}

\begin{deluxetable}{lccl}
\tablecolumns{4}
\tablewidth{\columnwidth}
\tablecaption{Stellar Population Synthesis Fits to BCGs\label{tab:sps}}
\tablehead{\colhead{Cluster} & \colhead{$\MLVSPS$} & \colhead{$N_{\textrm{filt}}$} & \colhead{Photometry source}}
\startdata
MS2137 & 2.05 & 10 & \emph{HST} ACS and WFC3 (CLASH) \\
A963 & 2.31 & 4 & SDSS DR8 \\
A383 & 2.26 & 7 & \emph{HST} ACS and WFC3 (CLASH) \\
A611 & 2.24 & 5 & \emph{Subaru} and \emph{HST} WFC3-IR \\
A2537 & 2.32 & 4 & SDSS DR8 \\
A2667 & 2.04 & 5 & \emph{HST} WFPC2, ACS, NICMOS \\
 & & & (proposal IDs 8882 and 10504) \\
A2390 & 1.80 & 5 & SDSS DR8
\enddata
\tablecomments{Stellar mass-to-light ratios $\MLVSPS$ are derived from SPS fits assuming a Chabrier IMF. $N_{\textrm{filt}}$ denotes the number of filters used in the fit. The luminosities $\LV$ are given in Table~\ref{tab:bcgphot} and include any internal dust extinction.}
\end{deluxetable}

Interpreting stellar dynamics in the BCG requires a model for the distributions of luminous tracers and mass. The dPIE parameterization\footnote{Also referred to as a PIEMD, or pseudo-isothermal elliptical mass distribution.} is particularly appropriate, since it is analytically convenient, widely used in lensing studies, and provides good fits to observed galaxies. It is characterized by two scale radii $r_{\textrm{core}}$ and $r_{\textrm{cut}}$, and the three-dimensional (3D) density is defined by
\begin{equation}
\rho_\textrm{dPIE}(r) = \frac{\rho_0}{(1+r^2/r_{\textrm{core}}^2)(1+r^2/r_{\textrm{cut}}^2)}.
\end{equation}
The analytic properties of the profile and the introduction of ellipticity are discussed by \citet{Eliasdottir08}. The spherical radius enclosing half of the light is $r_h \approx r_{\textrm{cut}}$, while the projected effective radius is $R_e \approx \frac{3}{4}r_{\textrm{cut}}$ in the limit $r_{\textrm{core}} / r_{\textrm{cut}} \ll 1$. We fit dPIE profiles to the BCGs in our sample using \HST~imaging obtained in reduced form from the Hubble Legacy Archive, selecting observations around 6000~\AA, which is close in wavelength to the absorption features used to derive kinematics (Section~\ref{sec:kinematics}). In A2390 we opted to use a F850LP observation instead, due to a prominent central dust feature, although this had little effect ($\sim 8\%$) on the derived radius. The filters and instruments used are listed in Table~\ref{tab:bcgphot}.

The background level in the \HST~images was adjusted based on blank sky regions far from the BCG. A noise map was constructed based on the background and shot noise from the BCG. Light from other galaxies in the field was carefully excluded using large elliptical masks generated from \code{SExtractor} parameters and then manually tuned. The geometric parameters of ellipticity, position angle (P.A.), and center were first determined by fitting an $R^{1/4}$ profile to the 2D data using \code{Galfit} \citep{GALFIT}. We then extracted elliptical isophotes and fit the 1D surface brightness profile in the inner $20\arcsec$ to a dPIE model using a custom code, accounting for the \HST~PSF. MS2137 and A383 present gradients in P.A., and the BCG geometry contributes to the modeling of their radial arcs. In these clusters, we thus fixed the P.A.~to that measured near these arcs.

Figure~\ref{fig:bcg_sb} demonstrates that this procedure produces goods fits to the data, particularly within the radial range most critical for the dynamical modeling (solid lines). In the inner $10\arcsec$, rms residuals are typically $5\%$. At larger radii, some BCGs have a cD-type upturn in their surface brightness profile that is not well fit with a single-component model \citep[e.g.,][]{Gonzalez05}. This causes errors in the total luminosity and radii, but these are correlated such that the surface luminosity density within $\simeq 10\arcsec$ is well fixed. This is all that is necessary for our dynamical and lens models, given that the kinematic data are confined to $R < 5\arcsec$ in all but one case (A383), and the mass budget is always DM-dominated beyond a few arcseconds.

Varying the background level produced $5\%-10\%$ systematic variations in $r_{\textrm{cut}}$. In five clusters we additionally fit a redder band (F850LP, F125W, or F160W) in Advanced Camera for Surveys (ACS) or WFC3 imaging to investigate trends with color. In three cases the derived radii agree to $<7\%$, within the systematic errors, while in the remaining pair (A611 and A383) the radii are $\simeq 20\%$ smaller in the redder band. Even in these cases, the color gradients are minimal ($<0.1$~mag) within $R \lesssim 7\arcsec$, so the differences mainly reflect gradients beyond $\sim R_e$. While the redder data likely better trace the stellar mass, the dynamics are DM-dominated at these large radii. We therefore considered it more important to accurately model the tracers and adopted the measurements at $\simeq 6000$~\AA. This choice is justified further in Section~\ref{sec:systematics}.

\subsection{Stellar population synthesis}
\label{sec:sps}

We additionally fitted SPS models to the BCG colors. Since the BCG is often saturated in our Subaru imaging, we also rely on photometry from the SDSS or \emph{HST} imaging. The SDSS colors are based on model magnitudes, while colors in \emph{HST} imaging are measured in apertures with radii $\simeq 2\farcs5$ that avoid other galaxies, local dust features, and arcs. (This aperture corresponds to roughly the radial extent over which the stellar mass dominates.) The \code{kcorrect} code \citep{Blanton07} was used to fit SPS models from which a $k$-correction to the rest-frame $V$-band luminosity $\LV$ was computed (Table~\ref{tab:sps}). The luminosity was scaled to match total flux of the dPIE model and corrected for Galactic extinction. We assigned errors of 10\% to all photometric measurements in the fitting process -- much larger than the random errors -- to account for systematic errors in the photometry and models.

These SPS models fits also provide an estimate of the stellar mass-to-light ratio $\MLVSPS = M_*/L_{\textrm{V}}$ appropriate for a \citet{Chabrier03} initial mass function (IMF). Following standard practice, the stellar masses refer to the current mass in stars and do not include any gas lost during stellar evolution \citep[e.g.,][]{Treu10,Cappellari12}.
The photometric data and derived $\MLVSPS$ ratios are listed in Table~\ref{tab:sps}.
Overall the $\MLVSPS$ estimates are quite uniform, with an rms scatter of only 9\%. Reassuringly, the BCGs with the lowest estimates (A2667, A2390, MS2137) are those that show the strongest emission lines (Section~\ref{sec:kinematics}) and the most prominent cooling cores. The far-infrared emission detected by \emph{Herschel} in A2390 and A2667 also indicates that these systems host some ongoing star formation \citep{Rawle12}.


By perturbing the photometric measurements by their errors, we estimate the typical random uncertainty in $\MLVSPS$ is about 0.07~dex. Systematic uncertainties were estimated by comparing measurements derived from a variety of codes. Firstly, we used \code{FAST} \citep{Kriek09} to construct grids of both \citet[][BC03]{BC03} and Charlot \& Bruzual (private communication, 2007, CB07) models with exponentially declining star formation histories.  The range of parameters was restricted appropriately for massive ellipticals: ages $t$ with $9.5 < \log t/\textrm{yr} < 10$, star formation timescales $\tau$ with $8 < \log \tau/\textrm{yr} < 9.5$, dust attenuation with $0 < A_V < 0.5$~mag, and solar metallicity. Mean stellar masses were estimated by marginalizing over the likelihood surface.
(Simply taking the best-fitting model elevated $\log M_*$ by $\simeq 0.05$~dex on average.)
Second, for A963 and A611 we are able to compare to the MPA/JHU catalog of SDSS galaxies (DR7; \citealt{Kauffmann03}). Finally, in addition to the above comparisons involving our BCG sample, we also used \code{kcorrect} to fit massive ellipticals at $0.15 < z < 0.35$ with four-band photometry observed in the SLACS survey. The resulting stellar masses were compared to those of \citet{Auger09}, which were based on carefully constructed priors.
In all of the above comparisons, we find systematic mean offsets of $< 0.06$~dex compared to the masses derived using \code{kcorrect}. This level of uncertainty is typical given the current state of SPS. We conclude that our stellar mass scale is close to that of other authors who use similar data.

\section{BCG kinematics}\label{sec:kinematics}

In this section, we present long-slit spectroscopic observations of the BCGs in our sample and the spatially resolved stellar kinematics derived from them. As we demonstrate below, the data are of sufficient quality to measure stellar velocity dispersions to typical radial limits of $\approx 10-20$~kpc, while the slit width and seeing limit the resolution on small scales to $\approx 3$~kpc. The stellar kinematic data thus probe the mass distribution from the smallest scales, where stars dominate the mass, out to radii where DM is dynamically significant. In combination with lensing, they provide a long lever arm with which to study the inner mass distribution.

\begin{figure*}
\centering
\includegraphics[width=0.40\linewidth]{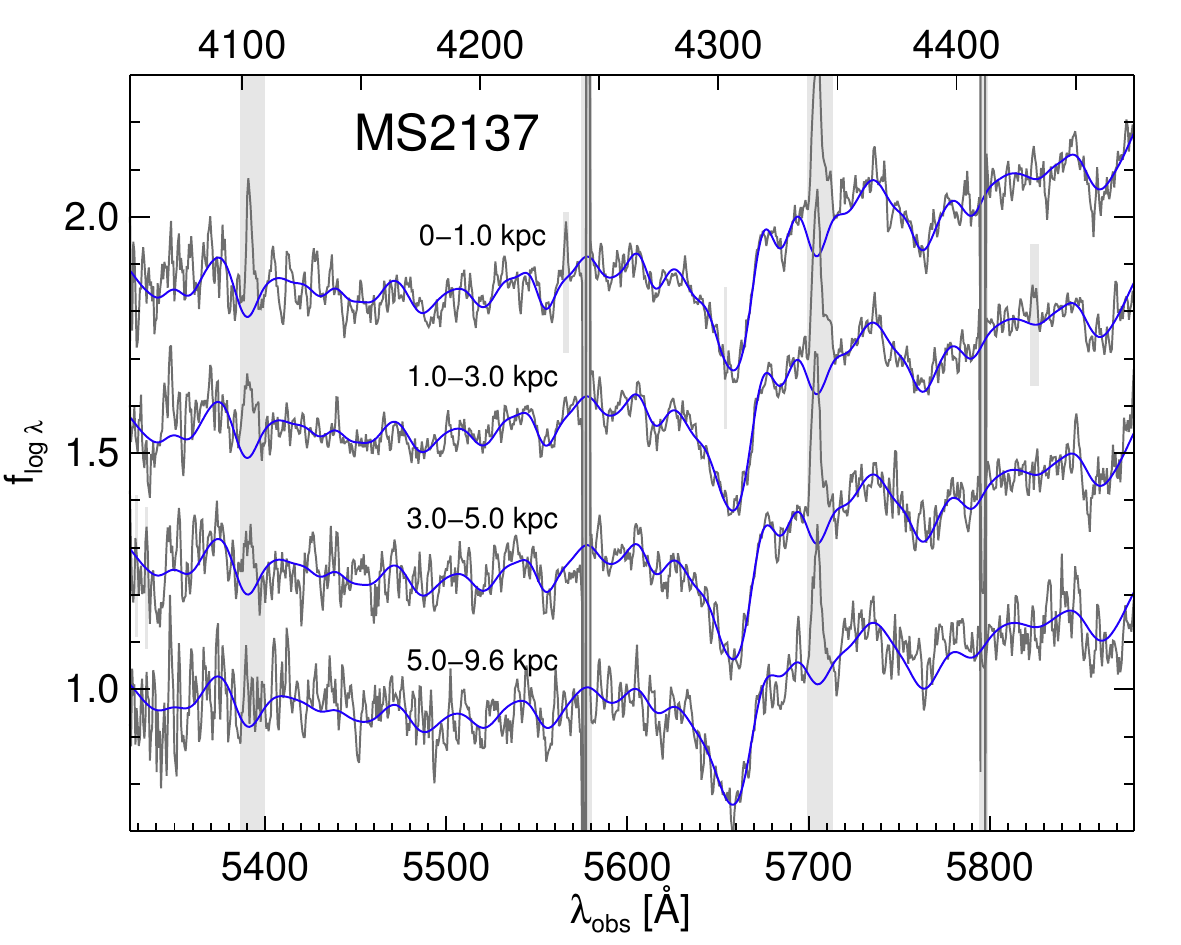}
\includegraphics[width=0.40\linewidth]{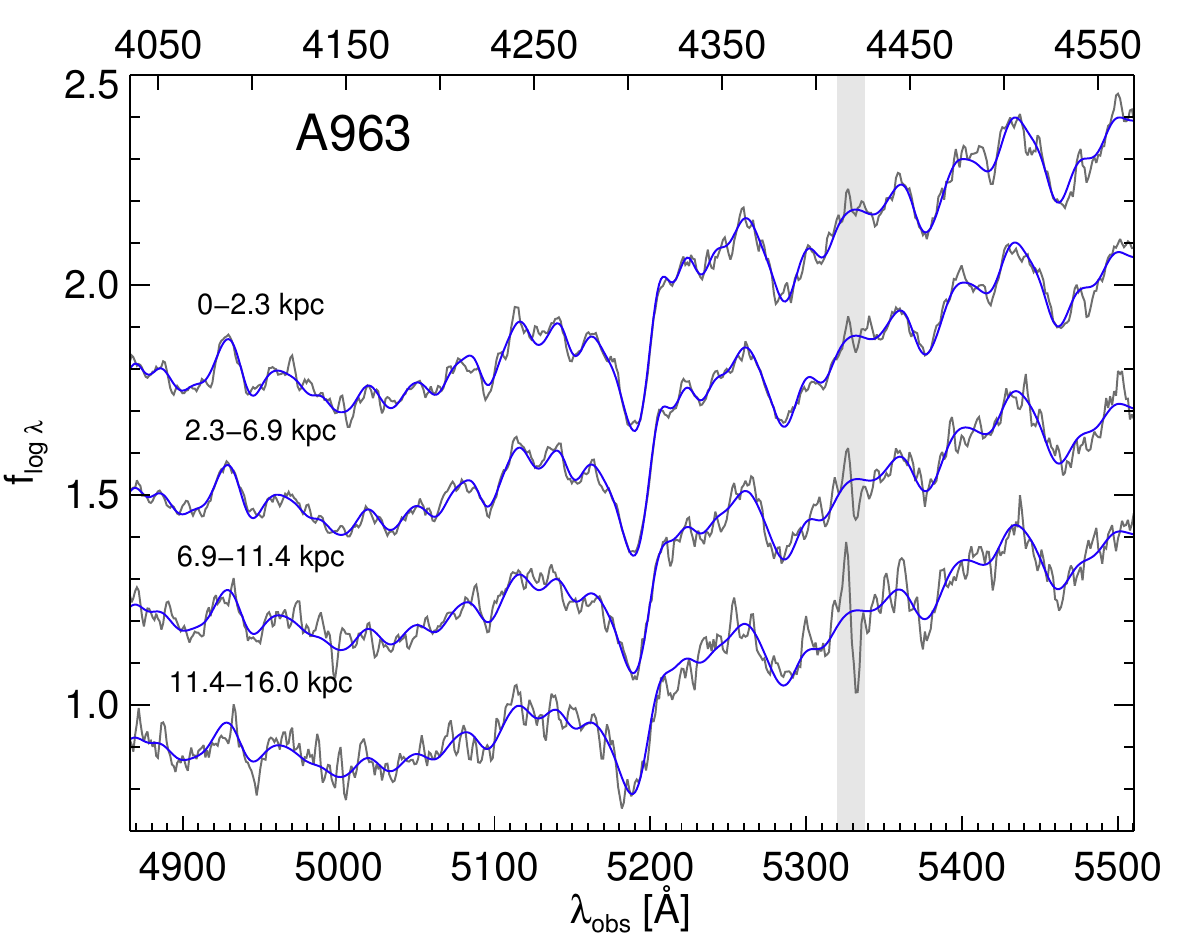} \\
\includegraphics[width=0.40\linewidth]{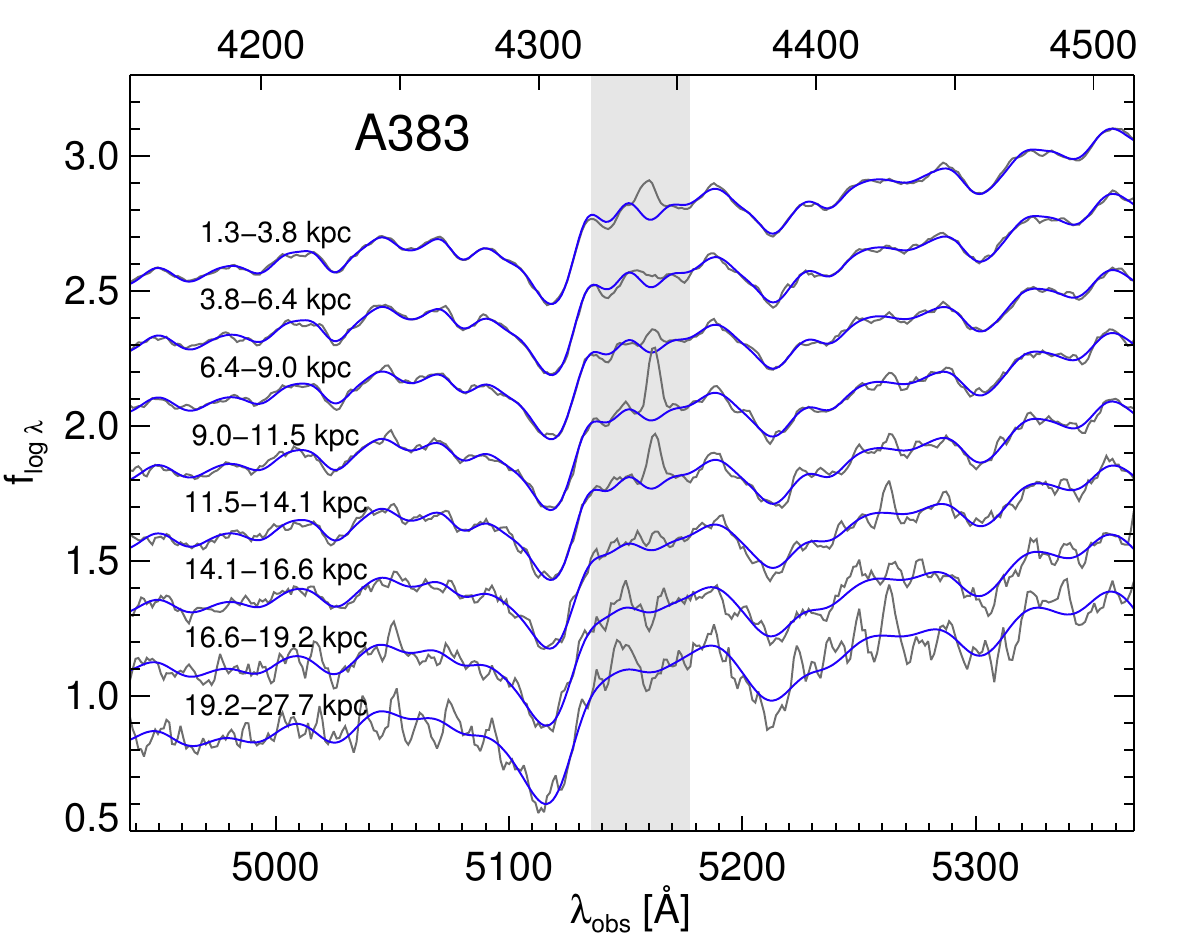}
\includegraphics[width=0.40\linewidth]{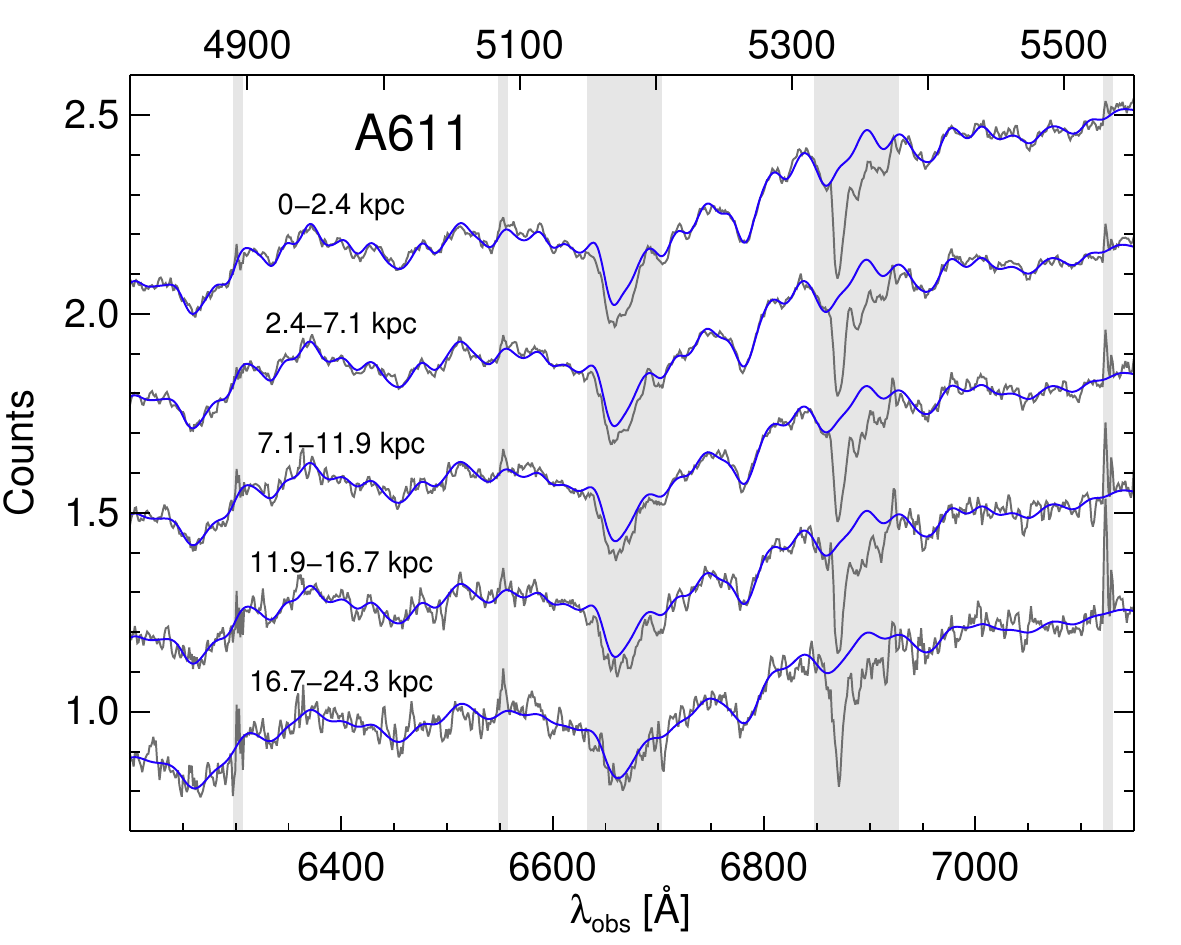}  \\
\includegraphics[width=0.40\linewidth]{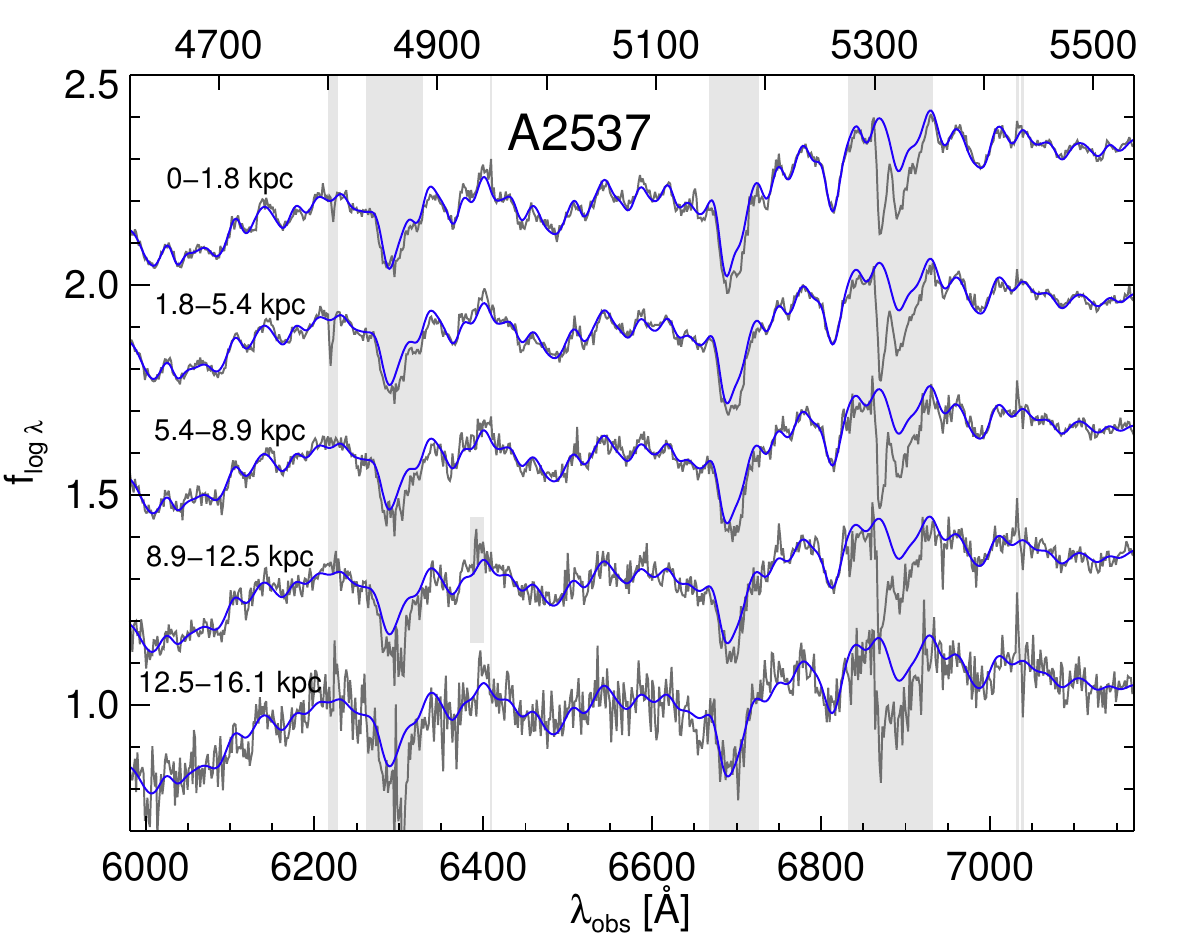}
\includegraphics[width=0.40\linewidth]{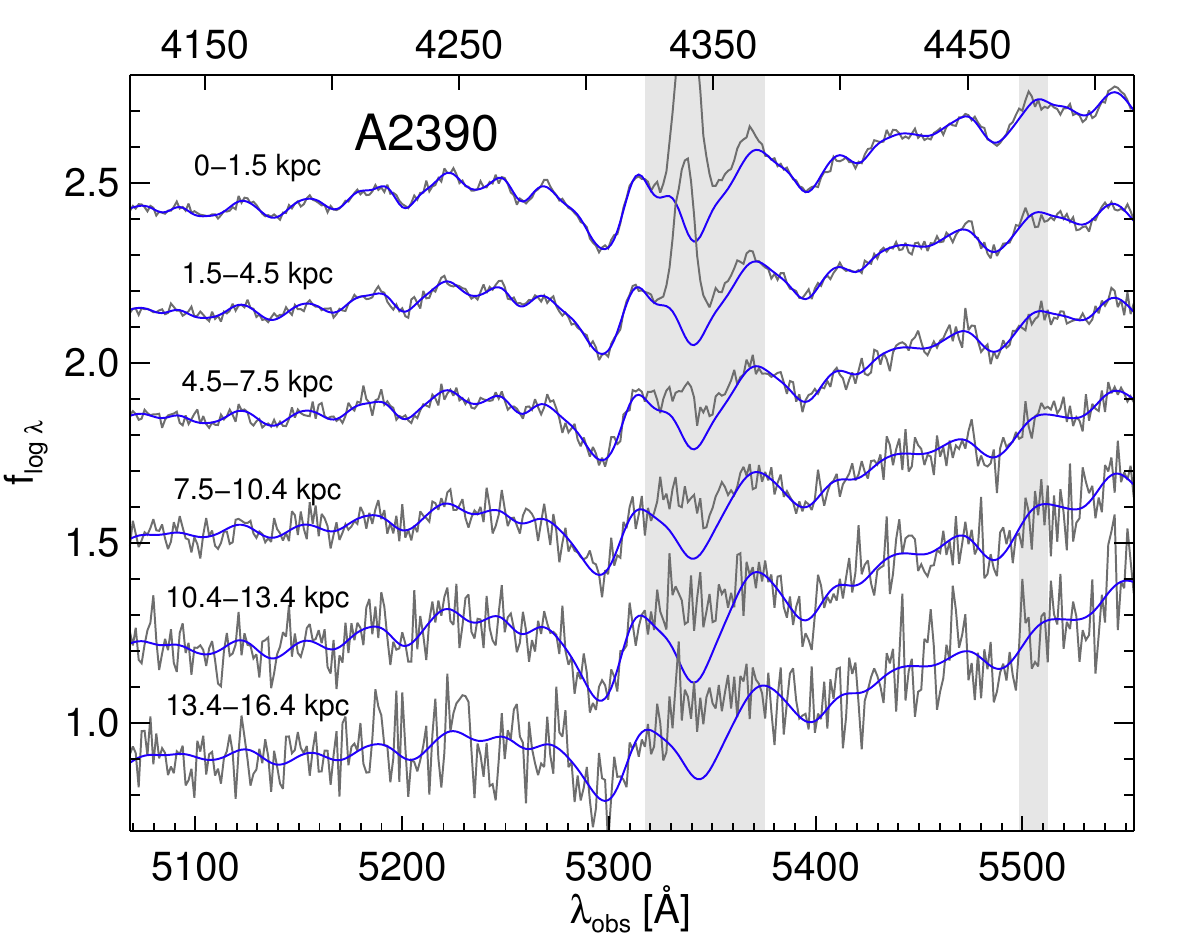} \\
 \includegraphics[width=0.80\linewidth]{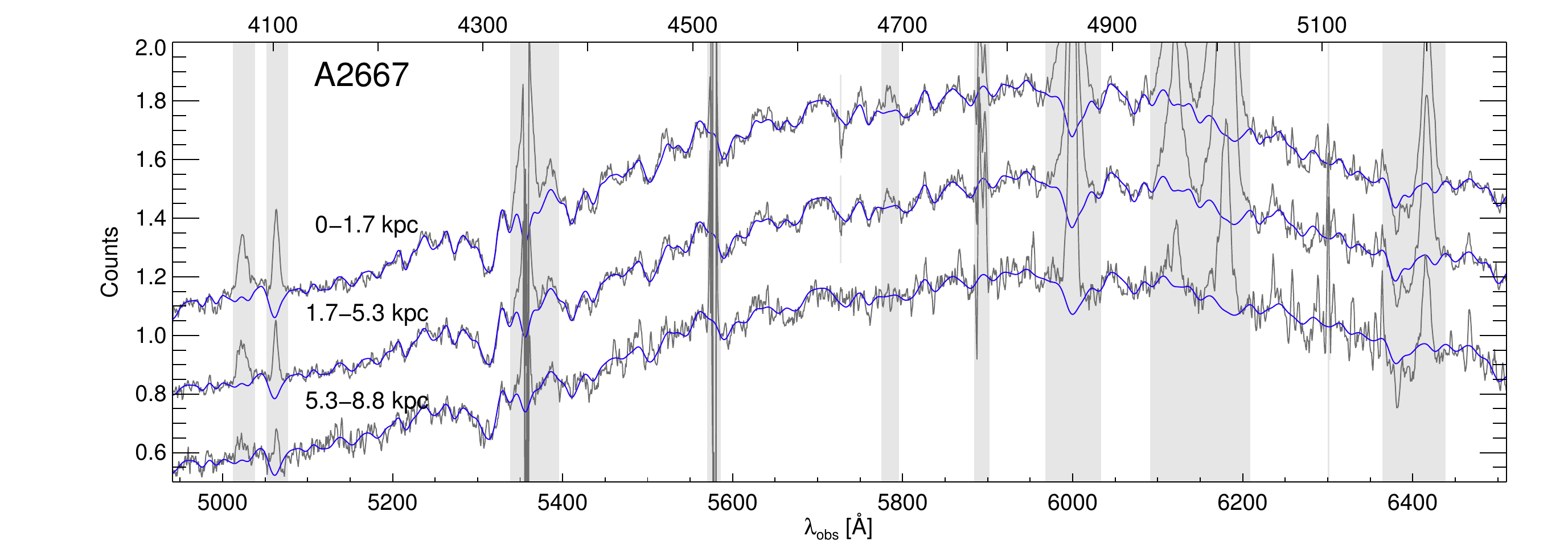}
\caption{Spatially resolved spectra of BCGs with fits used to measure kinematics. Gray lines show the data, and the fitted models are shown in blue. Each spatial bin is normalized to a median flux density of unity. The bins are then offset vertically for clarity. The top axis indicates the rest-frame wavelength. Gray bands denote masked pixels as described in the text. In A2390, A963, and A383 the Mg $b$/Fe spectral region was also observed and fitted, but only the $G$-band region is shown here. Symmetric spatial bins on either side of the BCG center are co-added for display purposes where possible, although fits were performed separately. (This was not done in A2390 due to its low-level rotation.) Spectra have been lightly smoothed with a 2~\AA~boxcar.\label{fig:specfits}}
\end{figure*}

\subsection{Observations and reduction}
\label{sec:vdobs}

We undertook spectroscopy of the BCGs using the Keck I and II and Magellan Clay telescopes, as recorded in Table~\ref{tab:speclog}. Total exposure times ranged from roughly 2 to 7 hr.  Five clusters were observed using the Low-Resolution Imaging Spectrometer (LRIS) on Keck I \citep{Oke95} using the 600 mm${}^{-1}$ grism blazed at 4000~\AA~in the blue arm and the 600 mm${}^{-1}$ grating blazed at 7500~\AA~in the red arm. A2537 and A2390 were observed through slitmasks in order to simultaneously secure redshifts of multiply imaged sources and of cluster members. The A383, A611, and A963 BCGs were observed using a long slit. In A383, we additionally observed a slitmask designed to cover gravitational arcs. The A611 and A383 observations were first presented in N09 and N11, respectively. MS2137 was observed using the Echelle Spectrograph and Imager (ESI; \citealt{Sheinis02}) on the Keck II telescope, as presented by \citet{Sand02}. Finally, A2667 was observed using LDSS-3 at the Magellan Observatory. In all but one case, the slit (Table~\ref{tab:speclog}) was aligned close to the major axis of the BCG, with some minor deviations tolerated to include gravitational arcs. For MS2137 the slit was instead aligned along the radial arc near the minor axis, although its isophotes are nearly circular.

The long-slit spectra were reduced with \code{IRAF} using standard techniques for bias subtraction, flat fielding, wavelength calibration, trace rectification, and sky subtraction as previously discussed in N09 and \citet{Sand04}.  For this work we have re-reduced the order of the ESI spectrum containing the $G$ band in MS2137 using similar methods. Multi-slit data were reduced using the software developed by \citet{Kelson03}. The wavelength-dependent instrumental resolution was measured via unblended sky lines or arc lamps and fitted with a low-order polynomial. The typical resolutions of the blue and red LRIS spectra are $\sigma = 159$ and 115~km~s${}^{-1}$, respectively, while the ESI and LDSS-3 observations have resolutions of $\sigma = 32$ and 84~km~s${}^{-1}$. These are much smaller than the velocity dispersions encountered in BCGs, so the uncertainties of a few km~s${}^{-1}$ in resolution have a negligible $\simeq 1\%$ effect on the derived dispersions.

\begin{deluxetable*}{lccccccc}
\tablecolumns{8}
\tablewidth{\linewidth}
\tablecaption{Spectroscopic observations\label{tab:speclog}}
\tablehead{\colhead{Cluster} &
\colhead{Instrument} &
\colhead{Date} &
\colhead{Exposure (ks)} &
\colhead{P.A. (deg)} &
\colhead{Seeing ($\arcsec$)} &
\colhead{Slit Width ($\arcsec$)} &
\colhead{Mode}}
\startdata
MS2137 & Keck/ESI &  2001 Jul~28 & 6.7 & 0 & 0.8 & 1.25 & Cross-dispersed \\
A963  & Keck/LRIS & 2012 Apr~18 & 7.8 & $-15.5$ & 2.5 & 1.5 & Long-slit \\
A383  & Keck/LRIS & 2009 Oct~12--14 & 23.7 & 2 & 0.7 & 1.5 & Long- and multi-slit \\
A611  & Keck/LRIS & 2008 Mar~1 & 7.8 & 45 & 1.4 & 1.5 & Long-slit \\
A2537 & Keck/LRIS & 2009 Oct~12--14 & 14.4 & 125 & 0.8 & 1.5 & Multi-slit \\
A2667 & Magellan/LDSS-3 & 2007 Jul~15, 17 & 19.8 & 27.4 & 0.9 & 1.0 & Long-slit \\
A2390 & Keck/LRIS & 2009 Oct~12--14 & 14.4 & $-45$ & 0.8 & 1.5 & Multi-slit
\enddata
\end{deluxetable*}

The center of the BCG was shifted to the center of a pixel during the reduction processes so that spatially-binned spectra could be extracted symmetrically on either side of the center. Our analysis focuses on two spectral regions with strong absorption features appropriate for kinematic study: the $G$ band at $\lambda 4308$ and the \ion{Mg}{1} $b$ region containing Fe $\lambda 5270$,  Fe $\lambda5335$ and other weaker lines. For the LRIS observations, the spatial bins were determined by adding CCD rows until a minimum signal-to-noise ratio (S/N) of 20~\AA${}^{-1}$ was reached in the Mg $b$/Fe spectral region of the LRIS-R spectrum, suitable for reliable kinematic measurements. A minimum number of rows comparable to the seeing element was also required. In some cases the outermost bin constructed by this scheme was conservatively excluded due to contamination of the key absorption features by sky residuals. Bins likely contaminated by flux from interloping galaxies were also excluded; this includes the innermost bin in A383.

When possible (A963, A2390, A383) identical spatial bins were extracted in the spectral region around the $G$ band in the LRIS-B spectrum, which was facilitated by the equal pixel scale of the detectors. Although the formal S/N is lower at the $G$ band, we found these spectra could nonetheless be reliably followed due to the cleaner sky. For A2537 and A611, the LRIS-B spectra were not used owing to the coincidence of the $G$ band with the \ion{O}{1} $\lambda5577$ sky line and the dichroic transition, respectively. For the ESI spectrum of MS2137, we considered only the order containing the $G$ band, since the Mg $b$/Fe region was strongly affected by atmospheric absorption. For the LDSS-3 spectrum of A2667, we extracted the rest-frame 4000--5280~\AA~interval, which was covered continuously. Figure~\ref{fig:specfits} shows the extracted spectra.

\subsection{Kinematic measurement technique\label{sec:kinmeasurement}}

In each spatial bin, the velocity and velocity dispersion were measured by direct fitting of Gaussian-broadened, redshifted stellar spectra using the \code{pPXF} software \citep{Cappellari04}, accounting for the instrumental resolution. An additive continuum polynomial was included in the fit, with the order determined identifying that beyond which the fit quality in the highest-S/N bin did not improve significantly. The derived velocity dispersions were insensitive to reasonable choices of the continuum order to a precision of $\simeq 1\%-3\%$. For the spectra that were not flux calibrated (A2667 and A611), a first- or second-order multiplicative polynomial was allowed to modulate the spectral shape. For flux-calibrated spectra this yielded no improvement in the fit, and the additional freedom was therefore excluded. Emission lines, regions of prominent sky subtraction residuals or absorption, and remaining defects were masked. Random uncertainties were assessed by shuffling the residuals in 5 pixel chunks, thus maintaining their correlation properties, adding these to the best-fitting model, and re-fitting the resulting spectra many times. This generally produced $1\sigma$ error estimates only slightly larger than those derived from the $\chi^2$ surface.

\begin{figure*}
\centering
\includegraphics[width=0.95\linewidth]{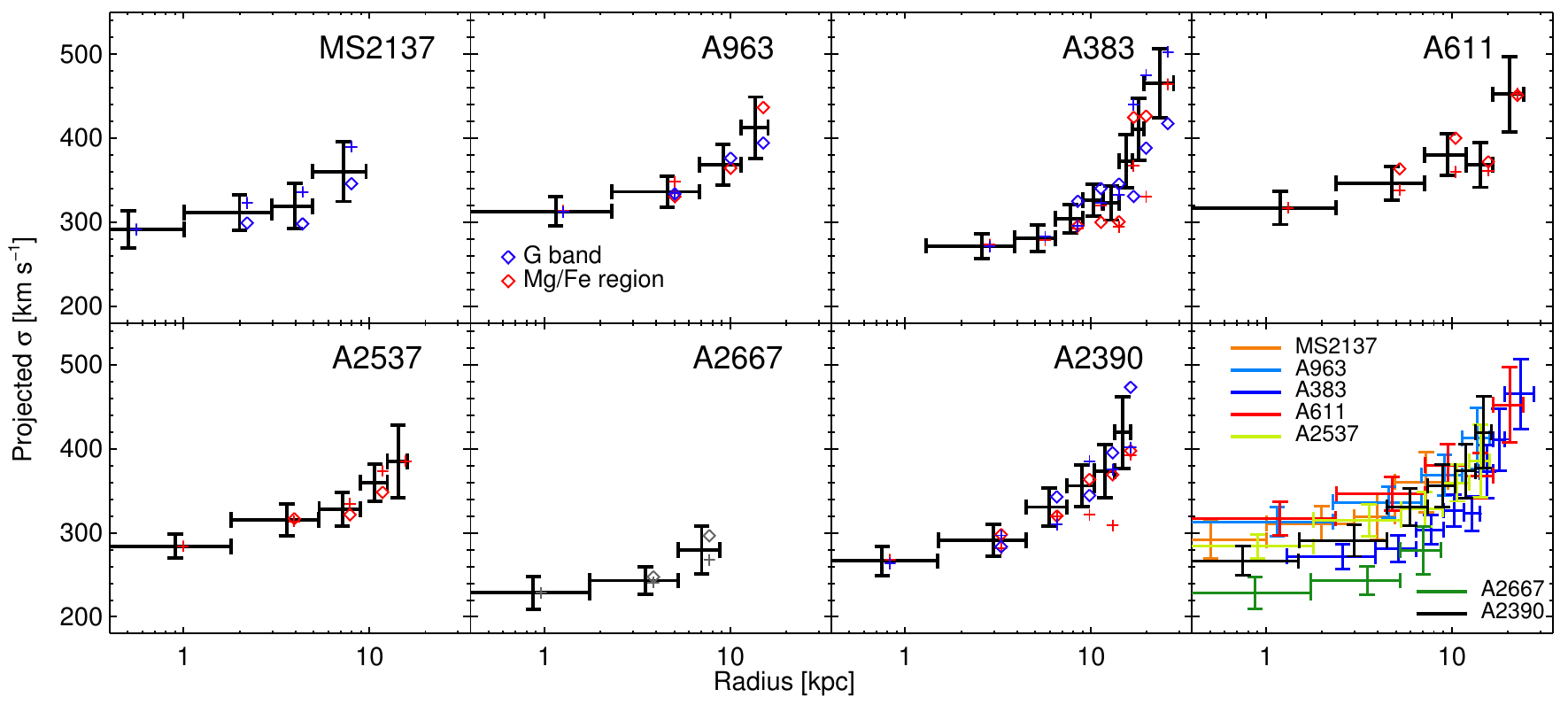}
\caption{Resolved stellar velocity dispersion profiles, with cross and diamond symbols denoting independent measurements on either side of the BCG center and colors denoting measurements in the spectral regions indicated in the caption. Radii are measured along the slit (i.e., are not circularized). Points with error bars show the weighted mean measurements, with errors including a systematic estimate as described in the text. The final panel combines these measurements for the full sample.\label{fig:vdprof}}
\end{figure*}

The stellar templates used to fit the BCG spectra were constructed from the MILES library \citep{MILES}. By default we allowed \code{pPXF} to build an optimal template from a linear combination 203 MILES stars with spectral types G5--K5 and luminosity classes III and IV, appropriate for old stellar populations. The template was determined using the spatially integrated spectrum and was then used to fit each spatial bin. \ion{Mg}{1} $b$, which is enhanced in massive galaxies, was masked since it generally produced biased results, consistent with other studies \citep{Barth02}. The resulting templates produce excellent fits to the BCG spectra, as shown in Figure \ref{fig:specfits}.

We experimented with including a wider range of stellar templates, including all non-peculiar stars of spectral types A--K in the MILES library and a subset that excludes those with low metallicity. For the A2390 and A2667 BCGs, some A- and F-type stars were preferred, consistent with the likely star formation activity discussed in Section~\ref{sec:sps}.
Our inclusion of these earlier spectral types impacts the derived dispersions in these systems by $\lesssim 5\%$. We also constructed templates based on the Indo-US coud\'{e} library \citep{Valdes04}. Finally, we experimented with templates optimized to each bin, rather than constructing a single template based on the integrated light; this led to no noticeable systematic changes. Details of the template construction led to systematic changes in the derived velocity dispersions at the $3\%-5\%$ level. Based on our estimates of uncertainties related to the template and the continuum polynomial order, we assign a systematic uncertainty of 5\% to all velocity dispersions, consistent with previous studies.

\subsection{Velocity dispersion profiles}
\label{sec:vdprofiles}

We detected no significant rotation in all but one BCG. In A2390, the measured rotation of $44 \pm 13$~km~s${}^{-1}$ is negligible compared to the central velocity dispersion, with $(v/\sigma)^2 = 0.026$. In the remainder of our analysis, we thus focus only the velocity dispersions. When multiple measurements of the dispersion in the same radial bin were available, either from fits on either side of the BCG center or in different spectral regions, they were combined with a weighted mean to produce a more precise estimate. This is justified given that the agreement between independent measurements is very good overall: of the 87 pairs of overlapping measurements, 79\% agree within $1\sigma$ using the random error estimates only. In a few bins the spread among estimates appeared greater than could likely be explained by random errors only, and in these cases the error bars were inflated based on the spread in estimates. In all cases, 5\% was added in quadrature to the final uncertainty to account for the systematic effects described in Section~\ref{sec:kinmeasurement}.

\begin{deluxetable*}{lcclcc}
\tablecaption{Velocity dispersion profiles}
\tablecolumns{6}
\tablehead{\colhead{Cluster} & \colhead{Radial Bin (arcsec)} & \colhead{$\sigma$ (km~s${}^{-1}$)} & \colhead{Cluster} & \colhead{Radial Bin (arcsec)} & \colhead{$\sigma$ (km~s${}^{-1}$)}}
\startdata
MS2137 & $0 - 0.22$ & $292 \pm 22$    & A2537 & $0 - 0.41$ & $284 \pm 14$ \\
\ldots & $0.22 - 0.65$ & $311 \pm 21$ & \ldots & $0.41 - 1.22$ & $315 \pm 19$ \\
\ldots & $0.65 - 1.08$ & $319 \pm 27$ & \ldots & $1.22 - 2.03$ & $328 \pm 20$ \\
\ldots & $1.08 - 2.09$ & $360 \pm 36$ & \ldots & $2.03 - 2.84$ & $360 \pm 22$ \\
A963 & $0 - 0.68$ & $313 \pm 17$      & \ldots & $2.84 - 3.65$ & $385 \pm 43$ \\
\ldots & $0.68 - 2.03$ & $336 \pm 18$ & A2667 & $0 - 0.47$ & $228 \pm 19$ \\
\ldots & $2.03 - 3.38$ & $369 \pm 24$ & \ldots & $0.47 - 1.42$ & $243 \pm 16$ \\
\ldots & $3.38 - 4.73$ & $413 \pm 36$ & \ldots & $1.42 - 2.36$ & $279 \pm 28$ \\
A383 & $0.41 - 1.22$ & $272 \pm 15$   & A2390 & $0 - 0.41$ & $266 \pm 17$ \\
\ldots & $1.22 - 2.03$ & $281 \pm 16$ & \ldots & $0.41 - 1.22$ & $291 \pm 19$ \\
\ldots & $2.03 - 2.84$ & $304 \pm 17$ & \ldots & $1.22 - 2.03$ & $331 \pm 23$ \\
\ldots & $2.84 - 3.65$ & $326 \pm 19$ & \ldots & $2.03 - 2.84$ & $356 \pm 25$ \\
\ldots & $3.65 - 4.46$ & $323 \pm 20$ & \ldots & $2.84 - 3.65$ & $374 \pm 32$ \\
\ldots & $4.46 - 5.27$ & $373 \pm 31$ & \ldots & $3.65 - 4.46$ & $420 \pm 43$ \\
\ldots & $5.27 - 6.08$ & $411 \pm 37$ & & & \\
\ldots & $6.08 - 8.78$ & $465 \pm 41$ & & & \\
A611 & $0 - 0.55$ & $317 \pm 20$      & & & \\
\ldots & $0.55 - 1.65$ & $347 \pm 20$ & & & \\
\ldots & $1.65 - 2.75$ & $380 \pm 25$ & & & \\
\ldots & $2.75 - 3.85$ & $368 \pm 27$ & & & \\
\ldots & $3.85 - 5.61$ & $452 \pm 45$ & & &
\enddata
\tablecomments{Line-of-sight velocity dispersions are derived from averaging observations on either side of the BCG center and, in most cases, in multiple wavelength intervals, as described in Section~\ref{sec:vdobs}. Radii are measured along the slit, which is oriented near the major axis with the exception of MS2137; they can be circularized using the axis ratios in Table~\ref{tab:bcgphot}. Error bars include a 5\% systematic component added in quadrature.\label{tab:vddata}}
\end{deluxetable*}

The derived velocity dispersion profiles for each cluster are shown in Figure~\ref{fig:vdprof}, including the weighted mean estimate and the individual measurements described above. The data are listed in Table~\ref{tab:vddata}. In all seven clusters, the velocity dispersion rises with radius. This contrasts strikingly with massive field ellipticals, which show velocity dispersion profiles that are flat or slowly declining \citep[e.g.,][]{Carollo95,Gerhard01,Padmanabhan04}. Our data imply a strongly rising total mass-to-light ratio, which we will show in Section~\ref{sec:discussion} can be naturally explained by the cluster-scale halo. An alternative explanation for the rising dispersions is that the stellar orbits rapidly become more tangential at large radii. This can be tested using the detailed shape of stellar absorption lines in nearby systems, which would reveal ``peakier'' profiles at large radii if circular orbits dominate. Observations of local cD galaxies instead favor nearly isotropic or mildly radial orbits \citep{Carter99,Kronawitter00,Saglia00,Hau04}, which indicates that the rising dispersions are not an artifact of the orbital distribution but reflect the genuine dynamical influence of the cluster potential.

\subsection{Comparison to previous work}

We have reanalyzed the spectra of A611 and MS2137 presented in N09 and \citet{Sand02}, respectively, and obtained a new, deeper spectrum of A963 compared to \citet{Sand04}. The A383 spectrum and kinematic measurements are identical to N11, with the exception of a small adjustment ($<1\sigma$) to the outermost bin only. However, the velocity dispersion measurements in A611, MS2137, and A963 have changed systematically and significantly compared to the previously published values. While the earlier works (\citealt{Sand02,Sand04,Sand08}; N09)  indicated a flat or even declining (in the case of MS2137) dispersion profile in these clusters, we find a rising trend in common with the rest of the sample.

Given that multiple codes and techniques were used to reduce the present data, yielding very similar dispersion profiles (Figure~\ref{fig:vdprof}, final panel), the differences in these measurements appear unrelated to the data reduction itself. More likely they arise from improvements to the velocity dispersion measurement procedure. In particular, we now (and in N11) rely on large libraries of high-quality stellar spectra to construct templates, whereas earlier works were restricted to a relatively small number of stars observed with ESI. Furthermore, we now construct composite templates from linear combinations of these spectra, rather than taking a single star. This provides much higher-quality fits (Figure~\ref{fig:specfits}) with virtually no residual ``template mismatch.'' We have also tested the dispersion measurements in MS2137 using an independent code developed by M.~Auger and find identical results (A.~Sonnenfeld, private communication, 2012). The earlier suboptimal templates used in earlier works probably led to biases at higher $\sigma$ or lower S/N. Given the high quality of the data (the rising $\sigma$ can be seen by eye in many panels of Figure~\ref{fig:specfits}), the improved methodology, and the resulting uniformity of the dispersion profiles, we are confident in the present results.

\section{Modeling the Cluster Mass Distribution}
\label{sec:modeling}

Having introduced the observational data that form the basis of our analysis, we now describe the models and methods that we use to infer the cluster mass distribution. As in N09 and N11, our mass model consists of three components: the DM halo, the stars in the BCG, and the mass in other cluster galaxies. Each is described by one or more analytic models, introduced below, and the parameters of these models are constrained simultaneously using our full data set. 

Two flexible functional forms are adopted to describe the dark halo. In addition to length and density scaling parameters, each includes a third parameter that allows for variation in the shape of the density profile. In particular, they allow for deviations in the inner regions from the CDM density profiles produced in numerical simulations. As we described in Section \ref{sec:intro}, this is the region where the effects of baryons or non-standard DM should be the most pronounced. The generalized NFW profile (gNFW, \citealt{Zhao96}), given by
\begin{equation}
\rho_{\textrm{DM}}(r) = \frac{\rho_s}{(r/r_s)^\beta (1 + r/r_s)^{3-\beta}},
\label{eqn:gNFW}
\end{equation}
reduces to the NFW profile when $\beta = 1$, but the asymptotic inner slope $d \log \rho_{\textrm{DM}} / d \log r = -\beta$ as $r \rightarrow 0$ can be varied. When we fix $\beta = 1$ to fit NFW models, we refer to the virial mass $M_{200}$ as that within a sphere of radius $r_{200}$ that has a mean density equal to 200 times the critical density $\rho_{\textrm{crit}}$ of the universe at the cluster redshift. The concentration is then $c_{200} = r_{200} / r_s$. 

In order to verify that our results do not strongly depend on the functional form of the density profile, we have introduced a second parameterization that we refer to as a ``cored NFW'' (cNFW) model:
\begin{equation}
\rho_{\textrm{DM}}(r) = \frac{b \rho_s}{(1 + b r/r_s) (1 + r/r_s)^2}.
\label{eqn:corednfw}
\end{equation}
This is simply an NFW profile with a core introduced, i.e., with asympotically constant density as $r \rightarrow 0$. The scale of the core is controlled by the parameter $b$. A characteristic core radius can be defined as $r_{\textrm{core}}  = r_s / b$; at this radius, the density falls to half that of an NFW profile with equal $r_s$ and $\rho_s$. As $r_{\textrm{core}} \rightarrow 0$ ($b \rightarrow \infty$) the profile approaches the NFW form. We follow the \code{Lenstool} convention and use the parameter $\sigma_0^2 = \frac{8}{3} G \rho_s r_s^2$ in place of $\rho_s$. This is simply a defined scaling and should not be taken as the actual velocity dispersion.

We considered additionally using Einasto models, which have been shown to provide more accurate representations of halos in numerical simulations \citep[e.g.,][]{Merritt06,Navarro10,Gao12}. However, this form is not optimal for observational studies of the inner halo, because the behavior at large and small radii are strongly coupled: to explore flat inner profiles, one has to accept steep declines in the outer regions. By contrast, the large-radius behavior of the gNFW and cNFW density profiles are invariant. Further, Einasto profiles with the range of shape parameters seen in simulations can be approximated by gNFW profiles within $\simeq 10\%$ over the relevant range of radii.

The stellar mass in the BCG is modeled with a dPIE profile, introduced in Section~\ref{sec:sbprofiles}.\footnote{Note that there is no distinction between the halo of the BCG and that of the cluster, which would be observationally impossible and is not well defined theoretically.} The center, P.A., ellipticity, and scale lengths $r_{\textrm{core}}$ and $r_{\textrm{cut}}$ are fixed based on the fits to \HST~imaging described in that section. The only free parameter is then the stellar mass-to-light ratio $\MLV = M_*/L_{\textrm{V}}$, which we assume to be spatially invariant within the BCG. (This assumption is discussed further in Section~\ref{sec:systematics}.) We parameterize $\MLV$ relative to the values $\MLVSPS$ derived from our SPS fits, based on a Chabrier IMF (Section \ref{sec:sps}):
\begin{equation}
\log \alpha_{\textrm{SPS}} = \log \MLV / \MLVSPS
\end{equation}
\citep{Treu10}. We place a very broad uniform prior on $\log \alpha_{\textrm{SPS}}$, corresponding to a mass that is $1.5\times$ lighter than $\MLVSPS$ to a mass $2\times$ heavier than the $\MLVSPS$ inferred using a Salpeter IMF, where we take $\log M_{*,\textrm{Salp}}/M_{*,\textrm{Chab}} = 0.25$. The total allowed range in $\MLV$ is thus a factor of 5.3.

The final ingredient in the mass model is the dark and luminous mass in non-BCG cluster galaxies, which are significant perturbations in the strong-lensing analysis. The identification of these galaxies was described in Section~\ref{sec:galcat}. Their mass is modeled using dPIE profiles. The center, ellipticity, and P.A.~are fixed to that of the light, and for most of the cluster galaxies, the structural parameters are tied to scaling relations specific to each cluster (e.g., \citealt{Limousin07}; N09; \citealt{Richard10}):
\begin{equation}
\begin{aligned}
r_{\textrm{cut}} &= r_{\textrm{cut},*} (L_r/L_{r,*})^{1/2} \\
r_{\textrm{core}} &= r_{\textrm{core},*} (L_r/L_{r,*})^{1/2} \\
\sigma &= \sigma_* (L_r/L_{r,*})^{1/4}
\end{aligned}
\end{equation}
Following previous work (e.g., N09; \citealt{Richard10}), we place a Gaussian prior on $\sigma_*$ of $158 \pm 27$~km~s${}^{-1}$ based on the observed scaling relations in the SDSS \citep{Bernardi03}.
Based on the galaxy-galaxy lensing study of \citet{Natarajan09}, we allow $r_{\textrm{cut},*}$ to vary from 15 to 60 kpc. As those authors note, this is much larger than the optical radius of the galaxies, and our dPIE models therefore include galaxy-scale dark halos. Our analysis is insensitive to $r_{\textrm{core},*}$, which is thus fixed to 0.15~kpc.

These scaling relations are sufficient for the majority of cluster galaxies. In some cases, however, the position of a multiple image can be strongly influenced by a nearby galaxy. In these situations, the galaxy is freed from the scaling relations and modeled individually. These galaxies are indicated in Figure~\ref{fig:slimages}. It is sufficient to free either $\sigma$ or $r_{\textrm{cut}}$, since their effects are degenerate, and in practice we usually fix $\sigma$ based on the \citet{Bernardi03} results and vary $r_{\textrm{cut}}$. We note one peculiar case, that of galaxy P1 in A2537 (Section~\ref{sec:A2537_SL}). We found that individually optimizing this perturber improved the modeling of the arc system composed of images families 1 and 2, although P1 is clearly deflected and located behind the cluster ($z_{\textrm{phot}} = 0.59 \pm 0.04$ in SDSS DR8). This suggests a possible interesting two-plane effect, which is beyond the scope of this paper to fully model. Nevertheless, we find that the inferred mass parameters are consistent with an $\simeq L_*$ galaxy, which agrees reasonably with the (demagnified) luminosity.

The ICM is not modeled as a distinct mass component in our analysis and is therefore implicitly incorporated into the halo. In the present paper we focus on the total density profile, so the separation of the DM and gas is not a concern. Based on the $\simeq 3$~kpc spatial resolution of our spectra, we also do not consider a supermassive black hole. Observations of local BCGs indicate this becomes dynamically significant only at smaller scales $\lesssim 1$~kpc \citep[e.g.,][]{Kelson02}.

\begin{deluxetable}{lcc}
\tablecolumns{3}
\tablewidth{\columnwidth}
\tablecaption{Prior Distributions Used in the Cluster Mass Models\label{tab:priors}}
\tablehead{\colhead{Parameter} & \colhead{Units} & \colhead{Prior}}
\startdata
\sidehead{\emph{Cluster-scale dark matter halo}}
$\epsilon$ (pseudoellipticity) & \ldots & $U(\ldots)^\dagger$ \\
P.A. & deg & $U(\ldots)^\dagger$ \\
$r_s$ & kpc & $L(50, 1000)$ \\
$\sigma_0$ & km~s${}^{-1}$ & $L(500,3500)$ \\
$\beta$ (gNFW models) & \ldots & $U(0.01,1.5)$ \\
$b$ (cored NFW models) & \ldots & $L(1,1000)$ \\
\sidehead{\emph{Stellar mass in BCG}}
$\log \alpha_{\textrm{SPS}}$ & \ldots & $U(-0.176, 0.551)$ \\
\sidehead{\emph{Cluster galaxy scaling relations}}
$\sigma_{*}$ & km~s${}^{-1}$ & $G(158 \pm 27)$ \\
$r_{\textrm{cut},*}$ & kpc & $U(15,60)$ \\
\multicolumn{3}{l}{Individually-optimized galaxies each add an additional} \\
\multicolumn{3}{l}{parameter as discussed in the text.} \\
\sidehead{\emph{Weak-lensing shear calibration}}
$m_{\textrm{WL}}$ & \ldots & $G_{2\sigma}(0.89 \pm 0.05)$ \\
\cutinhead{Additional parameters for individual clusters}
\sidehead{A611}
Redshift of source 3 & \ldots & $U(1,2)$ \\\hline
\sidehead{A2667: second NFW clump at $R \simeq 1.4$~Mpc}
$\Delta x$ & arcsec & $G(7 \pm 45)$ \\
$\Delta y$ & arcsec & $G(370 \pm 45)$ \\
$\epsilon$ & \ldots & $U(0, 0.3)$ \\
P.A. & deg & $U(0, 180)$ \\
$M_{200}$ & $\msol$ & $L(10^{13}, 10^{15})$ \\
$\textrm{ln}~c_{200}$ & \ldots & $G(\textrm{ln}(4) \pm 0.4)$ \\
Redshift of source 3 & \ldots & $U(1, 4.5)$ \\
Redshift of source 4 & \ldots & $U(1, 4.5)$ \\[0.15cm]\hline
\sidehead{A2390}
$\Delta x$ & arcsec & $G(0 \pm 1.5)$ \\
$\Delta y$ & arcsec & $G(0 \pm 1.5)$ \\\hline
\sidehead{A383 (see Section \ref{sec:A383} and N11)}
$q_{\textrm{DM}}$ & \ldots & $U(1,2.5)$\\
$q_*$ & \ldots & see N11 \\
$m_{\textrm{X}}$ & \ldots & $G_{2\sigma}(0.9 \pm 0.1)$
\enddata
\tablecomments{$U(x,y)$ denotes a uniform prior over the interval bounded by $x$ and $y$. $L(x,y)$ denotes a prior that is uniform in the logarithm. $G(\mu \pm \sigma)$ denotes a Gaussian prior with mean $\mu$ and dispersion $\sigma$, while $G_{2\sigma}$ denotes a Gaussian prior truncated at $2\sigma$. Positions $\Delta \textrm{x}$ and $\Delta \textrm{y}$ are given relative to the BCG; positive values indicate west and north, respectively. Position angles are measured east of north. ${}^\dagger$ The intervals were determined based on initial lensing fits; see the text for the special case of A963.}
\end{deluxetable}

\subsection{Additional mass components}
\label{sec:additionalcomponents}

In A2667 the weak lensing map (Figure~\ref{fig:kappa}) shows a clear second clump located $\simeq 1.4$~Mpc north of the BCG, which is likely in the foreground (Section~\ref{sec:wlresults}). Due to the large separation, this mass is unimportant for our strong lensing and dynamical analysis, but it must be considered for weak lensing. We therefore added a second dark halo to the model near the position indicated in the 2D mass map, as listed in Table~\ref{tab:priors}. Since the internal structure is not well constrained by the shear data, an NFW profile is assumed with a broad log-normal prior on $c_{200}$. The mean of this prior was taken to be 4, appropriate to the virial mass of $\log M_{200} / \msol = 14.7$ inferred from the full modeling discussed below, although adopting an even broader prior did not significantly affect the results. 

We experimented with adding a second mass clump to the west of the BCG in A2390, based on the extension of galaxies and X-ray emission on $\simeq 100$~kpc scales discussed in Section~\ref{sec:thesample}, but found that this did not improve the quality of the fit to the lensing data and substantially lowered the Bayesian evidence. We therefore consider a single dark clump to be sufficient. In A2537 the curvature of the arcs suggests a possible additional mass clump to the north of the BCG, which is given further credence by the multimodal dynamical structure described in Section~\ref{sec:thesample}. We experimented with adding a second clump and found that it did improve the Bayesian evidence when only strong lensing constraints are fit, but not with the full data set. The inferred mass was small ($\simeq 1 \times 10^{13} \msol$), and correspondingly the most relevant parameters for our study (halo mass and concentration, inner slope, $\MLV$) change little. Therefore, we retain a single dark clump when fitting this cluster also.

\subsection{Inferring mass models from data}
\label{sec:inferringmodels}

Our analysis is based on the \code{Lenstool} code \citep{Kneib93,Jullo07}, which has been widely used for studying strong lenses. For this project, we have added components to \code{Lenstool} that incorporate weak lensing and stellar kinematic constraints. The inference method is fully Bayesian. The prior distributions we adopted are listed in Table~\ref{tab:priors}. For the key parameters (i.e., those describing the DM halo and $\MLV$) we chose uninformative priors that are broad and flat. A Markov Chain Monte Carlo (MCMC) method is used to explore the large parameter space \citep{Jullo07}. We checked for convergence of the MCMC chains by inspecting their traces, running means, and auto-correlation functions.

For each model proposed by the MCMC sampler, a likelihood is computed based on the full data set. Since we assume the errors in our measurements are independent and Gaussian, this is equivalent to summing $\chi^2$ terms based on the strong lensing, weak lensing, and stellar velocity dispersion constraints:
\begin{equation}
\chi^2 = \chi^2_{\textrm{SL}} + \chi^2_{\textrm{WL}} + \chi^2_{\textrm{VD}}.
\end{equation}

The strong lensing analysis is conducted in the image plane, with
\begin{equation}
\chi^2_{\textrm{SL}}  = \sum\limits_{i} \frac{(x_i - x_i^{\textrm{obs}})^2 + (y_i - y_i^{\textrm{obs}})^2}{\sigma_{\textrm{pos}}^2},
\end{equation}
where $(x_i, y_i)$ and $(x_{i}^{\textrm{obs}}, y_{i}^{\textrm{obs}})$ are the predicted and observed positions, respectively, of a single image, and the sum runs over all multiple images \citep[see][]{Jullo07}. In two clusters somewhat different techniques were used. In A383, $\chi^2_{\textrm{SL}}$ was calculated in the source plane when we include kinematic data, due to the slower two-integral dynamics we compute only in this system (Section~\ref{sec:A383}). We verified this has a minimal effect on the results. Secondly, in A963 the merging images that form the tangential arc could not be clearly separated (Section~\ref{sec:A963SL}). We therefore identified a symmetry point and required that the critical line pass through it, with a positional uncertainty of $0\farcs2$. We also imposed Gaussian priors of $\epsilon = 0.21 \pm 0.02$ (the pseudoellipticity introduced below) and P.A.~$ = (86 \pm 3)^{\circ}$, based on the shape of the isophotes at the radius of the tangential arc, since the break point provided to \code{Lenstool} cannot constrain them.

The uncertainty in the image positions $\sigma_{\textrm{pos}}$ is a key quantity when combining strong lensing with other data sets. Although compact images can in principle be located in \HST~imaging with an astrometric precision of $\lesssim 0\farcs05$, cluster lens models are generally not able to reproduce image positions to better than $\sigma \simeq 0\farcs2-0\farcs3$, with a scatter of up to $\sim 3''$ in the best-studied clusters \citep[e.g.,][]{Limousin07}. This is likely partly due to perturbations by unmodeled substructures, either in the cluster or along the l.o.s.~\citep{Jullo10}. An additional factor is that simply parameterized models are not perfect representations of real or simulated clusters. This is particularly important when combining diverse data: since strong-lensing constraints are exquisitely precise, assigning a very small positional uncertainty can fully constrain the model. Given that strong lensing, weak lensing, and stellar kinematics contribute comparably to the logarithmic radial extent of our study, it is important not to overly concentrate the weight of the data in one radial interval.

We find that $\sigma_{\textrm{pos}} = 0\farcs5$ strikes an appropriate balance and adopt this for our analysis (except see Section~\ref{sec:alignment} on A2390, for which we take $\sigma_{\textrm{pos}} = 1\farcs0$). For the same reason, we have generally not imposed the detailed substructure of arcs as constraints on the model. One tool to evaluate $\sigma_{\textrm{pos}}$ empirically is the Bayesian evidence ratio. We compared the evidence obtained using $\sigma_{\textrm{pos}} = 0\farcs3$ and $1\farcs0$ relative to our default $\sigma_{\textrm{pos}} = 0\farcs5$ in fits to the full set of lensing and kinematic data.\footnote{In A963, for which the data consist of a critical line position, we instead varied the error in its position to $0\farcs1$ and $0\farcs5$ from our default $0\farcs2$. A2390 was excluded in this comparison due to the special treatment described in Section~\ref{sec:alignment}.} In all clusters $\sigma_{\textrm{pos}} = 1\farcs0$ is disfavored, with a decisive total evidence ratio $\Sigma(\ln E_{1\farcs0} / E_{\textrm{default}}) = -50.6$. The clusters are divided over whether $\sigma_{\textrm{pos}} = 0\farcs3$ is favored over our default $0\farcs5$. In total we find $\Sigma(\ln E_{0\farcs3} / E_{\textrm{default}}) = 7.0$, which indicates that the smaller uncertainty is somewhat preferred. We found, however, that the key parameter inferences do not shift significantly (see Section~\ref{sec:systematics}), while the error estimates slightly shrink when $\sigma_{\textrm{pos}} = 0\farcs3$ as expected. Thus, we have retained our more conservative $\sigma_{\textrm{pos}} = 0\farcs5$, but note that other reasonable estimates for the positional uncertainty yield very similar results.

Weak lensing constraints are incorporated by the term
\begin{equation}
\chi^2_{\textrm{WL}} = \sum\limits_{i} \frac{(g_{1,i} m_{\textrm{WL}} - g_{1,i}^{\textrm{obs}})^2 + (g_{2,i} m_{\textrm{WL}} - g_{2,i}^{\textrm{obs}})^2}{\sigma_g^2},
\end{equation}
where $(g_{1,i}^{\textrm{obs}}, g_{2,i}^{\textrm{obs}})$ is the observed reduced shear polar $g = \gamma / (1 - \kappa)$ for galaxy $i$, $(g_{1,i}, g_{2,i})$ is the model reduced shear at the angular position and photometric redshift of galaxy $i$, and the factor $m_{\textrm{WL}}$ incorporates our shear calibration. Based on the results in Section~\ref{sec:shear}, we assign a Gaussian prior of $0.89 \pm 0.05$ to $m_{\textrm{WL}}$. The uncertainty $\sigma_g$ is dominated by the intrinsic ellipticities of galaxies (``shape noise'') and is estimated using the standard deviation in shear measurements far from the cluster centers to be $\sigma_g = 0.32$.

Only the halo is considered in the weak-lensing modeling, since the mass is DM-dominated on $\gtrsim 100$~kpc scales. The ellipticity of the halo in the plane of the sky is incorporated using the ``pseudo-elliptical'' formalism of \citet{Golse02}, in which the ellipticity is introduced in the lens potential. Using their notation, we derive
\begin{align}
\gamma_{1,\epsilon} &= -|\gamma| \cos(2\phi_{\epsilon}) - \epsilon \kappa \\
\gamma_{2,\epsilon} &= -|\gamma| \sin(2\phi_{\epsilon}) \sqrt{1-\epsilon^2} \\
\kappa_{\epsilon} &= \kappa + \epsilon |\gamma| \cos(2\phi_{\epsilon}),
\end{align}
where $\gamma_{1,\epsilon}, \gamma_{2,\epsilon}$, and $\kappa_{\epsilon}$ are the shear components and convergence for the elliptical model, and $|\gamma|$ and $\kappa$ are the corresponding values for a circular lens. (See also \citealt{DumetMontoya12}.) As described by \citet{Golse02}, the pseudoellipticity $\epsilon$ is approximately the ellipticity of the potential and \emph{not} that of the surface mass density, which is about twice as large. The pseudoelliptical formalism is also used for the strong-lensing modeling. It is a reasonable approximation for the moderate ellipticities $\epsilon \lesssim 0.3$ encountered in our sample \citep{Sand08}.

Finally, we compute the l.o.s.~velocity dispersions $\sigma_{\textrm{los}}$ using the spherical Jeans equation. We assume the BCGs are completely pressure-supported, consistent with the lack of observed rotation (Section~\ref{sec:vdobs}):
\begin{equation}
\Sigma_* \sigma_{\textrm{los}}^2(R) = 2G \int_{R}^{\infty} \frac{\nu_*(r) M(r) \mathcal{F}(r)}{r^{2-2\beta_{\textrm{aniso}}}}\,\mathrm{d}r.
\end{equation}
By default we consider isotropic orbits with $\beta_{\textrm{aniso}} = 0$ and $\mathcal{F}(r) = \sqrt{r^2-R^2}$ 
\citep{Cappellari08}. Here $\nu_*$ and $\Sigma_*$ are the density and surface density profiles of the stellar tracers, as measured in Section~\ref{sec:sbprofiles}, and $M(r)$ is the total mass (stars and DM) enclosed within a radius $r$. 
In A383, axisymmetric two-integral dynamical models are used due to the significant l.o.s.~elongation of this cluster. These are described fully in N11 (and see Section~\ref{sec:A383}).

The observational effects of seeing and the slit width are included following \citet{Sand04}. The model $\sigma_{\textrm{los}}$ are spatially binned to match the extraction apertures used for the data. These constraints are incorporated as
\begin{equation}
\chi^2_{\textrm{VD}} = \sum\limits_{i} \frac{(\sigma_{i} - \sigma_{i}^{\textrm{obs}})^2}{\Delta_i^2},
\end{equation}
where $\sigma_i$ and $\sigma_{i}^{\textrm{obs}}$ are the model and observed l.o.s.~dispersions in bin $i$, respectively, and $\Delta_i$ is the uncertainty. 

As discussed by \citet{Sand04,Sand08}, a spherical treatment is a good approximation to the dynamics of the galaxies in our sample, which have a mean axis ratio $\langle b/a \rangle = 0.8$. Furthermore, detailed local studies find that massive, non-rotating ellipticals are intrinsically close to spherical and have low anisotropy \citep[e.g.,][]{Gerhard01,Cappellari07}. We discuss the effects of introducing mild orbital anisotropy into our dynamical models in Section~\ref{sec:systematics}.

\subsection{Alignment between the halo center and the BCG}
\label{sec:alignment}

In order to locate the center of the DM halo, we fit the lensing data with gNFW-based models in which the center of the halo was allowed to vary from that of the BCG, taking a Gaussian prior with $\sigma = 3''$ along each axis. Since we are concerned only with an astrometric measurement, we adopted a lower $\sigma_{\textrm{pos}} = 0\farcs3$ for these fits only. The inferred offsets between the centers of the halos and BCGs are given in Table~\ref{tab:sample}. They are typically $\simeq 1-4$~kpc with a $1\sigma$ uncertainty of $\simeq 1-3$~kpc, roughly consistent with the typical offset between the BCG and X-ray centroid. Given that the offsets are small and often not significant, we have fixed the center of the halo to that of the BCG in the following analysis. This allows for a consistent lensing and dynamical analysis. We note also that the P.A.~of the DM halo is close to the BCG light in all cases, never differing by more than $14^{\circ}$ in projection. (Given that BCGs often exhibit $\simeq 15^{\circ}$ gradients in P.A., such small differences are not completely well defined.)

The one exception to the above is A2390. While its lensing and kinematic data can be well fit when the halo center is fixed to the BCG, the resulting models demand an unusually high $\MLV$ ($\log \alpha_{\textrm{IMF}} > 0.42$ at 95\% confidence). Given the possible complexities in the mass distribution in A2390 described in Section~\ref{sec:thesample}, we considered it prudent to increase the freedom in this model and allow the center of the halo is to vary slightly from the BCG. We took a Gaussian prior having $\sigma = 1\farcs5$, based on the lensing analysis described above. The positional uncertainty $\sigma_{\textrm{pos}}$ was also relaxed to $1\farcs0$. (Nonetheless, the best-fitting models still reproduce the image positions with a fidelity of $0\farcs5$, as shown in Section~\ref{sec:totdens}.)

\section{Comparison Between Lensing- and X-ray-Derived Mass Profiles: Constraining the Line-of-Sight Ellipticity}
\label{sec:losellip}

\begin{deluxetable*}{lccccccc}
\tablecolumns{8}
\tablewidth{\linewidth}
\tablecaption{NFW Parameters Derived from X-Ray and Lensing Analyses\label{tab:xray}}
\tablehead{\colhead{Cluster} & \multicolumn{3}{c}{X-Ray} & \multicolumn{4}{c}{Lensing (Strong + Weak)} \\
\colhead{} & 
\colhead{$r_s$ (kpc)} &
\colhead{$c_{200}$} &
\colhead{Source} & 
\colhead{$r_s$ (kpc)} &
\colhead{$c_{200}$} &
\colhead{$\log M_{200}/\msol$} &
\colhead{$r_{200}$ (kpc)}}
\startdata
MS2137 & $180^{+20}_{-20}$ & $8.19^{+0.54}_{-0.56}$ & S07 & $119^{+49}_{-32}$ & $11.03^{+2.81}_{-2.39}$ & $14.56^{+0.13}_{-0.11}$ & $1318^{+140}_{-107}$ \\
A963 & $390^{+120}_{-80}$ & $4.73^{+0.84}_{-0.77}$  & S07 & $197^{+48}_{-52}$ & $7.21^{+1.59}_{-0.94}$ & $14.61^{+0.11}_{-0.15}$ & $1430^{+127}_{-151}$ \\
A383 & $470^{+130}_{-100}$ & $3.8^{0.7}_{-0.5}$ & A08 & $260^{+59}_{-45}$ & $6.51^{+0.92}_{-0.81}$ & $14.82^{+0.09}_{-0.08}$ & $1691^{+128}_{-102}$ \\
A383 (prolate) & \ldots & \ldots & \ldots & $372^{+63}_{-51}$ & $4.49^{+0.50}_{-0.48}$ & $14.80\pm0.08$ & $1665^{+107}_{-95}$ \\
A611 & $320^{+200}_{-100}$ & $5.39^{+1.60}_{-1.51}$ & S07 &$317^{+57}_{-47}$ & $5.56^{+0.65}_{-0.60}$ & $14.92 \pm 0.07$ & $1760^{+97}_{-89}$ \\
A2537 & $370^{+310}_{-150}$ & $4.86^{+2.06}_{-1.62}$ & S07 & $442^{+46}_{-44}$ & $4.63^{+0.35}_{-0.30}$ & $15.12\pm0.04$ & $2050^{+65}_{-69}$ \\
A2667 & $700^{+479}_{-207}$ & $3.02^{+0.74}_{-0.85}$ & A03 & $725^{+118}_{-109}$ & $2.99^{+0.32}_{-0.27}$ & $15.16 \pm 0.08$ & $2164^{+137}_{-129}$ \\
A2390 & $757^{+1593}_{-393}$ & $3.20^{+1.59}_{-1.57}$ & A03 & $763^{+119}_{-107}$ & $3.24^{+0.35}_{-0.31}$ & $15.34^{+0.06}_{-0.07}$ & $2470^{+112}_{-123}$
\enddata
\tablecomments{All X-ray fits are to the total gravitating mass and have been standardized to the same cosmology. Sources: S07 = \citet{Schmidt07}, A08 = \citet{Allen08}, A03 = \citet{Allen03}. The A383 (prolate) row shows a fit to lensing and X-ray data using triaxial isodensity surfaces (Equation~\ref{eqn:3D}, and see N11); we report sphericalized NFW parameters in this case.}
\end{deluxetable*}

Before turning to the cluster mass distribution over the full radial range $3-3000$ kpc spanned by our complete data set, in this section we first consider fits based only on strong and weak lensing, excluding velocity dispersion constraints, and compare these to independent X-ray measures. As we describe below, the combination of projected and 3D mass measures allows us to constrain the l.o.s.~geometry of the clusters in our sample.

Lensing directly probes the gravitational potential projected along the l.o.s., whereas the ICM follows the 3D potential. Mock observations of simulated clusters show that to a remarkable degree, X-ray observations are able to recover spherically-averaged mass profiles with a scatter of only $\simeq 5\%-10\%$ \citep{Nagai07,Lau09,Meneghetti10,Rasia12}. This is true even when a spherical geometry is (incorrectly) imposed in the analysis. The same simulations show that X-ray masses are biased slightly low due to non-thermal pressure support arising primarily from bulk gas motions. This bias is generally estimated to be only $\simeq 10\%$, although this depends on the detailed physics included in the simulations and may be somewhat higher (see \citealt{Rasia12}). As argued in N11, when much larger discrepancies between X-ray- and lensing-derived masses are encountered in relaxed clusters, they most likely arise from elongation or compression of the mass distribution along the l.o.s.

By comparing projected (lensing- or Sunyaev-Zel'dovich-based) and nearly spherical (X-ray) mass measures, the l.o.s.~shape can be inferred \citep[e.g.,][]{Piffaretti03,Gavazzi05,deFilippis05,Sereno06,Morandi10,Morandi11,N11,Morandi12} assuming that the ICM is near equilibrium. This is important for the present analysis given that stellar kinematics reflect the 3D gravitational potential. Coupling projected (lensing) and kinematic measurements can thus lead to errors if the l.o.s.~shape of the cluster is highly aspherical and this is not taken into account via an independent probe, such as X-ray data (see discussions in N11 and \citealt{Gavazzi05}).\footnote{As described in N11, the dynamical problem is more complex due to velocity anisotropy. As a result, the problem is acute only if the dark halo and stellar tracers have significantly different l.o.s.~ellipticities and this is neglected. If the tracers and total mass are stretched equally along the l.o.s., the reduction in density is nearly balanced by the boosted velocities of stars moving along the major axis, and the projected velocity dispersions change fairly little.}

Assuming for simplicity that one of the principal axes of the cluster is along the l.o.s. ($z$-axis), the surface density derived from the lensing can be deprojected onto isodensity surfaces with coordinates
\begin{equation}
\label{eqn:3D}
r = \sqrt{(1-\epsilon_{\Sigma})x^2 + (1+\epsilon_{\Sigma})y^2 + (z/q)^2}.
\end{equation}
Here $\epsilon_{\Sigma}$ and $q$ parameterize the projected and l.o.s.~ellipticity, respectively. For a ``spherical'' deprojection, $q = 1$. Note that lensing precisely measures the projected ellipticity $\epsilon_{\Sigma}$ but does not itself constrain $q$.

\begin{figure}
\centering
\includegraphics[width=\linewidth]{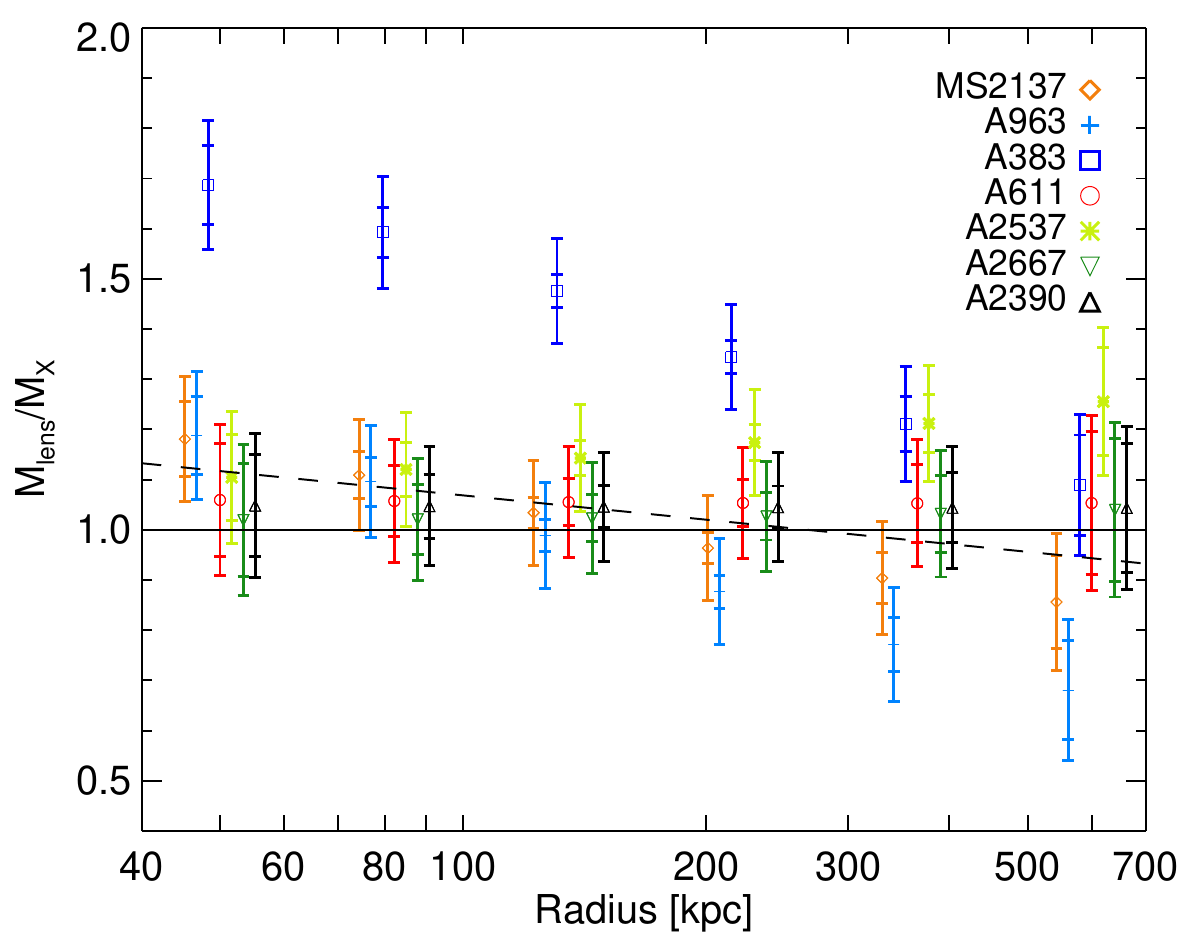}
\caption{Spherically enclosed masses $M_{\textrm{lens}}$ derived from strong- and weak-lensing analyses, assuming a spherical deprojection, are compared to those derived from published X-ray studies, $M_{\textrm{X}}$. The random and total (including a 10\% systematic estimate) errors are reflected in the inner and outer error bars, respectively. Note that measurements at various radii are not independent, as they are derived from two-parameter NFW models. The dashed line indicates the mean trend excluding A383 (Equation~\ref{eqn:MlensX}).\label{fig:lensingXray}}
\end{figure}

In order to compare our lensing results to X-ray analyses, we have compiled the results of several studies listed in Table~\ref{tab:xray}. X-ray studies typically adopt a parametric form for either the density or temperature profiles, and these studies adopted an NFW profile to represent the total density. For a clean comparison, it is thus appropriate to restrict to NFW models for the dark halo when fitting the lensing data in this section. Further, since X-ray studies generally do not separately model the BCG, we include only the dark halo in the lensing mass measurements below; this has a minor effect outside the innermost bin. Figure~\ref{fig:lensingXray} shows the ratio $M_{\textrm{lens}}/M_{\textrm{X}}$ of the spherically enclosed mass $M_{\textrm{lens}}$ derived from lensing by assuming a spherical deprojection, to the mass $M_{\textrm{X}}$ based on X-ray analyses. 
The inner error bars in Figure~\ref{fig:lensingXray} reflect the statistical uncertainty, which for the lensing mass is derived from the Markov chains. Estimating the uncertainty in the X-ray-based mass at a given radius cannot be done precisely with published NFW parameters, since the covariance is usually not given. We therefore estimated this using the full A383 mass profile provided by S.~Allen (private communication, 2011), including properly propagated errors, rescaling the errors based on the X-ray flux and exposure time as appropriate for Poisson-dominated formal errors. This is sufficiently accurate for our purposes given that systematic uncertainties are comparable. The larger error bars in Figure~\ref{fig:lensingXray} include an additional 10\% systematic contribution added in quadrature that reflects uncertainties in the \Chandra~temperature calibration \citep{Reese10}.

In general the agreement between the X-ray- and lensing-based masses assuming a spherical deprojection is very close, as Figure~\ref{fig:lensingXray} and Table~\ref{tab:xray} demonstrate. A383 is clearly discrepant, with $M_{\textrm{lens}} \gg M_{\textrm{X}}$; as discussed in Section~\ref{sec:A383} and N11, this can be explained by a prolate halo that is elongated along the l.o.s. For the remaining six clusters, however, the mean trend
\begin{equation}
M_{\textrm{lens}}/M_{\textrm{X}} = (1.07 \pm 0.01) - (0.16 \pm 0.04) \log r/100~\textrm{kpc}
\label{eqn:MlensX}
\end{equation} (dashed in Figure~\ref{fig:lensingXray}; errors are random only) is consistent with unity within the systematic uncertainty of $\approx 0.1$. None of these six clusters show systematic deviations larger than $|M_{\textrm{lens}}/M_{\textrm{X}} - 1| \gtrsim 0.2$ over scales of $50 - 600$~kpc. At $r \sim 100$~kpc, where strong lensing fixes the mass, the spherically deprojected mass $M_{\textrm{lens}}$ scales roughly $\propto q^{0.6}$ for an NFW profile with the range of $r_s$ encountered in our sample. Therefore, the similarity of the X-ray and lensing measures implies that $|q - 1| \lesssim 0.3$ in these systems, with the mean l.o.s.~ellipticity being smaller ($\langle q-1 \rangle \approx 0.1 - 0.2$). The asphericity will be yet smaller if some of the elevation of $M_{\textrm{lens}}/M_{\textrm{X}}$ is not due to geometry but to non-thermal pressure in the ICM, which is expected.

Strong-lensing--selected clusters as an ensemble are sometimes thought to be biased toward clusters elongated along the l.o.s., since this orientation boosts the lensing cross-section. Given that l.o.s.~elongation and non-thermal pressure support would both act to elevate $M_{\textrm{lens}} / M_{\textrm{X}}$, our results show that the clusters in our sample must be both close to hydrostatic equilibrium and not strongly elongated along the l.o.s.~(excepting A383). We note that our sample consists of fairly massive clusters, and that an orientation bias may be more stronger at lower masses. Since any compression or elongation along the l.o.s.~is constrained to be both small and consistent with null within the systematic uncertainties, $q = 1$ is fixed for the remainder of our analysis in all clusters except A383, which is discussed individually below.
The effect on our results of varying $q$ within the allowed limits is discussed in Section~\ref{sec:systematics}. The good agreement between lensing and X-ray masses further supports our contention that we have selected relaxed clusters (Section~\ref{sec:thesample}). We discuss the mass--concentration relation described by our sample in Section~\ref{sec:discussion}.

\subsection{The case of A383\label{sec:A383}}

As shown by N11 and independently confirmed by \citet{Morandi12}, A383 is significantly elongated nearly along the l.o.s. This is unique in our sample and necessitates a special treatment for A383 in several ways. Since the method was detailed in N11, only a summary is provided here. The l.o.s.~ellipticities of the dark halo $q_{\textrm{DM}}$ and BCG stars $q_*$ are included as additional parameters. An additional $\chi^2_{\textrm{X}}$ term incorporates the spherically-averaged mass profile derived from the X-ray analysis \citep{Allen08} into the likelihood, which constrains $q_{\textrm{DM}}$ as described above. Since the projected ellipticity ($1-b/a \simeq 0.1$) is much less than that along the l.o.s., the stellar dynamics can be approximated using a prolate axisymmetric model with the symmetry axis along the l.o.s. (Using a more sophisticated model of the ICM, \citet{Morandi12} also showed that the major axis is close to the l.o.s., with a separation of $21^{\circ} \pm 10^{\circ}$.) By accounting for the l.o.s.~geometry, N11 showed that the lensing, kinematic, and X-ray data can be brought into agreement.

\subsection{Comparison to previous lensing results}
\label{sec:previouslensing}

All of the clusters in this sample have been the subject of previous lensing analyses. Although a comparison with each of these earlier works is impractical, and in most cases our NFW parameters agree with previous determinations within the errors, we wish to note a few cases that have been the source of some confusion in the literature. First, in N09 we reported a high concentration ($c_{200} = 10.0 \pm 1.1$) for strong lensing-based fits in A611, which was in tension with weak-lensing estimates at the time and is higher than the present measurements. This is attributable to the revised spectroscopic redshift discussed in Section~\ref{sec:A611multiples}.

Second, \citet[][G05]{Gavazzi05} studied the mass distribution in MS2137 using strong and weak lensing
and reported substantial differences between lensing- and X-ray-based mass models. They inferred that a significant elongation along the l.o.s.~was a likely explanation. Our lensing results instead agree closely with a recent analysis by \citet{Donnarumma09}. They are also consistent with X-ray measurements by \citet{Schmidt07}, which is incompatible with the highly prolate shape ($q \approx 2$) suggested by G05. This likely arises from a numerical error in the G05 results (R.~Gavazzi, private communication, 2012).

\section{The total density profile}
\label{sec:totdens}

\begin{figure*}
\centering
\includegraphics[width=0.94\linewidth]{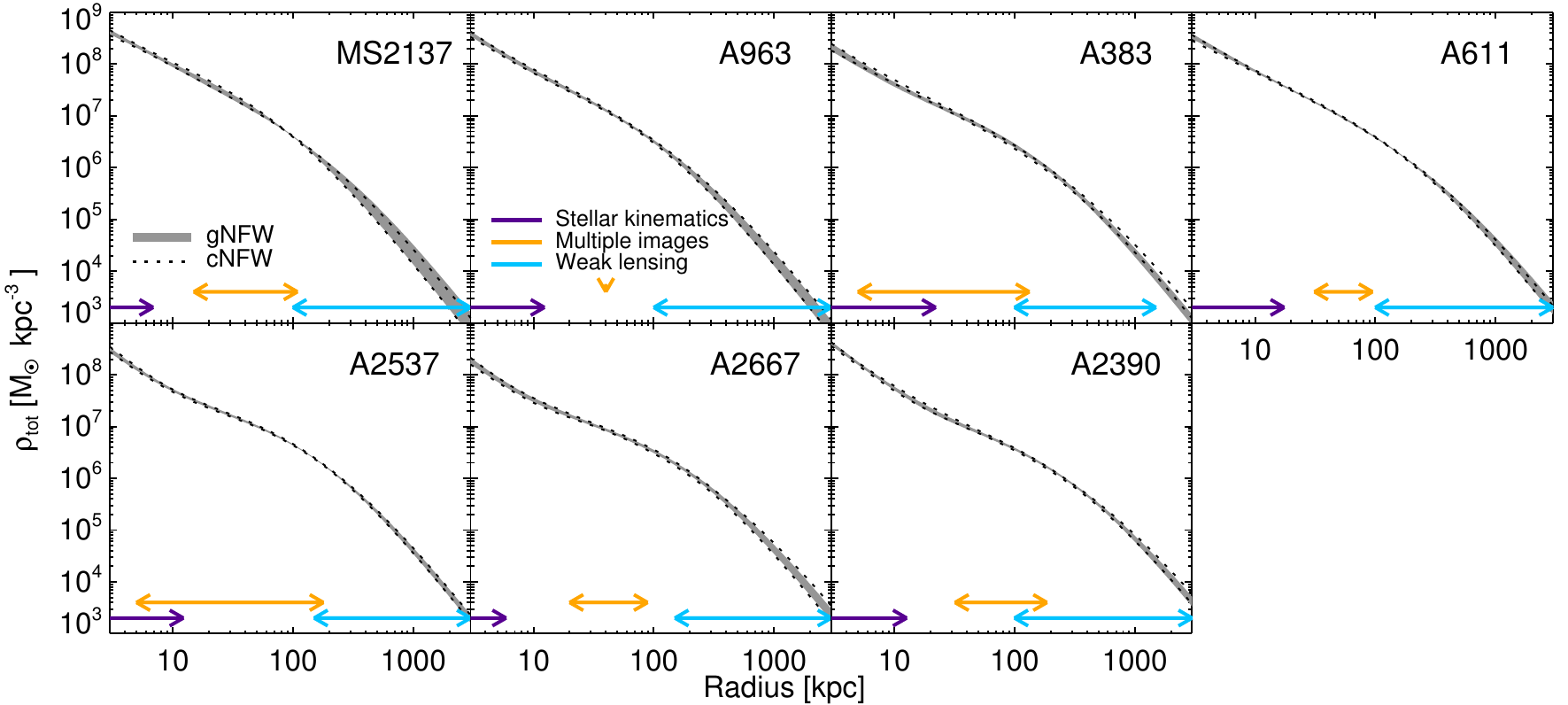}\\
\includegraphics[width=0.94\linewidth]{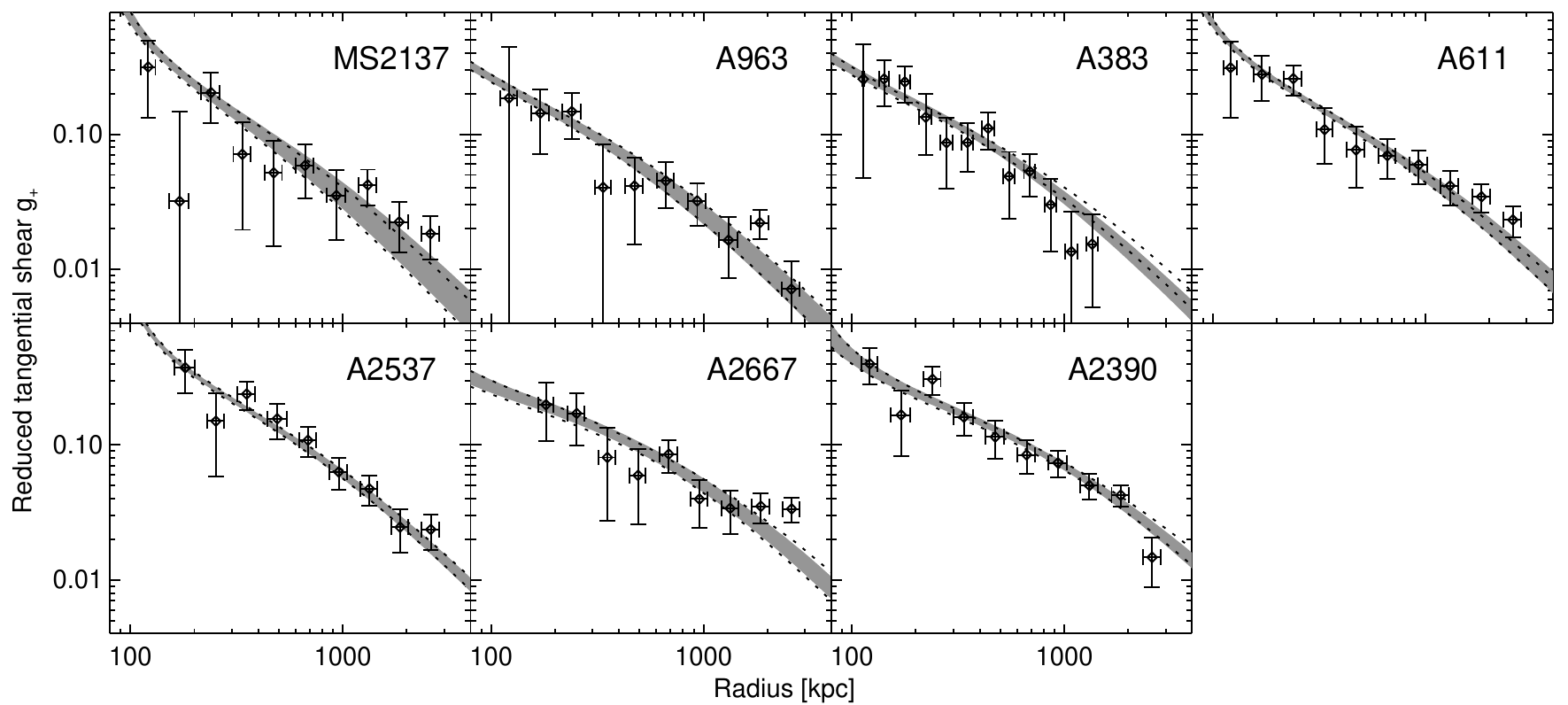}\\
\includegraphics[width=0.94\linewidth]{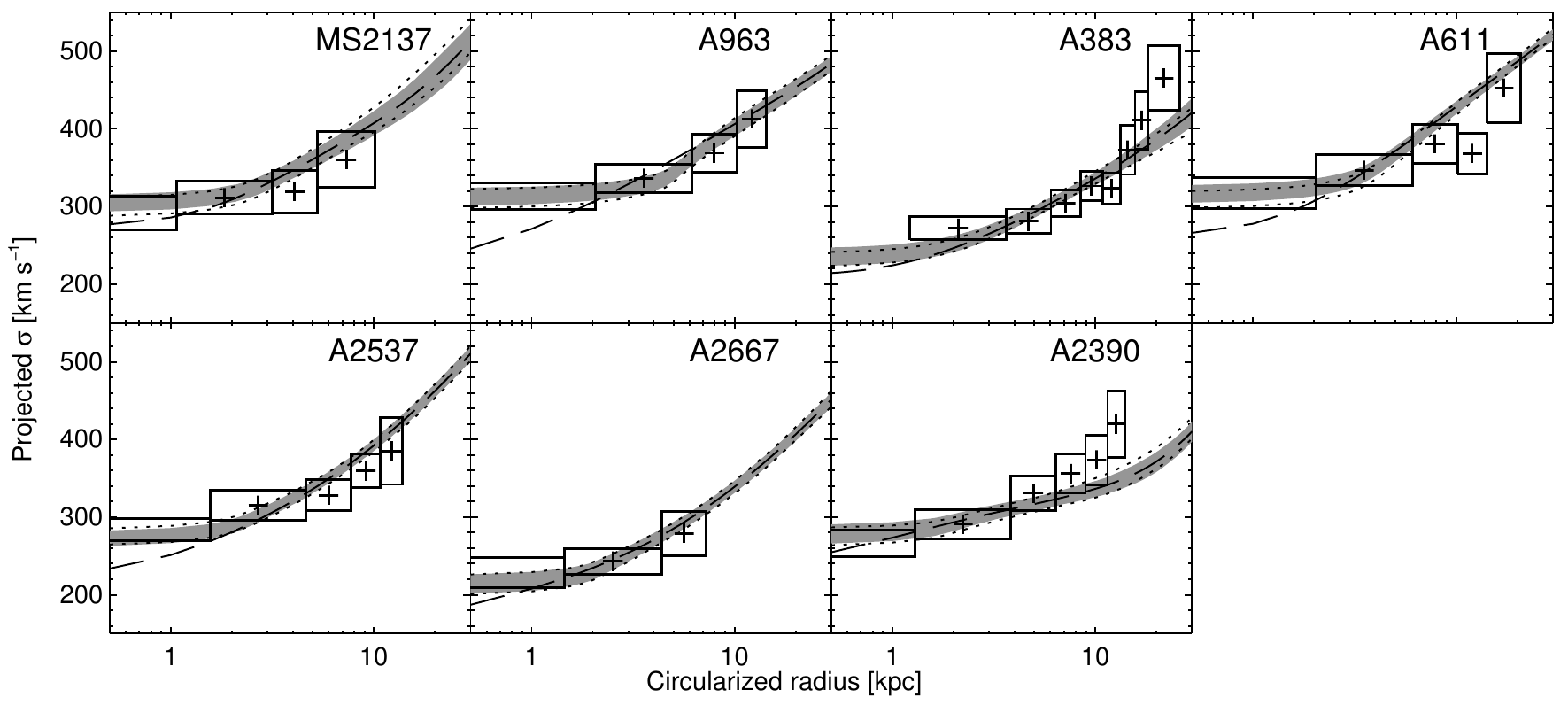}
\caption{Total density (top), tangential reduced shear (middle), and velocity dispersion (bottom panel) profiles for fits to lensing and stellar kinematic data. In all panels, the shaded region and dotted lines indicate the 68\% confidence intervals for the gNFW and cNFW models, respectively. \textbf{Top:} the radial intervals spanned by each data set are indicated. \textbf{Middle:} the shear averaged in circular annuli is shown for display purposes, although elliptical models are used throughout the quantitative analysis. For A2667, the shear from the second clump is subtracted as described in the text. \textbf{Bottom:} model dispersions (shaded and dotted) include the effects of seeing and the slit width; the dashed line shows the mean gNFW model excluding these effects. The extraction radii of the data have been circularized.\label{fig:totdens}}
\end{figure*}

In this section, we combine strong and weak lensing with resolved stellar kinematic measures within the BCG to constrain the radial density profiles of the seven clusters in our sample over $r \simeq 3 - 3000$~kpc. We remind the reader that the present paper is concerned with the \emph{total} density, inclusive of dark and baryonic matter, not that of the DM alone as in our earlier works \citep{Sand02,Sand04,Sand08,N09,N11}, which is discussed in Paper~II.

\subsection{Mass models and fit quality}

The top panel of Figure~\ref{fig:totdens} shows the total density profiles $\rho_{\textrm{tot}}(r)$ that are inferred using gNFW (solid) and cNFW (dotted) models for the halo. The colored bars at the bottom illustrate the radial extent of each data set, which taken together provide coverage over most of the 3 decades in radius plotted. Correspondingly, the mass models are tightly constrained over the entire range. Furthermore, the density profiles derived using gNFW and cNFW models (Section~\ref{sec:modeling}) are virtually identical. This demonstrates that within the range well constrained by the data, the derived density profiles do not strongly depend on the particular parameterization of the halo.

Given the simple parameterization of the mass distribution, it is important to verify that good fits are achieved to the wide range of data. The middle and bottom panels of Figure~\ref{fig:totdens} demonstrate that, in all cases, a statistically acceptable fit to the weak lensing and stellar kinematic data is obtained. The quality of the strong-lensing fits is shown in Figure~\ref{fig:slimages}, in which the positions of the multiple images in the best-fitting model are indicated as crosses. The image positions are typically matched within $0\farcs5$, which is fairly typical of other studies using similar models \citep[e.g.,][]{Richard10}. In some cases, the best models predict images that were not included as constraints because they could not be unambiguously identified (Section~\ref{sec:SL}), particularly when buried in cluster galaxy light, but no predicted counterimages lack a plausible identification when one should be observable.

In A2667, the modeled shear arising from the second mass clump located at $R \simeq 1.4$~Mpc has been subtracted from the data points in Figure~\ref{fig:totdens}. Nevertheless, the measured shear exceeds the model at $R \gtrsim 2$~Mpc, which may indicate a more complex mass distribution near the virial radius. The fit quality at smaller radii and the close agreement with X-ray measurements reassure us that the mass is well modeled within $\simeq 2$~Mpc

Table~\ref{tab:slquality} quantifies the quality of fit for the various sources of data in each cluster. For the weak-lensing data, the noise is easily characterized, since it is dominated by random shape errors; thus, these data are fit with $\chi^2_{\textrm{WL}} / N_{\textrm{WL}} \simeq 1$. For the strong lensing and velocity dispersion data, the mean reduced $\langle \chi^2 / N \rangle \simeq 0.6$. This indicates that the error bars may be conservatively overestimated by $\simeq 30\%$. However, the similarity of $\langle \chi^2 / N \rangle$ indicates that the relative weighting of the kinematic and strong-lensing data is appropriate. These data essentially set the density slope on small scales that we derive below. However, the weak-lensing data are essential when comparing to simulations, since they constrain the scale and virial radii and thus characterize the radial span over which the inner slope is measured in terms of these key theoretical scales. 

Considering the entire sample, we find no notable difference in the quality of the fit between the gNFW and cNFW models: $\Sigma (\chi^2_{\textrm{gNFW}} - \chi^2_{\textrm{cNFW}}) = 3.0$, while the total Bayesian evidence ratio $\Sigma (\ln E_{\textrm{gNFW}} / E_{\textrm{cNFW}}) = -3.5 \pm 3.1$. These indicate a slight preference for the cNFW models, but it is not very significant, as expected based on the similarly of the derived density profiles discussed above. In Paper II we impose a more informative prior on $\ML$, using results derived from the whole sample, and find that the evidence ratio is close to unity. We conclude that the data do not clearly prefer one of our flexible DM halo models over the other. For this reason, we focus on the gNFW models for the remainder of the paper.

\begin{deluxetable*}{lccccccccc}
\tablecolumns{10}
\tablewidth{\textwidth}
\tablecaption{Fit Quality to Strong Lensing, Weak Lensing, and Stellar Kinematic Data\label{tab:slquality}}
\tablehead{\colhead{} & \multicolumn{4}{c}{gNFW Models} & \multicolumn{4}{c}{cNFW Models} \\
\colhead{Cluster} & \colhead{$\sigma_{\textrm{img}}$} & \colhead{$\chi^2_{\textrm{SL}} / N_{\textrm{SL}}$} & \colhead{$\chi^2_{\textrm{WL}} / N_{\textrm{WL}}$} & \colhead{$\chi^2_{\textrm{VD}} / N_{\textrm{VD}}$} & \colhead{$\sigma_{\textrm{img}}$} & \colhead{$\chi^2_{\textrm{SL}} / N_{\textrm{SL}}$} & \colhead{$\chi^2_{\textrm{WL}} / N_{\textrm{WL}}$} & \colhead{$\chi^2_{\textrm{VD}} / N_{\textrm{VD}}$} & \colhead{$\ln E_{\textrm{gNFW}}/ E_{\textrm{cNFW}}$}}
\startdata
MS2137 & $0\farcs44$ & 8.6/16 & 13136.8/12670 & 1.6/4 & $0\farcs43$ & 8.1/16 & 13137.2/12670 & 1.4/4 & $-2.2$ \\
A963 & 0 & 0/2 & 32092.5/31132 & 0.3/4 & 0 & 0/2 & 32092.5/31132 & 0.2/4 & $0.1$\\
A383 & $0\farcs46$ & 2.5/18 & 7070.4/6936 & 7.4/8 & $0\farcs41$ & 2.0/18 & 7070.9/6936 & 7.4/8 & $0.7$ \\
A611 & $0\farcs60$ & 17.1/18 & 14140.9/14252 & 8.9/5 & $0\farcs58$ & 16.4/18 & 14141.5/14252 & 9.0/5 & $-2.1$ \\
A2537 & $0\farcs65$ & 27.4/24 & 12812.3/12912 & 3.2/5 & $0\farcs66$ & 28.0/24 & 12813.3/12912 & 2.4/5 & $0.2$ \\
A2667 & $0\farcs29$ & 4.4/18 & 18734.3/18526 & 0.2/3 & $0\farcs27$ & 3.7/18 & 18732.2/18526 & 0.2/3 & $-2.3$ \\
A2390 & $0\farcs53$ & 3.7/16 & 16319.5/16186 & 3.9/6 & $0\farcs60$ & 4.6/16 & 16319.9/16186 & 2.0/6 & $1.0$
\enddata
\tablecomments{$\chi^2$ for the best-fitting gNFW and cNFW models are shown, with degrees of freedom defined as follows: for weak lensing, $N_{\textrm{WL}}$ is twice the number of background galaxies, corresponding to the two shear components; for velocity dispersion data, $N_{\textrm{VD}}$ is the number of extracted spatial bins; for strong lensing $N_{\textrm{SL}} = 2(N_{\textrm{img}} - N_{\textrm{src}})$, where $N_{\textrm{img}}$ is the number of images and $N_{\textrm{src}}$ is the number of distinct sources. Note that $\chi^2_{\textrm{SL}}$, $N_{\textrm{SL}}$, and $\sigma_{\textrm{pos}}=0\farcs5$ (the uncertainty in the image positions) are presented on a \emph{per coordinate} basis, whereas following common practice, the rms error in the model image-plane positions $\sigma_{\textrm{img}}$ refers to the total. Thus, $\chi^2_{\textrm{SL}} / N_{\textrm{SL}} \simeq 1$ when $\sigma_{\textrm{img}} \simeq \sqrt{2} \sigma_{\textrm{pos}}$. In A963 the single strong lensing constraint (Section~\ref{sec:inferringmodels}) can be fitted exactly. In A383 X-ray data are also fit  (Section~\ref{sec:A383}), with $\chi_{\textrm{X}}^2 / N_{\textrm{X}} = 5.8 / 5$ and $\chi_{\textrm{X}}^2 / N_{\textrm{X}} = 4.1 / 5$ for the gNFW and cNFW models. The final column gives the natural logarithm of the Bayesian evidence ratio; the typical sampling error in this quantity, estimated from repeated MCMC runs, is 1.0.}
\end{deluxetable*}

\begin{figure}
\centering
\includegraphics[width=0.99\linewidth]{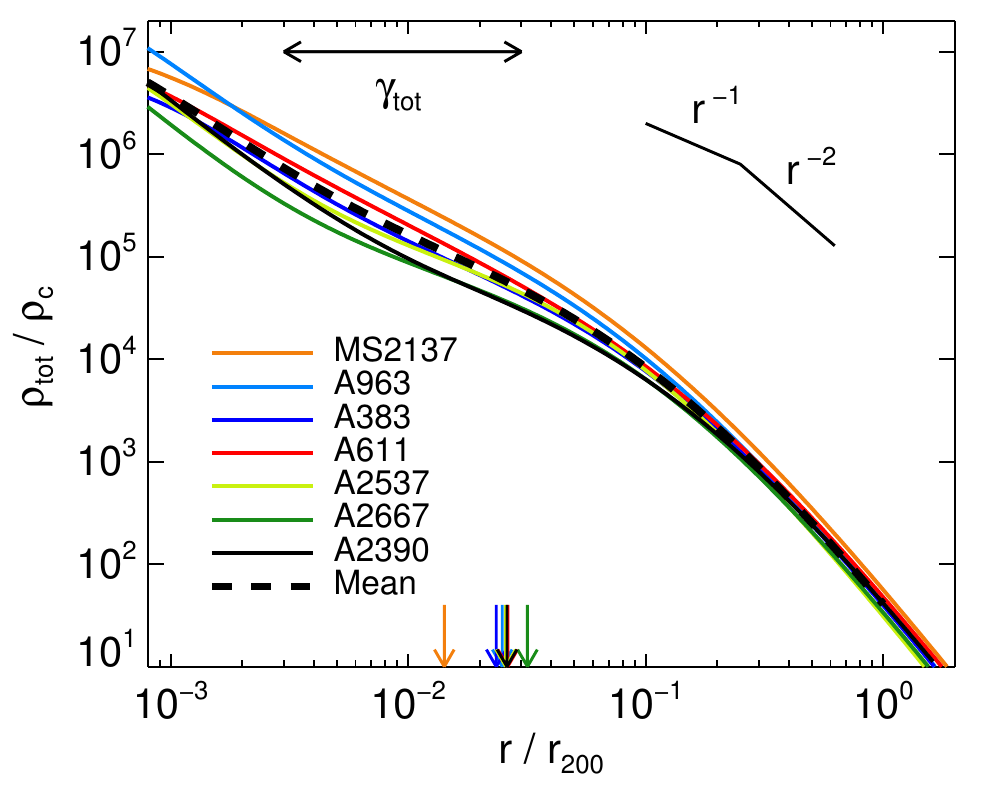} \\
\includegraphics[width=0.99\linewidth]{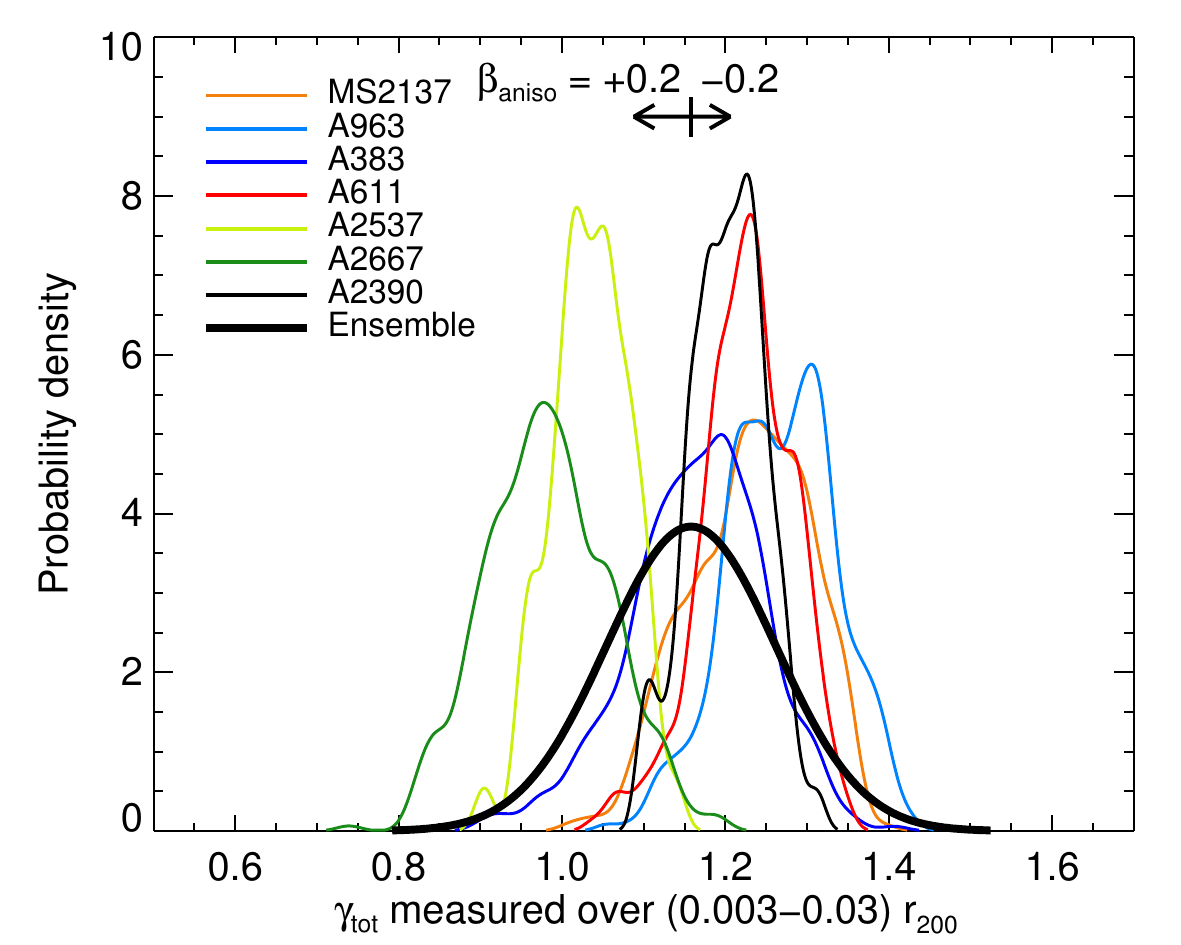}
\caption{\textbf{Top}: spherically averaged profiles of the \emph{total} density, normalized by the virial radius $r_{200}$ (Table~\ref{tab:xray}) and the critical density $\rho_c(z_{\textrm{clus}})$. Uncertainties are shown in Figure~\ref{fig:totdens}. The range over which the inner slope $\gamma_{\textrm{tot}}$ is defined is shown at the top of the panel. Arrows at the bottom indicate the three-dimensional half-light radii $r_h$ of the BCGs. \textbf{Bottom:} marginalized probability densities for the inner slope $\betatot$ of the \emph{total} mass distribution, measured over $(0.003 - 0.03)r_{200}$. The thick curve shows the inferred parent Gaussian distribution, as described in the text. The top of the panel indicates the effects of introducing mild orbital anisotropy (Section~\ref{sec:systematics}). \label{fig:totdenscombined}}
\end{figure}

\begin{figure}
\centering
\includegraphics[width=\linewidth]{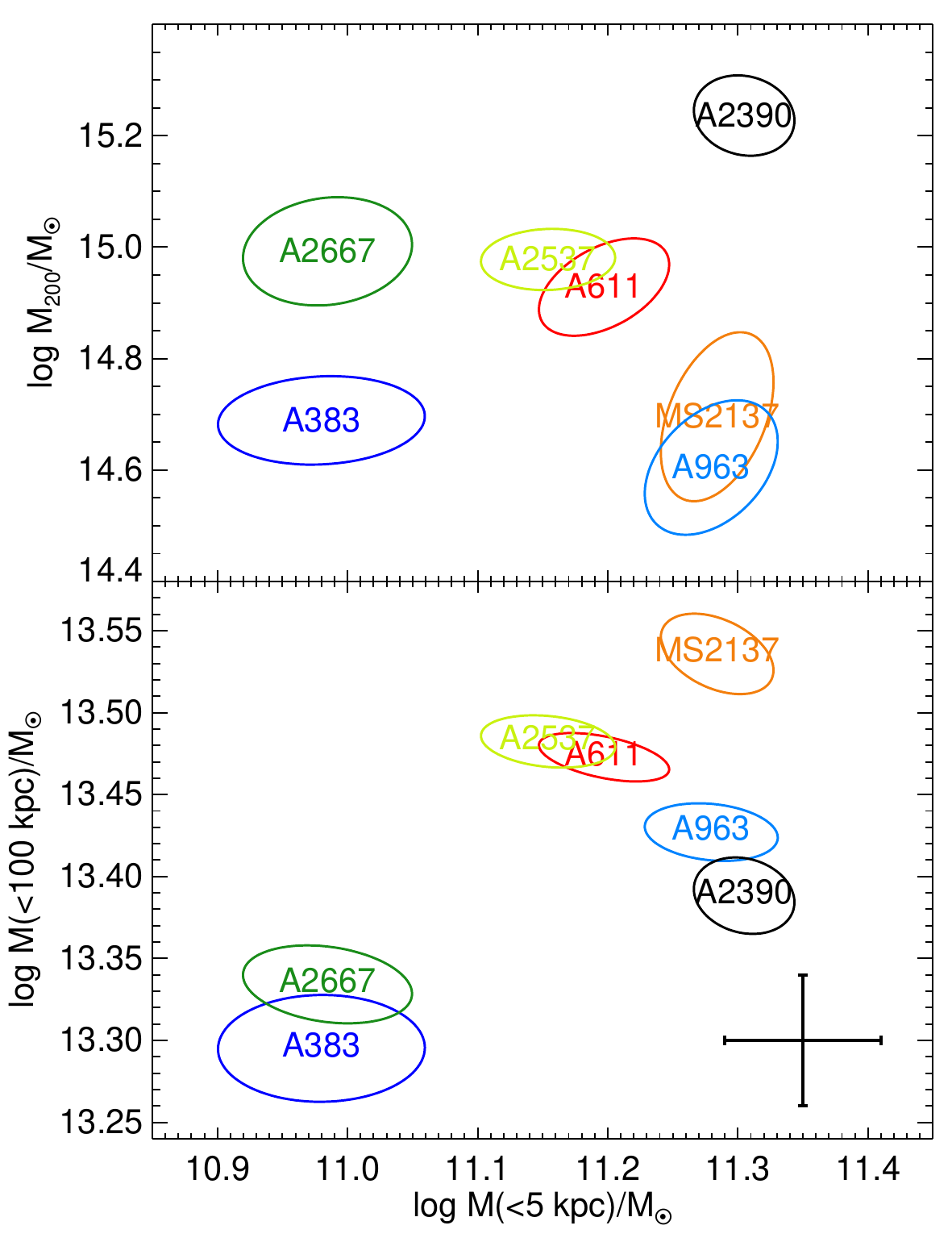}
\caption{Comparison of mass contained within the central 5~kpc, comprising mostly stars, to that within the virial radius (top, showing no correlation) and within 100~kpc (bottom, showing a positive correlation), which are both dominated by DM. Error ellipses ($1\sigma$) indicate the formal model uncertainty. Error bars in the bottom panel estimate the systematic uncertainties due to orbital anisotropy (see Section~\ref{sec:systematics}) and projection effects (Section~\ref{sec:losellip}). \label{fig:inner_mass_ratio}}
\end{figure}

\subsection{The total inner density slope}
\label{subsec:totslope}

The top panel of Figure~\ref{fig:totdenscombined} shows that the density profiles of these clusters are similar in their inner regions. At very small radii $\lesssim 0.003 r_{200} \approx 5$~kpc, the density profiles often steepen. As we describe in \PaperII, this is where the density becomes strongly dominated by stars. However, outside this innermost region the slopes of the total density profiles are quite comparable. To quantify this similarity, we introduce a measure of the \emph{total} inner slope $\betatot = -d \log \rho_{\textrm{tot}} / d \log r$. Since the BCG and the DM halo are modeled as distinct components, $\betatot$ is not a directly inferred parameter. We define it by fitting a line in the $\log r - \log \rho_{\textrm{tot}}$ plane over the interval $r/r_{200} = 0.003 - 0.03$, illustrated at the top of the panel, with errors derived by repeating this for many models in the Markov chains.\footnote{Grid points are logarithmically spaced and equally weighted.} For the median $r_{200}$ in our sample (Table~\ref{tab:xray}), the corresponding interval is $5-53$~kpc, or typically $\approx 0.2-2R_e$ in terms of the effective radius $R_e$ of the BCG. The endpoints of this range are well constrained by stellar kinematics and strong lensing, and therefore $\betatot$ is observationally robust.

The bottom panel of Figure~\ref{fig:totdenscombined} shows the probability distributions of $\betatot$, which is well constrained for each cluster, with a typical formal $1\sigma$ uncertainty of 0.07. In order to characterize the mean inner slope and its scatter, we assume that the distribution of $\betatot$ in the parent population of massive, relaxed galaxy clusters is Gaussian. Following the formalism described by \citet{Bolton12}, we then infer a mean $\langle \betatot \rangle = 1.16 \pm 0.05 {}^{+0.05}_{-0.07}$ (errors are random and systematic, respectively, with the latter described below) and an intrinsic cluster-to-cluster scatter of $\sigma_{\gamma} = 0.10^{+0.06}_{-0.04}$. For comparison, an NFW profile with concentration $c_{200} = 4.5$ typical of our sample has a slope $\beta = 1.10$ ($\rho_{\textrm{DM}} \propto r^{-\beta}$) over the same radial interval. Interestingly, the mean \emph{total} density slope in our sample is therefore consistent with that expected of \emph{CDM-only} halos, with fairly small scatter. We return to this point in Section~\ref{sec:discinnerslope}, where we make comparisons to numerical simulations. In Paper~II, we investigate correlations of $\betatot$ with other properties.

\begin{figure}
\includegraphics[width=\columnwidth]{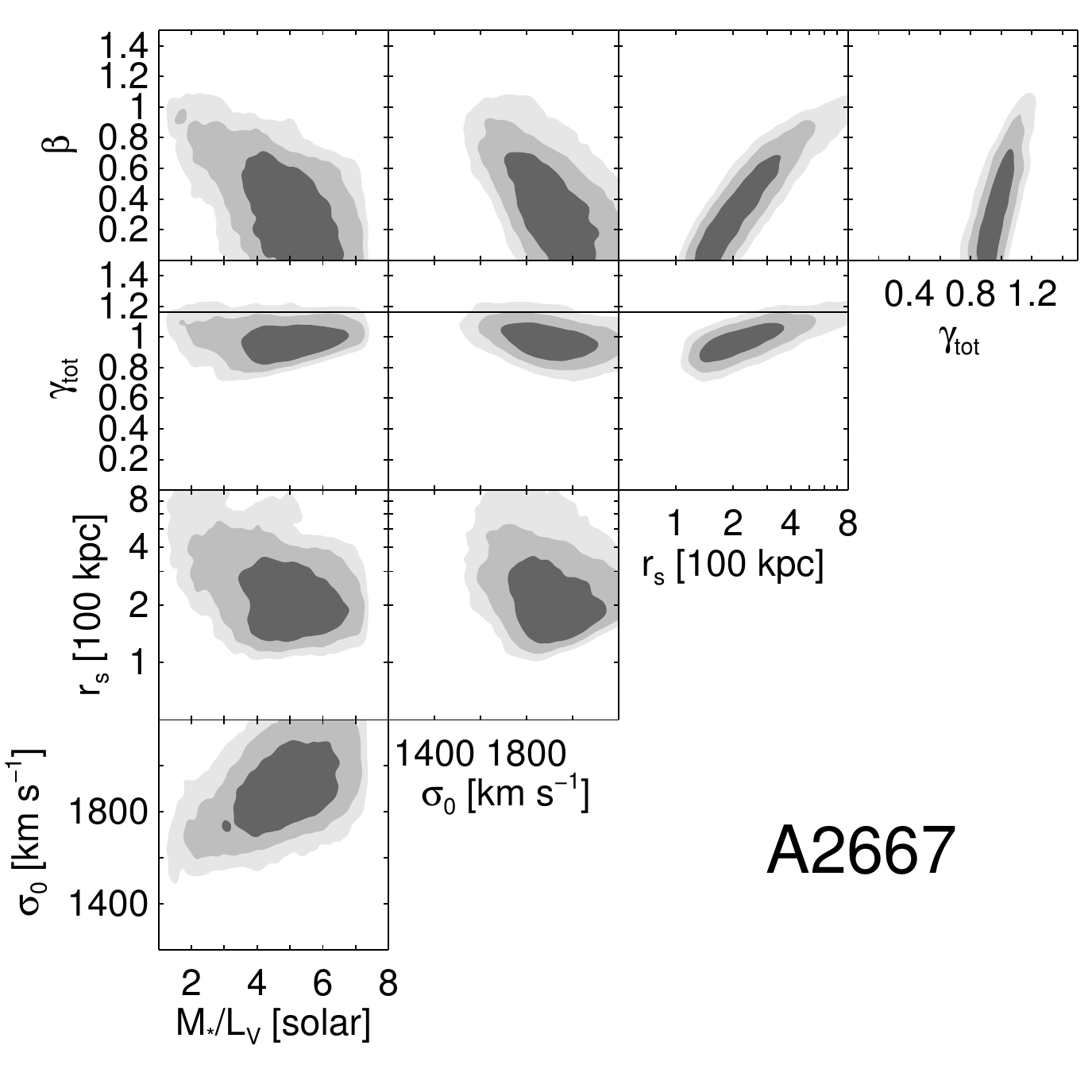}
\caption{Degeneracies among key parameters describing the radial density profile in A2667 for gNFW-based fits. Contours indicate the $68\%$, $95\%$, and $99.5\%$ confidence regions. Similar plots for the other clusters are presented in the Appendix. Note that $\gamma_{\textrm{tot}}$ is derived from the other parameters and is not independent. The horizontal line indicates the mean $\gamma_{\textrm{tot}}$ among the whole sample.\label{fig:degen_example}}
\end{figure}

A conservative approach is to view the intrinsic scatter in the inner slope $\sigma_{\gamma}$ as an upper limit: $\sigma_{\gamma} < 0.13$ (68\% CL). This is because systematic errors may contribute additional scatter in the measurements (Section~\ref{sec:systematics}) beyond that reflected in the formal errors, which would imply that the true physical scatter is smaller. The results presented here are not very sensitive to the precise radial interval over which the slope is measured. Taking $r/r_{200} = 0.005 - 0.03$ or $0.003 - 0.05$, for example, only shifts $\langle \betatot \rangle$ within its $1\sigma$ uncertainty.

Figure~\ref{fig:inner_mass_ratio} illustrates the uniformity of the inner mass distribution via a different metric, demonstrating a connection between the mass on very small scales of 5~kpc and the mass of the cluster core within 100~kpc. In \PaperII~we show that stars typically compose 75\% of the mass within 5~kpc, whereas the mass on 100~kpc scales is almost entirely DM. Despite this and the small range in these masses within our sample -- each roughly a factor of two -- we detect a probable correlation (Pearson correlation coefficient $r = 0.70$, two-sided $P = 0.08$).
The top panel of the figure shows that, in contrast, there is no correlation when the virial mass of the cluster is considered instead. As we discuss further in Section~\ref{sec:discussion}, this can be understood if the innermost regions of the present BCG and cluster halo were in place at early times and changed little in mass subsequently, with accretion mostly adding mass to the outer regions to grow the BCG and the cluster halo.

The key degeneracies among the parameters relevant to the radial density profile ($\beta$, $r_s$, $\sigma_0$ for the DM halo and $M_*/L_V$ for the BCG) are illustrated in Figure~\ref{fig:degen_example}. The best-constrained parameter is $\gamma_{\textrm{tot}}$, which is easily understood based on two physical reasons: first, $\gamma_{\textrm{tot}}$ refers to a slope measured over a fixed radial interval, unlike the inner gNFW slope $\beta$ which is approached only asymptotically; second, measuring the total density profile requires no separation of the dark and luminous components. Clearly, measurements of the inner DM slope $\beta$ could be improved using additional information on $M_*/L_V$ beyond that which can be inferred on a cluster-by-cluster basis. This is the subject of the Paper II.

\subsection{Systematic uncertainties}
\label{sec:systematics}

Before turning to the physical interpretation of these results, we first review the sources of systematic uncertainty in our analysis. Our dynamical models are based on isotropic stellar orbits. Prior studies \citep[e.g.,][and see references in Section~\ref{sec:vdprofiles}]{Carter99,Gerhard01,Cappellari07} have shown this to be a good approximation for luminous, non-rotating ellipticals in their central regions, with a possible tendency toward slightly radial orbits. We reran our analysis using a constant $\beta_{\textrm{aniso}} = +0.2$ (radial bias) or $-0.2$ (tangential bias) in the dynamical calculations, where $\beta_{\textrm{aniso}} = 1 - \sigma_{\theta}^2/\sigma_r^2$ characterizes deviations from isotropy \citep[e.g.,][]{BinneyTremaine}. The mean shifts in $\betatot$ were $-0.07$ and 0.05, respectively. This could be a common bias among the whole sample. Variable anisotropy could also introduce spurious scatter in the measured $\betatot$ at the same level; in that case, the true physical scatter would be less.

Since we measure kinematics well within the effective radii of the BCGs, taking $|\beta_{\textrm{aniso}}| = 0.2$ corresponds to changes in $\sigma_{\textrm{los}}$ by $\simeq 5\%-10\%$ for the same mass distribution. This is larger than the systematic errors of $\lesssim 5\%$ in the measurements themselves (Section~\ref{sec:kinmeasurement}), and therefore the resulting errors are less than from those from anisotropy. Furthermore, most of the systematic measurement errors are probably not correlated across all BCGs. Errors arising from the spherical dynamical treatment are expected to be similarly small \citep[e.g.,][]{Kronawitter00,Jiang07} for nearly-round systems like our sample.

Spherical masses estimates derived from lensing will be biased if the cluster is elongated or compressed along the l.o.s.~In Section~\ref{sec:losellip}, we found a mean tendency for the lensing mass to exceed that derived from X-ray measurements by 7\% at 100~kpc. Although this is consistent with zero within the uncertainties in the X-ray calibration, a 7\% bias in the spherically averaged mass profile would shift $\langle \betatot \rangle$ by only $-0.03$. Cluster-to-cluster variation with $|q-1| \lesssim 0.3$ (Section~\ref{sec:losellip}) could introduce scatter of $\sigma_{\gamma} \lesssim 0.08$; accounting for this would again lower the inferred intrinsic scatter in $\betatot$.

Our analysis assumes that the stellar mass in the BCGs follows the light measured at $\simeq 6000$~\AA, i.e., that $\MLV$ does not vary with radius. Color gradients indeed appear to be small in the majority of the sample (Section~\ref{sec:sbprofiles}), but two BCGs (A611 and A383) show a stronger gradient. We take A383 as an example. Assuming that the near-infrared light measured in the F160W filter is a better proxy of the stellar mass, we applied a radial gradient to the model stellar mass profile based on the ratio of the F160W and F606W fluxes. For the same tracers, the velocity dispersions should change by $\lesssim 4\%$, less than the systematic uncertainty in the measurements. This is because the $M_*/L_{\textrm{F606W}}$ gradient becomes significant only at large radii where DM is dominant. We also tested the impact of the BCG size $r_{\textrm{cut}}$ by perturbing it by its 10\% uncertainty in A2537 and repeating our analysis, accounting for the correlated change in $L_{\textrm{V}}$. This led to no significant shifts in $\MLV$, $\beta$, or $b$ (see also the discussion in \citealt{Sand04}).

To assess the impact of the strong lensing positional uncertainty $\sigma_{\textrm{pos}}$ on our findings, we reran our analysis with $\sigma_{\textrm{pos}} = 0\farcs3$ (see Section~\ref{sec:inferringmodels}). This had very little effect on $\betatot$, typically shifting the inferred values in individual clusters by $\lesssim 0.02$, with no net bias on $\langle \betatot \rangle$. We conclude that our results are robust to reasonable changes in the weighting of the lensing data. 

Considering the combination of the above uncertainties, we estimate that on a cluster-by-cluster basis there is an additional error of $\simeq 0.10$ in $\betatot$ beyond the formal random estimates, which are comparable in magnitude. Not all of this error budget is coherent across the full sample: the largest source of global systematic bias is likely the orbital distribution. Thus, we take the uncertainty $\Delta \betatot = -0.07, +0.05$ arising from orbital anisotropy as the systematic error in the mean $\langle \betatot \rangle$.

Finally, we wished to explore the impact on our dynamical analysis if the BCG is not precisely at rest in the center of the halo. As discussed in Section~\ref{sec:thesample}, the X-ray centroid and the lensing center are generally quite close to the optical center of the BCG. However, small offsets of a few kpc are not excluded. In order to assess how the stellar dynamics could be affected by small-scale oscillations around the center of the cluster potential, we performed some simple numerical simulations using the parallel $N$-body code \code{FVFPS} \citep{Londrillo03,Nipoti03}. The BCG is modeled as a single-component equilibrium isotropic $\gamma$ model \citep{Dehnen93} with $\gamma=1.5$, scale radius $a=23.5$ kpc (i.e., 3D half-mass radius $r_{\textrm{half}} \simeq 40$ kpc), and total mass $M_*=1.5\times10^{12}M_{\odot}$, representative of the BCGs in the observed sample. The galaxy was realized with $N\simeq2\times10^5$ particles following the same procedure as \citet{Nipoti03,Nipoti09}, using a softening parameter $\varepsilon=0.03a$. At the beginning of each simulation the galaxy is placed at a distance $r_{\rm offset}$ from the center of a fixed gravitational potential representing the cluster DM halo, either at rest or in a circular orbit. We explored two halo models: a steep halo ($\gamma = 1$, $a = 352$~kpc) that approximates an NFW profile with $\rho_s = 1.52 \times 10^6 \msol$~kpc${}^{-3}$ and $r_s = a$ (see Equation \ref{eqn:gNFW}) within the scale radius, and a shallow halo ($\gamma = 0.5$, $a = 226$~kpc) that approximates a gNFW profile with $\beta = 0.5$, $\rho_s = 5.37 \times 10^6 \msol$~kpc${}^{-3}$, and $r_s = a$. The two models were chosen to nearly match at $r > 100$~kpc but differ in their inner slope.

In the halo with the steeper NFW-like cusp, we found that small displacements -- even up to 40~kpc -- are highly unstable. Even when initially set on a circular orbit, the BCG quickly falls to the halo center within 350~Myr. During this time the isophotes are clearly disturbed, which is inconsistent with the galaxies in our sample. In the halo with a shallower density cusp, on the other hand, we found that stable oscillations with an amplitude of $\simeq 5$~kpc are possible. During these oscillations, the central velocity dispersion varies from that attained by the same system with a stationary BCG (at the cluster center) by only a few percent. We conclude that small offsets between the BCG and cluster center do not pose a significant problem for our Jeans analysis. Furthermore, if the small offsets are genuine, they appear to imply a DM cusp with $\beta \lesssim 1$. 

\section{Discussion}
\label{sec:discussion}

\begin{figure}
\centering
\includegraphics[width=\linewidth]{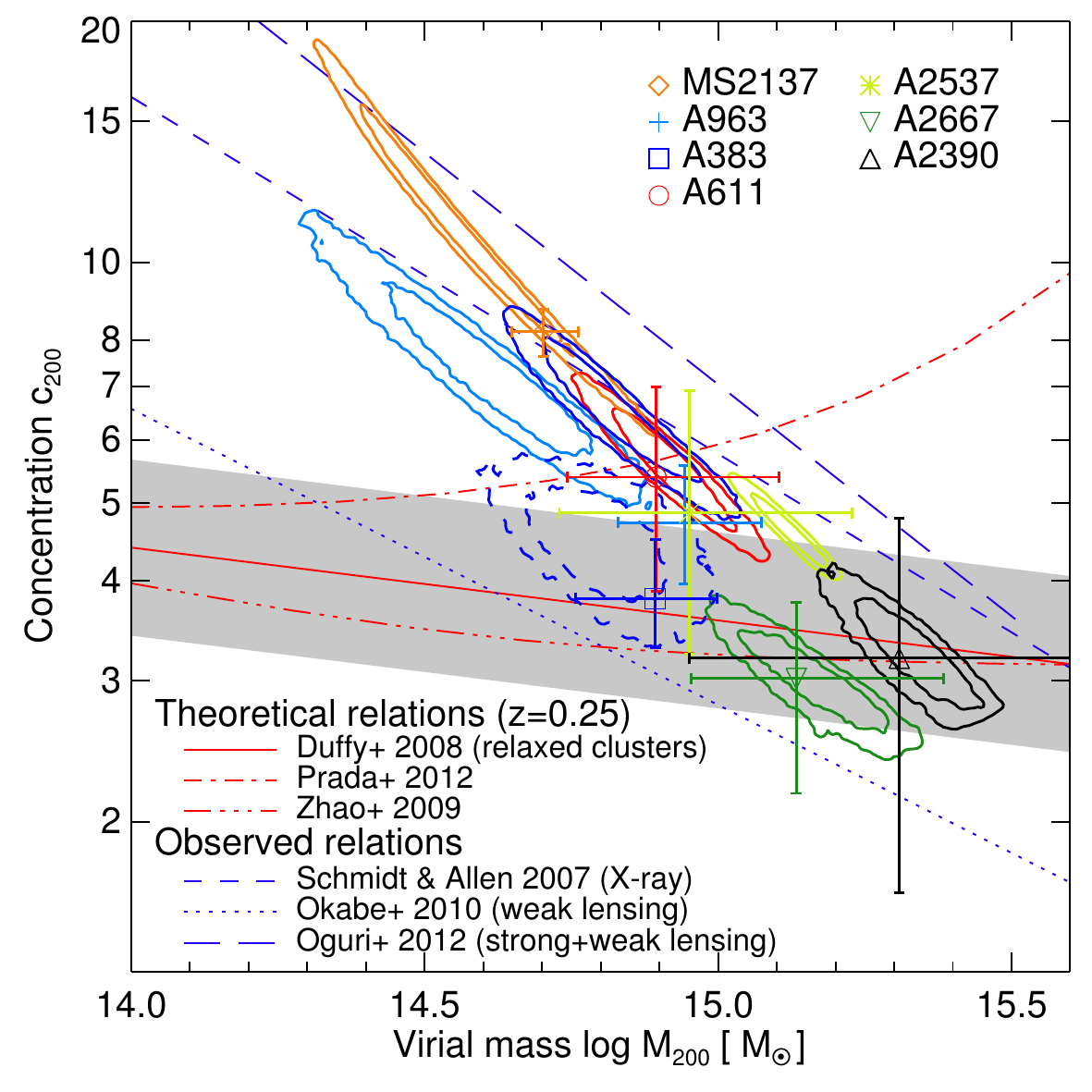}
\caption{Mass--concentration relation based on strong+weak lensing (contours; 68\% and 95\% confidence) and X-ray (points with marginalized $1\sigma$ error bars) analyses for the full sample. Empirical \citep{Schmidt07,Okabe10,Oguri12} and theoretical relations (\citealt{Prada12,Zhao09,Duffy08}, with shading indicating the $1\sigma$ scatter) are shown for comparison, standardized to the same overdensity. Dashed contours for A383 show the effect of adopting a prolate halo, which brings the lensing and X-ray measures into agreement (Section~\ref{sec:A383}). \label{fig:mass_concentration}}
\end{figure}

We now consider the physical implications of our results and compare our measured density profiles to recent simulations. After discussing the mass--concentration relation, we turn to evidence for a uniform total inner density slope and compare to both DM-only simulations and those that include baryons. We conclude with a discussion of the processes that may be responsible for establishing this observed uniformity.

\subsection{The mass--concentration relation}
\label{sec:massconc}

Figure~\ref{fig:mass_concentration} shows the mass--concentration relation for our sample, which was derived from NFW fits to the gravitational lensing data in Section~\ref{sec:losellip}. Halo concentrations are generally expected to vary inversely with mass, due to lower background densities at the later epochs in which more massive halos assemble \citep[e.g.,][]{Bullock01,Wechsler02}. The more massive clusters ($M_{200} \gtrsim 10^{15} \msol$) in our sample have concentrations in line with the predictions of most numerical simulations, although we note that current simulations do not have the necessary volume to provide good statistics in this regime. The exception is \citet{Prada12}, who surprisingly have reported an \emph{increasing} concentration at higher masses.\footnote{Figure~\ref{fig:mass_concentration} represents an extrapolation to higher masses than are contained in the simulations on which their model is calibrated.} However, as we move toward lower mass the concentrations become significantly higher than CDM simulations. MS2137, in particular, has a quite high concentration inferred from both lensing and X-ray measurements, which has long been recognized \citep{Gavazzi03}. The effect is to produce a significantly steeper slope in the mass--concentration relation compared to CDM simulations. Interestingly, the steep slope defined by our sample agrees well with measurements by \citet[][X-ray]{Schmidt07}, \citet[][weak lensing]{Okabe10}, and \citet[strong and weak lensing]{Oguri12}.

Lensing-based concentrations could potentially be biased high for two reasons. Firstly, projection effects can cause an upward bias if the major axis of the cluster is near the l.o.s. This is an unlikely to be a major effect in our sample given the overall good agreement between the lensing- and X-ray-based measures (Section~\ref{sec:losellip}). Secondly, more concentrated clusters -- particularly among the lower-mass systems -- are more likely to reach the critical surface density for forming multiple images, which is a necessary condition for entering our sample. Simulations of this potential bias suggest that the population of cluster lenses may have $\simeq 10\%-35\%$ higher concentrations on average \citep{Hennawi07,Fedeli07,Meneghetti10}, but that highest concentrations seen in MS2137 and other clusters \cite[e.g.,][]{Broadhurst08,Zitrin11b} are still not explained. Baryon cooling is also generally expected to increase cluster concentrations by only $\lesssim 20\%$ \citep[e.g.,][]{Duffy10,Mead10}. Larger samples of lenses \citep[e.g.,][]{CLASH} and close comparisons with X-ray observations \citep[e.g.,][]{Morandi10} should allow the significance of these trends to be verified or otherwise in the near future.

\subsection{The uniformity of the total inner mass distribution and comparison to simulations}
\label{sec:discinnerslope}

While the mass--concentration relation has a significant intrinsic scatter of $\sigma_{c_{200}} \simeq 25\%$ (\citealt{Neto07}, and higher when measured only in projection), the shape of the density profile is expected to be more uniform \citep[e.g.,][]{Gao12}. Thus, if the goal is a precise measure of the shape of the mass profile, i.e., its logarithmic slope, sample size is secondary to the density and radial extent of observational constraints. The combination of data sets we have presented provides precise constraints over the full range of radial scales, and thus forms an excellent basis for detailed study of the density profile, particularly in the inner regions.

\begin{figure*}
\centering
\includegraphics[width=0.49\linewidth]{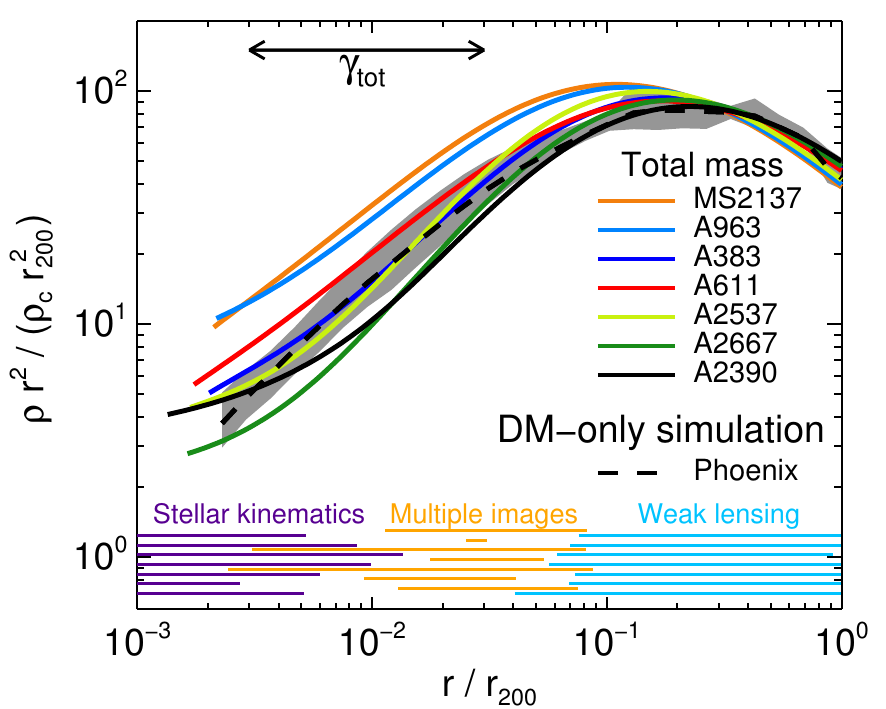} \hfill
\includegraphics[width=0.49\linewidth]{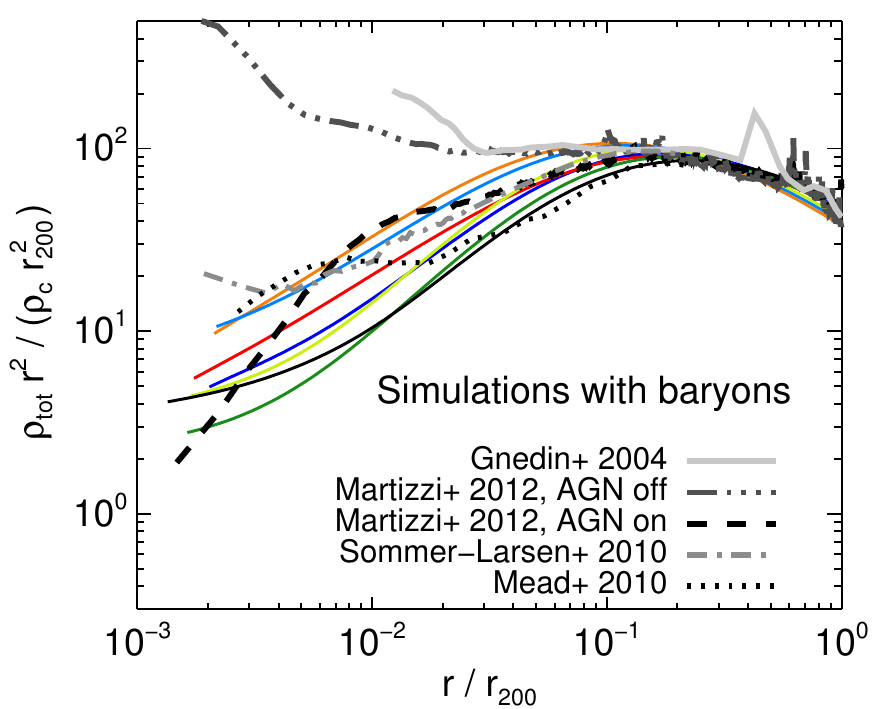}
\caption{\textbf{Left:} total scaled density profiles for the full sample (colored lines) are compared to simulated clusters -- containing only DM -- from the Phoenix project \citep{Gao12}. The dashed line shows the mean of the seven simulated Phoenix clusters, while the gray band outlines the envelope they define. Observed profiles are plotted down to 3~kpc. The radial range spanned by each data set is indicated at the bottom, and the interval over which $\betatot$ is defined is shown at the top of the panel. Note that the density has been multiplied by $r^2$ to reduce the dynamic range; thus, an isothermal slope $\rho \propto r^{-2}$ is horizontal. \textbf{Right:} the observed total density profiles (thin lines, as in left panel) are compared to several hydrodynamical simulations that include baryons, cooling, and feedback. The \citet{Gnedin04} results are taken from their Figure 2,  the \citet{SommerLarsen10} curves refer to their Coma ``Rz2'' simulation, and the \citet{Mead10} results are for their C4 simulation with cooling, star formation, and AGN feedback.\label{fig:gao}}
\end{figure*}

The slope of the total density profile at small radii is very similar within our sample (Section~\ref{sec:totdens}). In Figure~\ref{fig:gao} we compare the measured density profiles, scaled by the virial radius $r_{200}$, to recent numerical simulations. In the left panel these are overlaid on spherically-averaged density profiles from the Phoenix project \citep{Gao12}, the highest-resolution suite of $N$-body simulations of clusters to date. The typical convergence radius of $2.9~h^{-1}$~kpc is well matched to our observations, as is the mass range $M_{200} = 0.6 - 2.4 \times 10^{15}~h^{-1} \msol$. In the following comparisons we omit Phoenix-G and H, which are the latest clusters to assemble and remain in a unrelaxed state to $z = 0$, inconsistent with the properties of the observed sample. This leaves seven simulated clusters. The range of density profiles they span is illustrated by the gray band in the left panel of Figure~\ref{fig:gao}.

Remarkably, the observed \emph{total} density profiles closely parallel the Phoenix clusters that contain \emph{only dark matter}, despite the fact that the stellar mass in the BCG contributes noticeably within $\simeq 30$~kpc ($\simeq 0.02r_{200}$, comparable to $R_e$). Since our parametric models for the DM halo have the same large-radius behavior as the NFW profile, similar behavior at $r/r_{200} \gtrsim 0.3$ is guaranteed. At smaller radii, however, the agreement is not trivial, since it results from a combination of the concentration and inner slope ($r_s$ and $\beta$ or $b$) of the halo and the contribution of stellar mass ($\MLV$). The high concentrations of MS2137 and A963 cause them to appear shifted leftward of the Phoenix clusters in this plot, but even in these cases the \emph{slope} of the density profile is similar. The bottom of the panel indicates the radial intervals over when the models are constrained by the various data sets.\footnote{Minor ``wiggles'' appearing $r/r_{200} \approx 10^{-2}$ should not be overinterpreted given that we lack constraints  there and the mass model parameterization is simple.}

The similarity of the observations to DM-only simulations suggests that the net effect of adding baryons to the cluster core should mainly be to displace DM such that the total density does not change much at radii $\gtrsim 5-10$~kpc. In the right panel of Figure~\ref{fig:gao} we compare our results to several hydrodynamical simulations that include baryons, cooling, and star formation (\citealt{Gnedin04}, G04; \citealt{SommerLarsen10}, SL10; \citealt{Martizzi12}, M12). In general many such simulations suffer from a well known ``overcooling'' problem (see discussions and solutions in, e.g., G04, \citealt{McCarthy10,Puchwein10,Teyssier11}), in which the inability to suppress late cooling leads to the formation of far too much stellar mass at the cluster center. The build-up of baryons then leads to a significant contraction of the halo, increasing the central DM density. Thus in the G04 and M12 ``AGN off'' simulations, the central densities are much too high; even the density of DM alone (not plotted) exceeds the measured \emph{total}. SL10 estimated the effects of overcooling through an ad hoc simulation in which late-forming stars were slowly removed following $z = 2$ (their ``Rz2'' runs). This ameliorates the problem but still leaves a steeper total density slope than observed, with $\gamma_{\textrm{tot}} = 1.5$.

M12 performed a very high-resolution simulation that included feedback from an AGN. Interestingly, the AGN is effective not only at quenching late star formation but also at removing DM from the center. The latter is accomplished through several mechanisms that M12 discuss, including rapid fluctuations in the potential due to expulsion of gas during AGN outbursts.\footnote{This is similar to the mechanism suspected of producing cores in dwarf galaxies, fueled in that case by supernovae \citep{Pontzen12}.} The process is rather too effective, as it results in a 10~kpc stellar core that is much bigger than the largest observed example \citep{Postman12}. Still, this work points to a possibly important role for the supermassive black hole in shaping the small-scale DM distribution. \citet{Mead10} also found that the inclusion of AGN feedback results in more realistic total density profiles in their simulations. Although overcooling is significantly alleviated, the total density slope remains somewhat steeper ($\betatot = 1.7$) than we observe. We note that, except for SL10, the simulated clusters discussed here are less massive ($M_{200} \simeq 1-4 \times 10^{14} \msol$) than the observed sample. High-resolution simulations of more massive clusters are needed to make a more detailed comparison.

\begin{figure*}
\centering
\includegraphics[width=0.48\linewidth]{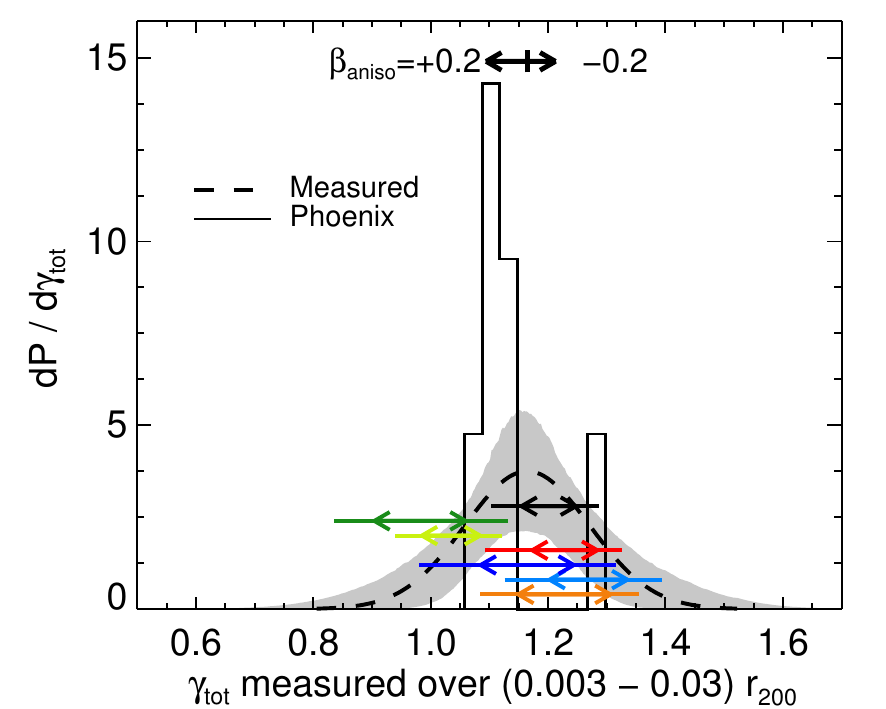} \hfill
\includegraphics[width=0.48\linewidth]{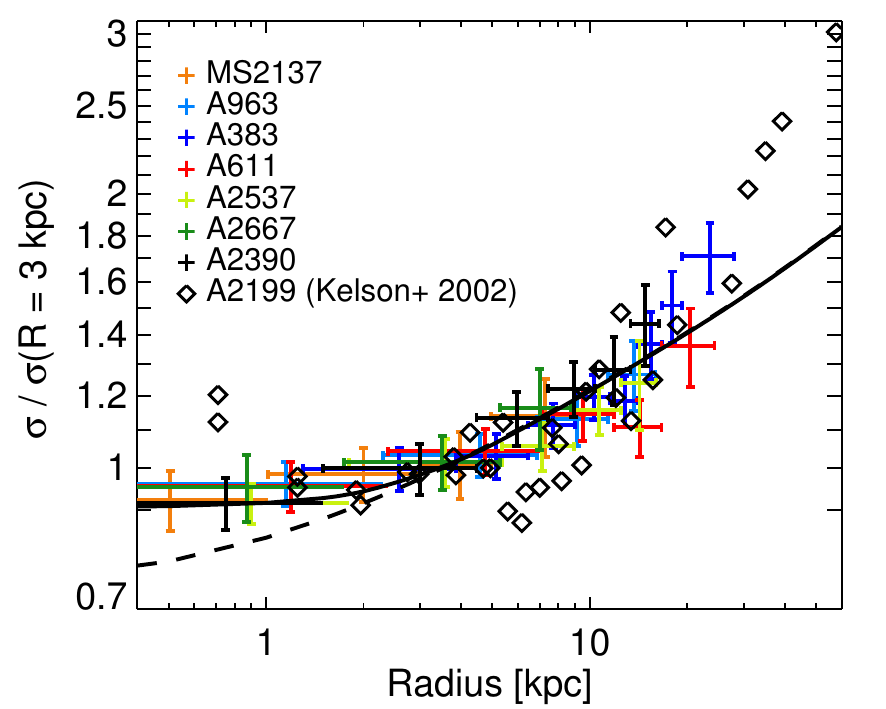}
\caption{\textbf{Left:}tThe total inner density slope $\betatot$ is shown for each cluster (colored lines, following the legend in Figure~\ref{fig:gao}) and compared to the slopes derived for the Phoenix clusters (DM-only simulations, black histogram). For the individual observed clusters, inner arrows and the full length of each line indicate the 68\% and 95\% confidence intervals, respectively. The inferred Gaussian parent population (Section~\ref{subsec:totslope}) is shown by the dashed curve, with the gray band showing the 68\% confidence region. The effects of mild orbital anisotropy on the observational results are illustrated at the top of the panel. \textbf{Right:} velocity dispersion data, as in Figure~\ref{fig:vdprof}, normalized for each cluster by $\sigma_{\textrm{los}}$ at $R = 3$~kpc by interpolating the measurements. Diamonds indicate measurements in NGC 6166, the BCG of A2199 ($z=0.03$), from \citet{Kelson02}. The solid line shown the mean of the mass models for the full sample, normalized in the same way, while the dashed line shows the same if observational effects (slit width and seeing) are excluded. \label{fig:vdprofcomp}}
\end{figure*}

Currently, the observed total density profiles appear more similar to those in CDM-only simulations than to results from hydrodynamical simulations, although the inclusion of AGN feedback in high-resolution simulations is producing much improved results. The similarity of the total density slope is quantified further in the left panel of Figure~\ref{fig:vdprofcomp}, which compares the $\betatot$ measurements of individual clusters, along with their inferred parent distribution (dashed; see Section~\ref{sec:totdens}), to the inner slopes of the Phoenix clusters defined in the same manner. The mean slope $\betatot = 1.13 \pm 0.02$ in the CDM-only Phoenix simulations agrees well with measured total density slope: $\langle \betatot \rangle = 1.16 \pm 0.05~\textrm{(random)}~{}^{+0.05}_{-0.07}~\textrm{(systematic)}$. The intrinsic scatter in $\betatot$ is possibly larger in the observations, but this cannot be asserted with much certainty due to the systematic limitations discussed in Section~\ref{sec:systematics}. We stress again that these are DM-only simulations and that their relevance to the total mass in real clusters over this range of radii is surprising. Other lensing \citep[e.g.,][]{Umetsu11,Morandi11,Zitrin11,Morandi12} and X-ray \citep{Lewis03,Zappacosta06} studies have also concluded that the total density is remarkably close to an NFW profile; we review these further in Paper~II.

The uniformity of the total inner mass distribution is further supported by the striking homogeneity in the shapes of the velocity dispersion profiles. The right panel of Figure~\ref{fig:vdprofcomp} plots these profiles normalized to the observed dispersion at $R = 3$~kpc. With this single scaling, the velocity dispersion profiles for all seven clusters are mutually consistent within their uncertainties. In this figure we also compare to the BCG of the nearby cluster A2199 ($z = 0.03$; \citealt{Kelson02}). Where the data overlap they are consistent with our sample, except at $\lesssim 1$~kpc where the black hole is probably dynamically significant (note that we cannot resolve these scales due to the slit width and seeing).\footnote{The \citet{Kelson02} data are higher than our models at radii $\gtrsim 25$~kpc, beyond the outer limits of our velocity dispersion measures but within the range of $\simeq 30-100$~kpc where strong-lensing constrains the mass. This could indicate that the dynamical structure becomes less homogeneous near $R_e$.}  Although rising $\sigma$ profiles in cD galaxies have been observed since \citet{Dressler79}, there has been some uncertainty \citep[e.g.,][]{Fisher95,Dubinski98,Carter99,Hau04} about the frequency of this phenomenon, which is the expected response of stars to the central cluster potential. Our observations suggests that it is ubiquitous in BCGs that are well aligned with the centers of relaxed clusters.

\subsection{A physical picture}
\label{sec:picture}
In $\Lambda$CDM-based models the formation of BCGs is expected to occur relatively late and be dominated by dry (dissipationless) merging \citep[e.g.,][]{DeLucia07}. Since NFW profiles are simply the product of collisionless collapse and merging, one interpretation of our findings is that the processes that set the inner density profile in clusters are primarily gravitational. Understanding how the total density profile remains similar to that expected of CDM alone is not trivial. \citet{Loeb03} and \citet{Gao04} hypothesized that repeated merging might drive the total collisionless (stars and DM) density toward an NFW-like profile, noting that this could solve two puzzles: the lack of very high-dispersion galaxies with $\sigma_e \gtrsim 400$~km~s${}^{-1}$, and our own earlier observations that the DM density profile is shallower than the NFW form in cluster cores \citep{Sand02,Sand04}. As a starting point, based on both analytic arguments and CDM simulations, they showed that the mass in the central regions of present-day massive clusters changes very little at $z \lesssim 6$, but the identity of these particles changes considerably. The particles arriving in mergers displace those already present, maintaining the central density.

In reality we expect the progenitors of the BCG and the infalling galaxies to have been compressed due to baryon loading (e.g., \citealt{Blumenthal86}, and see references in Section~1). Indeed, the total density profiles we derive do appear to steepen slightly in the inner $\simeq 5-10$~kpc. In \PaperII, we show that this is the regime in which the stellar density exceeds that of DM. Although it is difficult to pinpoint this scale precisely, it is certainly well within the present effective radius (median $\langle R_e \rangle = 34$~kpc), where stars begin to contribute non-negligibly to the total density. Furthermore, this scale bears a striking similarity to sizes of the most massive galaxies at high redshift, which many observations now indicate are quite compact \citep[e.g.,][]{Trujillo06,vanDokkum08,Williams10,Newman12}. For example, a simple extrapolation of the observed stellar mass--size relation at $z \approx 2.5$ \citep{Newman12} would yield a size of $R_e \sim 2-6$~kpc for likely progenitors.\footnote{We caution that low-$z$ BCGs do not lie on a simple extrapolation of the trend defined by lower-mass ellipticals \citep[e.g.,][]{vonderLinden07}, but the situation for the very most massive galaxies at high-$z$ is uncertain due to the small volumes surveyed.} Indirect support for this comes from our observation (Figure~\ref{fig:inner_mass_ratio}) that the mass contained in the inner 5~kpc -- mostly stars, but only a small fraction of the stars in the present BCG -- is correlated with the mass of the cluster core within 100~kpc, which is also expected to be in place by $z \approx 3$ and change relatively little subsequently (G04, Figure 1).
Interestingly, color gradients in BCGs (when present) occur mostly at $R \gtrsim 10$~kpc, while the innermost regions are more homogeneous in both color and luminosity \citep{Postman95,Bildfell08}.

This suggests a picture in which stars in the innermost $\simeq 5-10$~kpc are formed early within the BCG progenitor, where dissipation establishes a steep stellar density profile, while subsequent dry merging of infalling satellites mostly adds stars to the outer regions of the BCG in a manner that nearly maintains the total density. This requires that the stars and DM arriving in mergers displace a roughly equal amount of existing DM. Simulations indeed indicate that stars arriving in minor (low mass ratio) mergers, which dominate the accretion history of very massive galaxies, are primarily added to the outskirts of the BCG \citep[e.g.,][]{Naab09,Laporte12}. However, the precise effect of these mergers on the DM already in place is not clear. Using dissipationless $N$-body simulations, several authors have shown that dynamical friction of the infalling satellites on the halo can ``heat'' the cusp and reduce the central DM density \citep[e.g.,][and see references in Section~\ref{sec:intro}]{ElZant04,Nipoti04}, and that this can more than overcome the deeper central potential that results from the central build-up of baryons. This process is sensitive to the nature of the satellites \citep[e.g.,][]{Ma04,Jardel09}, and a fully realistic treatment has been lacking in cluster simulations to date. Satellites will bring in their own DM, counteracting this central depletion. Tightly bound galaxies are more effective, since they are more resistant to stripping and so survive longer. \citet{Laporte12} point out that the compact stellar configuration observed in high-$z$ massive galaxies is significant in this context, while \citet{Martizzi12} show that infalling central black holes are also important in their simulations.

In this scheme there is little room for additional contraction or steepening of the mass profile, and the relevant physics is primarily dissipationless. In contrast, a major focus in the theoretical literature has been the ``adiabatic contraction'' (AC) formalism \citep{Blumenthal86} and its modified versions \citep[e.g.,][]{Gnedin04,Gnedin11}, which predict the \emph{steepening} of the inner DM profile resulting from the slow cooling and central condensation of baryons. In contrast to the scheme described above, in which the orbital energy lost by infalling stellar clumps is transferred to the halo, the energy lost by baryons is radiated and lost from the system; thus, the AC model emphasizes the role of \emph{dissipation} in forming the BCG \citep[e.g.,][]{Lackner10}. This model takes no account of mergers at all. The scenario advanced above argues that while one may be able to fit the results of simulations by introducing additional parameters to the AC model, this does not necessarily mean that the most relevant underlying physics are being accurately described \citep[e.g.,][]{Duffy10}. The relevance of AC in describing the results of earlier cosmological simulations with gas probably reflects the known overcooling problem that cause the effects of dissipation to be overstated. As discussed in Section~\ref{sec:discinnerslope}, the inclusion of AGN appears to solve most of this problem and may additionally lower the central DM density.

Given that several baryonic mechanisms may play a role in altering the small-scale DM distribution (contraction from gas cooling, dynamical friction from infalling clumps, potential fluctuations due to AGN-driven gas outflows), continually improving simulations will be essential to better understand their relative importance, and our observations will provide a basis for detailed comparison. \citet{Martizzi12b} recently coupled a full cosmological cluster simulation with idealized simulations in order to isolate the most important physics for setting the inner DM profile. They found that dynamical friction from satellites initially flattens the DM cusp. Contraction from gas cooling becomes important at later times. At still later epochs, when the black hole is sufficiently massive, its effect on the central DM through fluctuations in the gas density is the dominant one.

An alternative possibility is that the central DM density is reduced relative to CDM simulations due to the nature of the DM particle. For example, a small self-interaction cross-section could produce moderately shallower DM density cusps \citep{Spergel00,Yoshida00,Rocha12,Peter12}, which could then leave room for baryons to boost the total density to an NFW-like profile. We discuss the likelihood of these scenarios further in Paper~II, where we measure the inner DM density slope. 

The results we present for BCGs are quite different from observations of massive field ellipticals, which uniformly show a total density slope within their effective radii that is nearly isothermal ($\rho_{\textrm{tot}} \propto r^{-2}$; \citealt{Treu06,Koopmans09,Auger09}).
The massive halos we consider are much less efficient at converting their baryons into stars \citep[e.g.,][]{Guo10,Leauthaud12}.
As a consequence, BCGs are much more DM-dominated, so it is not surprising that dissipation would play a lesser role in their formation. The greater importance of minor mergers in their assembly may also be important \citep{Naab09}. Thus, our results do not directly conflict with studies claiming that adiabatic contraction may be significant in lower-mass ellipticals \citep[e.g.,][]{Dutton11,Sonnenfeld12}. They do show if the currently discussed prescriptions for halo contraction are valid, they have a limited range of applicability that likely varies with star formation efficiency and assembly history. Although the isothermal and NFW limits have often been discussed as special configurations in the literature, we should be able to see an intermediate case in galaxy groups. Indeed, this may have already be observed by \citet{Spiniello11} and \citet{Sonnenfeld12}.

Finally, we emphasize that these results are fully consistent with our previous claims that the DM density slope in cluster cores is shallower than an NFW profile \citep{Sand02,Sand04,Sand08,N09,N11}, given that the subtraction of stellar mass from an NFW-like profile must yield one with a shallower inner slope. They also explain previously reported discrepancies between our results and independent analyses in the same clusters that are confined to radii where the stellar mass is negligible ($r \gtrsim R_e$, e.g., \citealt{Morandi12}). Quantifying the DM profile requires techniques for accurately separating the stellar and dark mass. We describe these in \PaperII.

\section{Summary}

We presented observations of a sample of seven massive, relaxed galaxy clusters at $\langle z \rangle = 0.25$. The data comprise 25 multiply imaged sources (21 with spectroscopic redshifts, of which 7 are original to this work) that present 80 images, weak-lensing constraints from multi-color imaging, and spatially resolved stellar kinematics within the BCGs. Taken together, these data from the \HST, Subaru, and Keck telescopes extend from $\simeq 0.002 r_{200}$ to beyond the virial radius, providing detailed constraints on the global mass distribution.

\begin{enumerate}
\item We find that the clusters in our sample are not strongly elongated along the l.o.s.~(except A383) and that their intracluster media are close to equilibrium, based on the agreement between mass profiles derived from independent lensing and X-ray observations (Section~\ref{sec:losellip}).

\item Physically motivated and simply parameterized models provide good fits to the full range of data. The inner logarithmic slope of the total density profile measured over $r/r_{200} = 0.003-0.03$ (on average, $5 - 55$~kpc) is remarkably uniform, with $\langle \betatot \rangle = 1.16 \pm 0.05~\textrm{(random)}~{}^{+0.05}_{-0.07}~\textrm{(systematic)}$ and an intrinsic scatter $\sigma_{\gamma} = 0.10^{+0.06}_{-0.04}$ ($\sigma_{\gamma} < 0.13$ at 68\% confidence).

\item Supporting the uniformity of the inner mass distribution, the extended stellar velocity dispersion profiles show a clear rise with radius and display a very homogeneous shape after a single scaling.

\item The shape of the \emph{total} density profile is in surprisingly good agreement with high-resolution simulations containing only CDM, despite a significant contribution of stellar mass within the BCG over the scales we measure. Hydrodynamical simulations including baryons, cooling, and feedback currently provide  poorer descriptions, although the inclusion of AGN in recent high-resolution simulations has resulted in a major improvement.

\item Our findings support a picture in which an early dissipative phase associated with star formation in the BCG progenitor establishes a steeper total density profile in the inner $\approx 5-10$~kpc -- comparable to the size of very massive, red galaxies at $z > 2$ -- while subsequent accretion of stars (still within the present effective radius) mostly replaces DM so that the total density is nearly maintained. 

\item These results are fully consistent with our earlier claims that the slope of the \emph{dark matter} profile is shallower than NFW-type cusp in the innermost regions of clusters \citep{Sand02,Sand04,Sand08,N09,N11}. In \PaperII~we turn to separating the dark and baryonic mass profiles.

\end{enumerate}

\acknowledgments
It is a pleasure to acknowledge insightful conversations with Annika Peter. We thank Liang Gao, Davide Martizzi, and Jesper Sommer-Larsen for providing tabulated results from their simulations. The anonymous referee is thanked for a helpful report.
R.S.E.~acknowledges financial support from DOE grant DE-SC0001101. Research support by the Packard Foundation is gratefully acknowledged by T.T. J.R.~is supported by the Marie Curie Career Integration Grant 294074.
Some of the data presented herein were obtained at the W. M. Keck Observatory, which is operated as a scientific partnership among the California Institute of Technology, the University of California and the NASA. The Observatory was made possible by the generous financial support of the W.M. Keck Foundation. The authors recognize and acknowledge the cultural role and reverence that the summit of Mauna Kea has always had within the indigenous Hawaiian community. We are most fortunate to have the opportunity to conduct observations from this mountain.
Based in part on observations made with the NASA/ESA \emph{Hubble Space Telescope}, obtained from the data archive at the Space Telescope Science Institute. STScI is operated by the Association of Universities for Research in Astronomy, Inc.~under NASA contract NAS 5-26555.
This research has made use of data obtained from the Chandra Data Archive and software provided by the Chandra X-ray Center (CXC). 
Based in part on data collected at Subaru Telescope, which is operated by the National Astronomical Observatory of Japan.

\clearpage
\bibliographystyle{apj}
\bibliography{paper1}

\clearpage
\begin{appendix}

\begin{deluxetable*}{lccccllccccl}
\tablecaption{Positions of multiple images\label{tab:slimages}}
\tablecolumns{12}
\tablewidth{\linewidth}
\tablehead{\colhead{Cluster} & \colhead{Image} & \colhead{$\Delta x$} & \colhead{$\Delta y$} & \colhead{$z_{\textrm{spec}}$} & \colhead{Source} & \colhead{Cluster} & \colhead{Image} & \colhead{$\Delta x$} & \colhead{$\Delta y$} & \colhead{$z_{\textrm{spec}}$} & \colhead{Source}}
\startdata
A611 & 1.1 & $13.0$ & $17.5$ & 1.49 & B12 & A2667 & 1.1 & $-4.0$ & $14.8$ & 1.0334 & S05 \\
\ldots & 1.2 & $-14.7$ & $-5.5$ & \ldots & \ldots & \ldots & 1.2 & $-8.3$ & $11.3$ & \ldots & \ldots \\ 
\ldots & 1.3 & $-12.7$ & $5.6$ & \ldots & \ldots & \ldots & 1.3 & $-16.2$ & $-0.4$ & \ldots & \ldots \\
\ldots & 1.4 & $3.2$ & $-8.9$ & \ldots & \ldots & \ldots & 2.1 & $-5.8$ & $13.8$ & \ldots & \ldots \\
\ldots & 1.5 & $2.2$ & $-6.6$ & \ldots & \ldots & \ldots & 2.2 & $-7.0$ & $12.9$ & 1.0334 & \ldots \\
\ldots & 2.1 & $-1.5$ & $16.0$ & 0.908 & R10 & \ldots & 2.3 & $-16.6$ & $-0.5$ & \ldots & \ldots \\
\ldots & 2.2 & $-10.9$ & $11.0$ & \ldots & \ldots & \ldots & 3.1 & $-11.6$ & $-9.0$ & --- & \ldots \\
\ldots & 2.3 & $-15.7$ & $3.4$ & \ldots & \ldots & \ldots & 3.2 & $-7.6$ & $-0.4$ & \ldots & \ldots \\
\ldots & 3.1 & $3.2$ & $15.0$ & --- & \ldots & \ldots & 3.3 & $14.8$ & $18.8$ & \ldots & \ldots \\
\ldots & 3.2 & $-2.0$ & $14.3$ & \ldots & \ldots & \ldots & 4.1 & $-11.4$ & $-16.4$ & --- & \ldots \\
\ldots & 3.3 & $-18.7$ & $-11.2$ & \ldots & \ldots & \ldots & 4.2 & $17.4$ & $13.2$ & \ldots & \ldots \\ 
\ldots & 3.4 & $7.6$ & $-3.3$ & \ldots & \ldots & \ldots & 4.3 & $5.3$ & $-7.4$ & \ldots & \ldots \\
A383 & 1.1 & $-1.5$ & $2.5$ & 1.01 & S04 & \ldots & 4.4 & $2.6$ & $-4.8$ & \ldots & \ldots \\
\ldots & 1.2 & $-0.9$ & $1.3$ & \ldots & \ldots & A2537 & 1.1 & $35.6$ & $11.9$ & 2.786 & This work \\
\ldots & 1.3 & $16.2$ & $-4.7$ & \ldots & \ldots & \ldots & 1.2 & $38.0$ & $7.2$ & \ldots & \ldots \\
\ldots & 2.1 & $6.9$ & $-14.0$ & 1.01 & S01 & \ldots & 1.3 & $14.3$ & $38.6$ & \ldots & \ldots \\
\ldots & 2.2 & $8.2$ & $-13.2$ & \ldots & \ldots & \ldots & 2.1 & $35.4$ & $12.8$ & 2.786 & This work \\
\ldots & 2.3 & $14.1$ & $-8.2$ & \ldots & \ldots & \ldots & 2.2 & $38.4$ & $6.4$ & \ldots & \ldots \\
\ldots & 3.1 & $14.6$ & $-14.7$ & 2.58 & N11, B12 & \ldots & 2.3 & $16.6$ & $37.3$ & \ldots & \ldots \\
\ldots & 3.2 & $16.5$ & $-14.4$ & \ldots & \ldots & \ldots & 3.1 & $-15.4$ & $-3.9$ & 1.970 & This work \\
\ldots & 3.3 & $5.8$ & $-22.0$ & \ldots & \ldots & \ldots & 3.2 & $11.3$ & $12.6$ & \ldots & \ldots \\
\ldots & 4.1 & $8.2$ & $-22.0$ & 2.58 & N11, B12 & \ldots & 3.3 & $-13.6$ & $28.5$ & \ldots & \ldots \\
\ldots & 4.2 & $17.4$ & $-17.3$ & \ldots & \ldots & \ldots & 3.4 & $16.7$ & $-24.8$ & \ldots & \ldots \\
\ldots & 4.3 & $17.9$ & $-15.5$ & \ldots & \ldots & \ldots & 3.5 & $-0.6$ & $1.0$ & \ldots & \ldots \\
\ldots & 5.1 & $1.6$ & $10.2$ & 6.027 & R11 & \ldots & 4.1 & $-22.6$ & $8.4$ & 3.607 & This work \\
\ldots & 5.2 & $-18.3$ & $-13.5$ & \ldots & \ldots & \ldots & 4.2 & $-19.0$ & $21.3$ & \ldots & \ldots \\
\ldots & $6^{\dagger}$ & 0.3 & $-14.6$ &1.826 & This work & \ldots & 4.3 & $0.0$ & $7.3$ & \ldots & \ldots \\
MS2137 & 1a & $2.6$ & $14.9$ & 1.501 & S02 & \ldots & 4.4 & $6.4$ & $15.3$ & \ldots & \ldots \\
\ldots & 1b & $-5.2$ & $13.7$ & \ldots & \ldots & \ldots & 4.5 & $17.7$ & $-33.1$ & \ldots & \ldots \\
\ldots & 1c & $-11.9$ & $-15.3$ & \ldots & \ldots & A2390 & 41a & $-4.8$ & $10.0$ & --- & \ldots \\
\ldots & 1d & $13.6$ & $-1.1$ & \ldots & \ldots & \ldots & 41b & $-3.4$ & $8.5$ & \ldots & \ldots \\
\ldots & 2a & $0.1$ & $6.8$ & 1.502 & S02 & \ldots & 51a & $-5.3$ & $-6.8$ & 0.535 &This work \\
\ldots & 2b & $-7.2$ & $-22.5$ & \ldots & \ldots & \ldots & 51b & $-8.7$ & $0.3$ & \ldots & \ldots \\
\ldots & 2c & $0.5$ & $3.3$ & \ldots & \ldots & \ldots & 51c & $-9.3$ & $1.3$ & \ldots & \ldots \\
\ldots & 3a & $4.7$ & $14.7$ & 1.501 & S02 & \ldots & B1 & $-9.1$ & $-9.9$ & 1.036 & This work\\
\ldots & 3b & $-11.7$ & $-15.0$ & \ldots & \ldots & \ldots & B2 & $-2.3$ & $-15.5$ & \ldots & \ldots \\
\ldots & 3c & $13.7$ & $-2.2$ & \ldots & \ldots & \ldots & H32a & $44.8$ & $19.7$ & 4.05 & P99 \\
\ldots & 3d & $-7.4$ & $12.7$ & \ldots & \ldots & \ldots & H32b & 49.5 & 9.4 & \ldots & \ldots \\
A963 & NA${}^{\dagger\dagger}$ & $-0.55$ & $12.18$ & 0.771 & E91 & \ldots & H32c & 46.4 & 13.5 & \ldots & \ldots \\
 & & & & & & \ldots & H51a & 20.0 & 4.0 & 4.05 & P99 \\
 & & & & & & \ldots & H51b & 24.8 & $-9.9$ & \ldots & \ldots \\
 & & & & & & \ldots & H51c & $-5.7$ & 32.9 & \ldots & \ldots
\enddata
\tablecomments{Positions are given relative to the BCG in arcseconds, with $\Delta x > 0$ and $\Delta y > 0$ representing offsets to the west and the north, respectively. ``---'' indicates that no spectroscopic redshift is available. ${}^{\dagger}$ Not used as a constraint; see Section~\ref{sec:A383lensing}. ${}^{\dagger\dagger}$ Location of break point used to constrain critical line position. Sources: E91: \citet{Ellis91}, P99: \citet{Pello99}, R10: \citet{Richard10}, R11: \citet{Richard11}, S01: \citet{Smith01}, S02: \citet{Sand02}, S04: \citet{Sand04}, S05: \citet{Sand05}, N11: \citet{N11}, B12: S.~Belli et al., in preparation.}
\end{deluxetable*}

\begin{figure}
\centering
\includegraphics[width=0.41\linewidth]{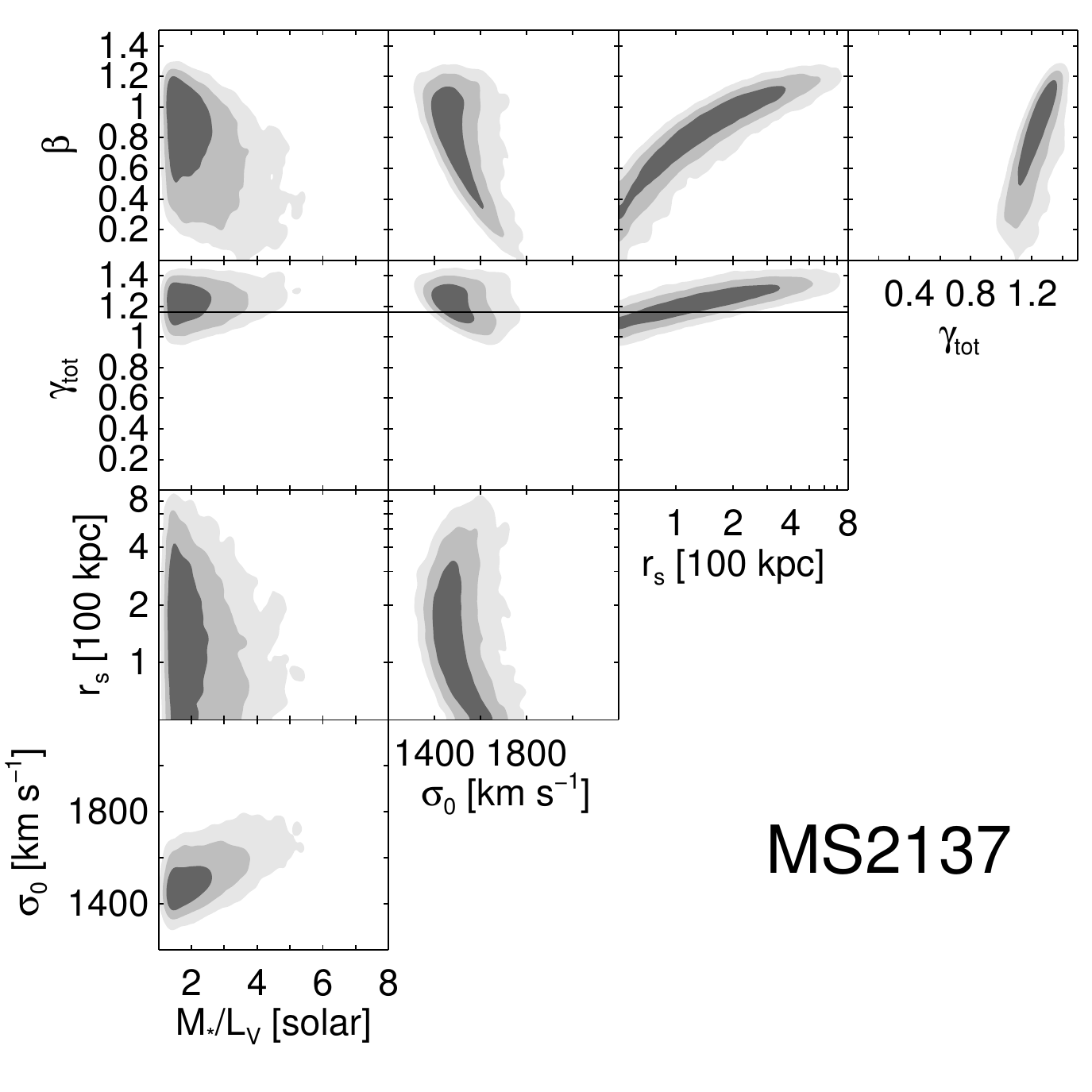}
\includegraphics[width=0.41\linewidth]{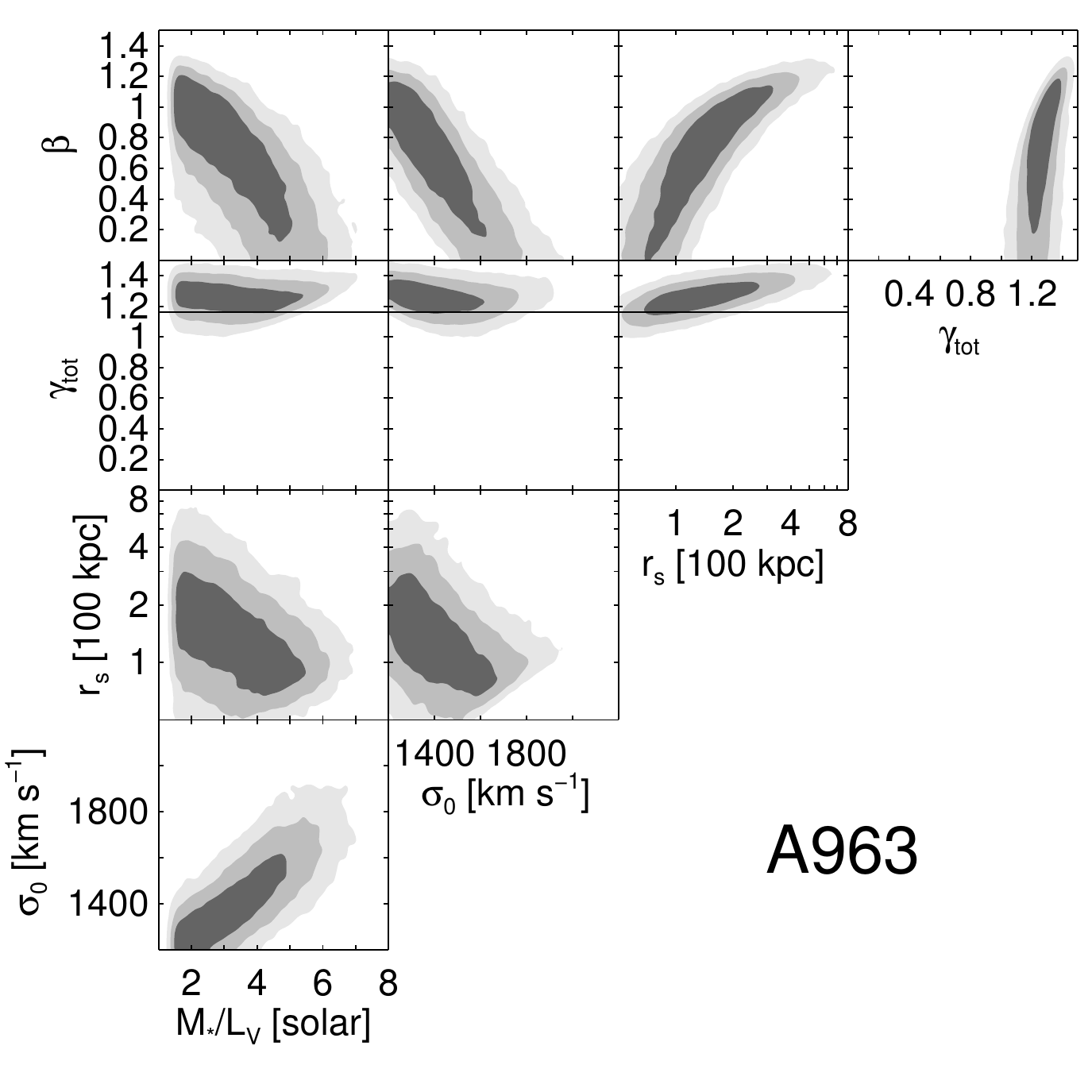} \\
\includegraphics[width=0.41\linewidth]{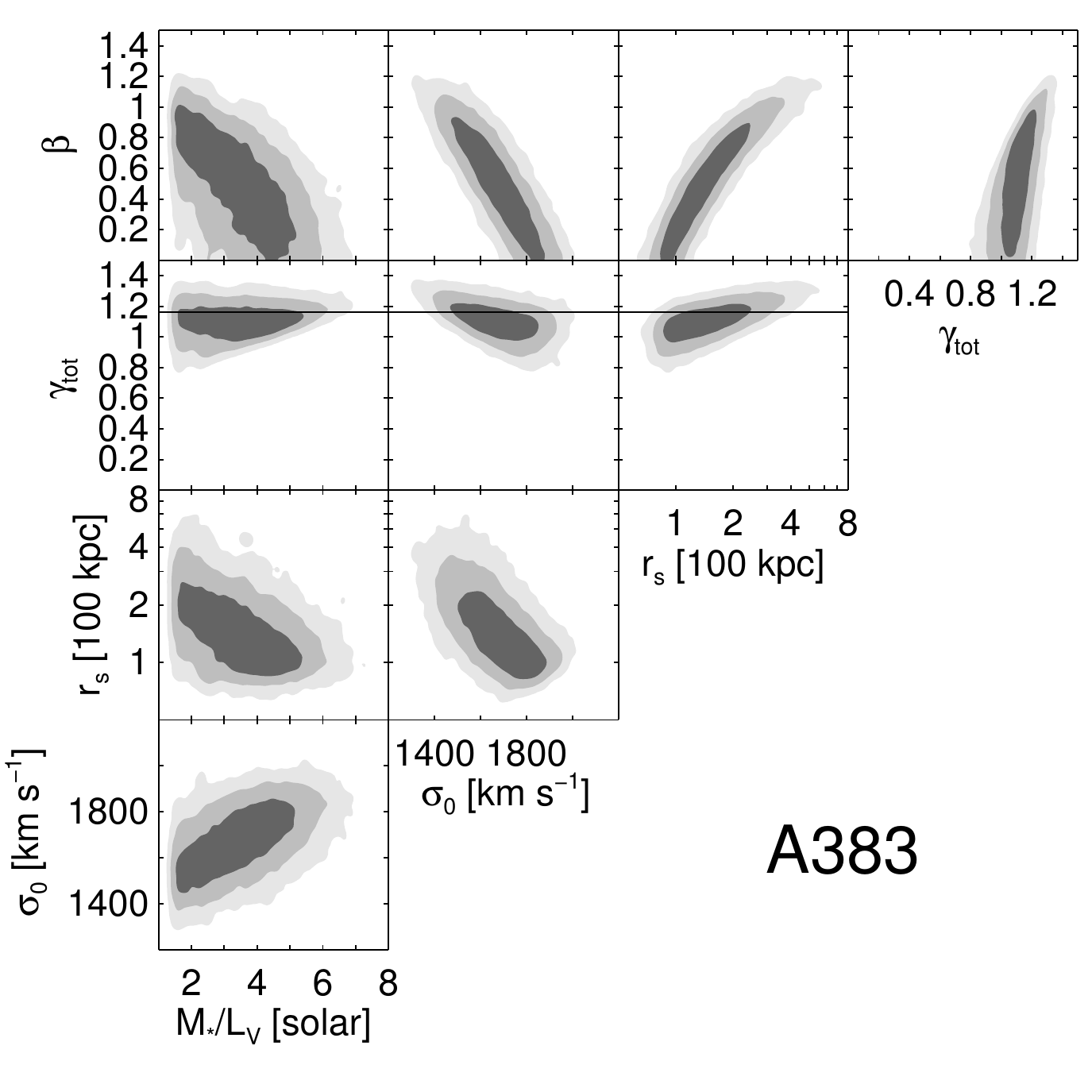} 
\includegraphics[width=0.41\linewidth]{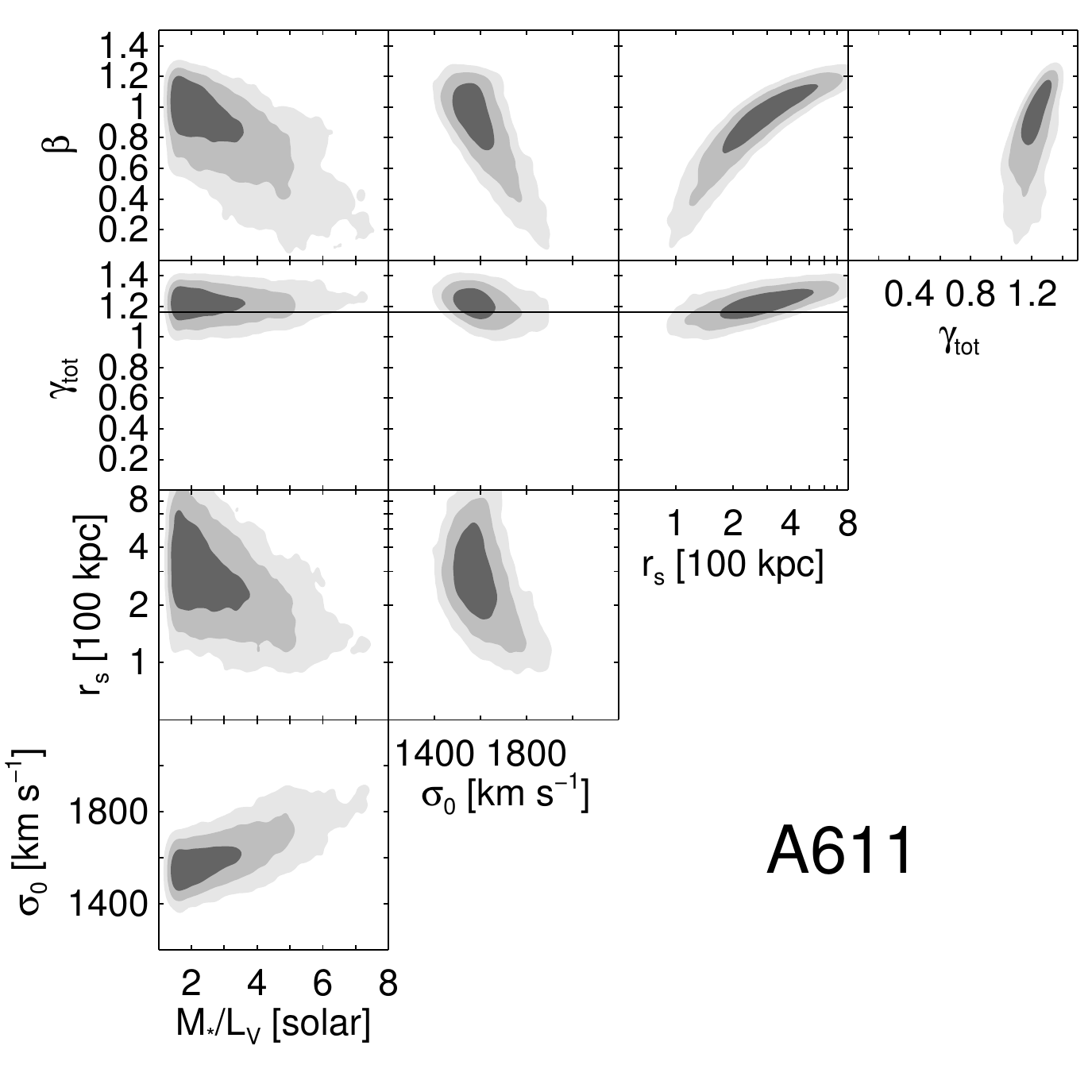} \\
\includegraphics[width=0.41\linewidth]{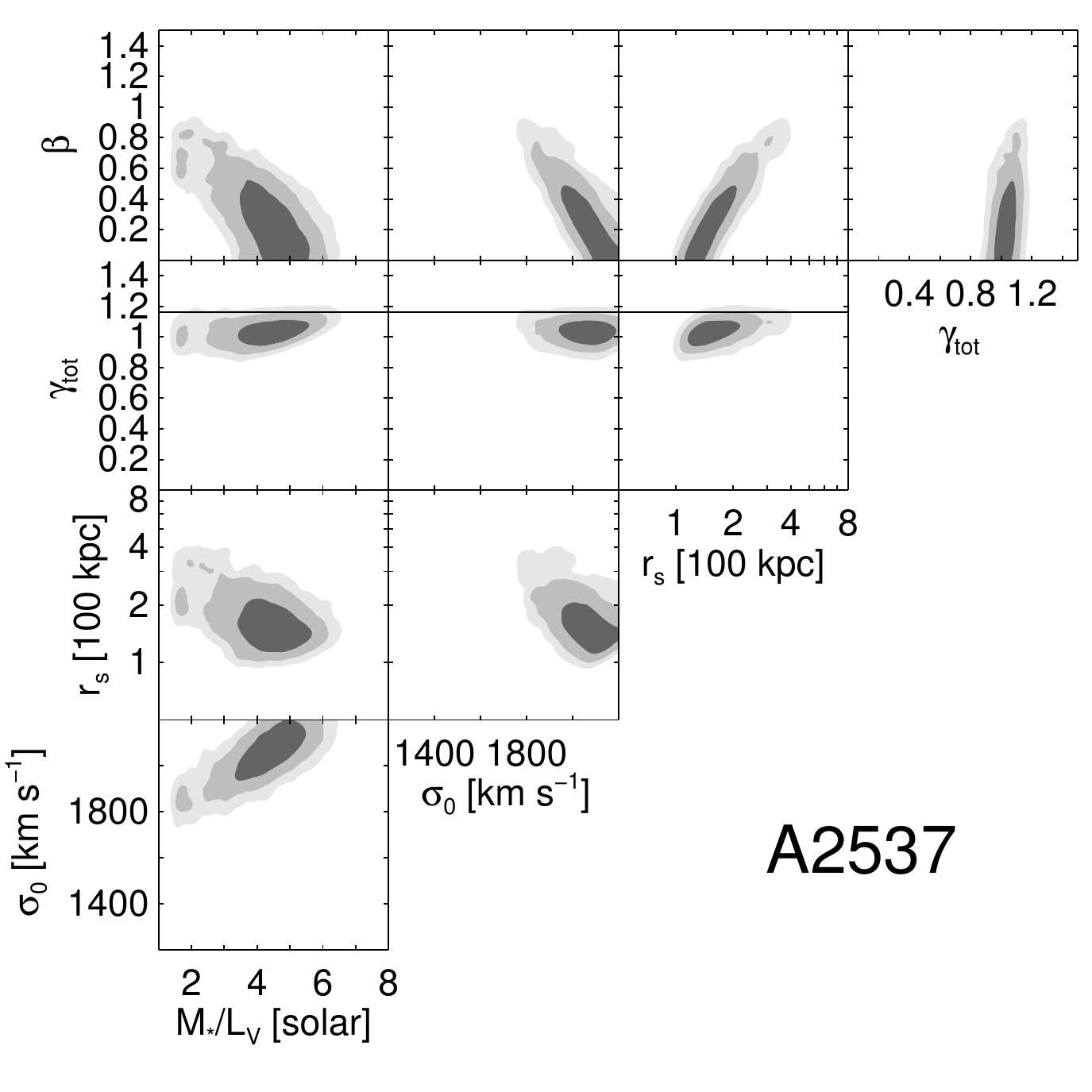}
\includegraphics[width=0.41\linewidth]{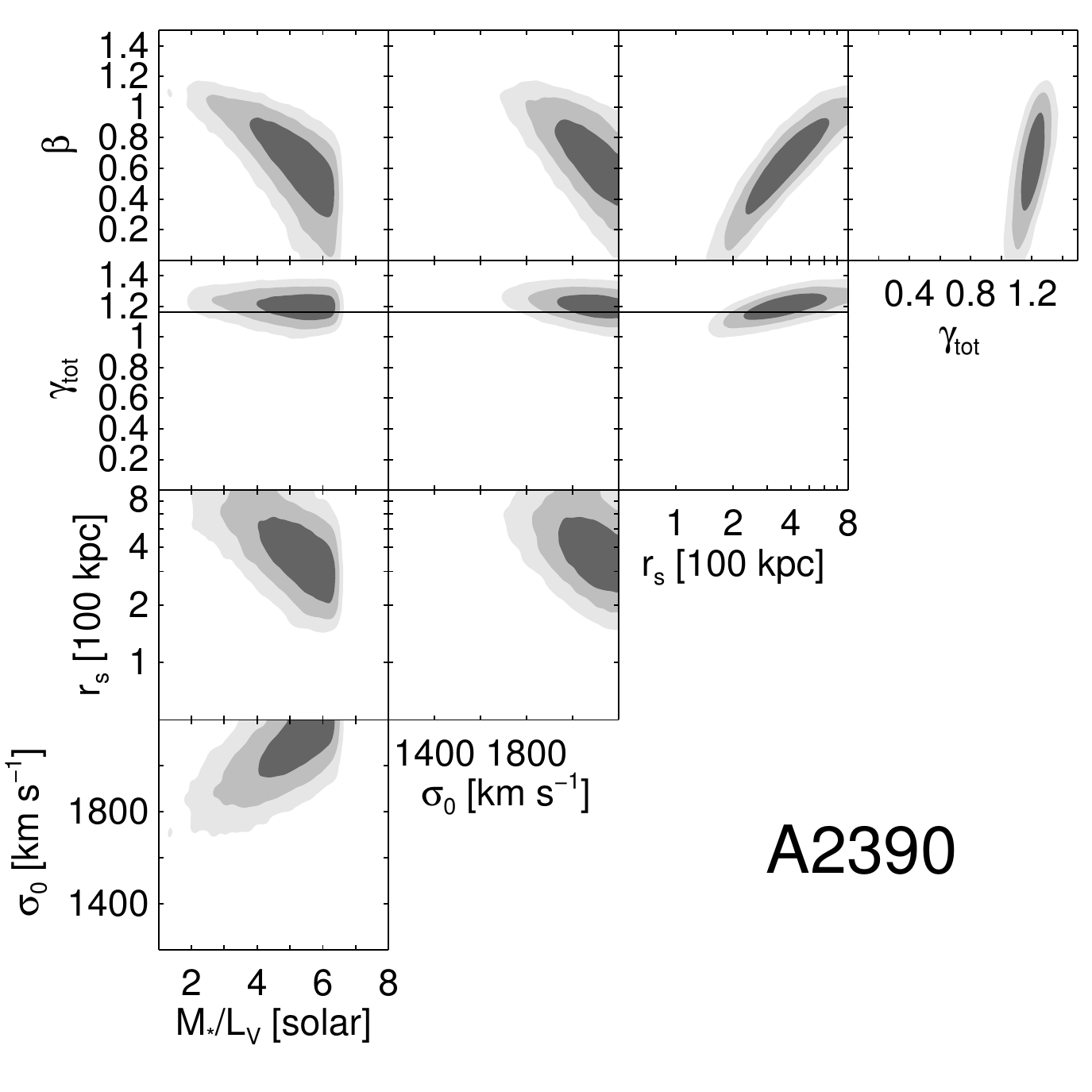}
\caption{Degeneracies among key parameters describing the radial density profile for gNFW-based fits; A2667 is shown in Figure~\ref{fig:degen_example}. Contours indicate the $68\%$, $95\%$, and $99.5\%$ confidence regions. Note that $\gamma_{\textrm{tot}}$ is derived from the other parameters and is not independent. The horizontal line indicates the mean $\gamma_{\textrm{tot}}$ among the whole sample. In A383, $\sigma_0$ and $r_s$ follow Equation~1 of \citet{N11} and have not been ``sphericalized'' using the measured line-of-sight ellipticity $q_{\textrm{DM}}$ (see Section~\ref{sec:A383}). \label{fig:fulldegen}}
\end{figure}

Table~\ref{tab:slimages} lists the angular positions and redshifts of the multiply imaged sources used in our strong-lensing analysis. Figure~\ref{fig:fulldegen} illustrates the main parameter degeneracies for all clusters in the sample except A2667, which was shown in Figure~\ref{fig:degen_example}.

\clearpage

\end{appendix}
\end{document}